\documentclass[aps,pra,notitlepage,reprint,floatfix,amsmath,amssymb]{revtex4-1}

\usepackage{siunitx}
\usepackage{xcolor}
\usepackage{graphicx}
\graphicspath{{figures/}}         

\usepackage[caption=false]{subfig} 
\usepackage{hyperref}
\usepackage{bbm}
\usepackage{mathrsfs}

\DeclareMathOperator{\ind}{\mathbbm{1}}

\newcommand{\Vie}[3]{\mathop{\mathbf{#1}_{#2}^{#3}}}
\newcommand{\Cie}[3]{\mathop{\mathcal{#1}_{#2}^{#3}}}

\let\Re\relax\DeclareMathOperator{\Re}{\mathrm{Re}}
\let\Im\relax\DeclareMathOperator{\Im}{\mathrm{Im}}

\begin{document}

\title{Single scattering of polarized light by correlated surface and volume disorder}
\author{J.-P. Banon$^{1,2}$}
\author{I. Simonsen$^{2,3}$}
\author{R. Carminati$^1$}
\affiliation{$^1$Institut Langevin, ESPCI Paris, CNRS, PSL University, 1 rue Jussieu, F-75005 Paris, France}
\affiliation{$^2$Surface du Verre et Interfaces, UMR 125 CNRS/Saint-Gobain, F-93303 Aubervilliers, France}
\affiliation{$^3$Department of Physics, NTNU -- Norwegian University of Science and Technology, NO-7491 Trondheim, Norway}

\date{\today}

\begin{abstract}
We study light scattering by systems combining randomly rough surface and volume dielectric fluctuations. We introduce a general model including correlations between surface and volume disorders, and we study the scattering properties within a single-scattering approach. We identify different regimes of surface and volume dominated scattering depending on length scales characterizing the surface and volume disorders. For uncorrelated disorders, we discuss the  polarization response of each source of disorder, and show how polarimetric measurements can be used to separate the surface and volume contributions in the total measured diffusely scattered intensity. For correlated systems, we identify two configurations of volume disorder which, respectively, couple weakly or strongly to surface scattering via surface-volume cross correlations. We illustrate these effects on different configurations exhibiting interference patterns in the diffusely scattered intensity, which may be of interest for the characterization of complex systems or for the design of optical components by engineering the degree of surface-volume correlations.
\end{abstract}

\maketitle 

\section{Introduction}

 The study of light scattering by disordered media has mainly been carried out in parallel on two separate fronts: one specialized on scattering by rough surfaces \cite{Simonsen2010}, and the other specialized on scattering by volume disorder made of discrete scatterers or fluctuations of the refractive index \cite{Ishimaru,Akkermans:Montambaux:all}. Some of the phenomena observed and predicted for surface scattering are also found for volume scattering and vice versa, a good illustrative example being coherent enhanced backscattering \cite{Kuga:84,Akkermans:1986,West:95}.

However, the study of systems combining both surface and volume disorders has remained relatively unexplored.
 A single-scattering theory for combined randomly rough surfaces and dielectric fluctuations confined to the vicinity of the surface was developed by Elson in order to explain the discrepancies of polarimetric measurements for metallic rough surfaces compared to the expected results from pure surface scattering theories \cite{Elson:84}.
  Numerical studies beyond single scattering in two dimensions then followed for treating either the case of an individual object or the case of a set of randomly positioned scatterers buried below a rough surface \cite{Pak:1993,Madrazo:97,Zhang:98,Giovannini:98,Sentenac:02}.
A heuristic summing rule for the intensity of the diffusely scattered light was proposed by Sentenac and coworkers \cite{Sentenac:02,Guerin:07}.
As a consequence, a splitting rule was formulated, which states that the diffusely scattered intensity for the combined surface and volume disordered medium can be obtained as the sum of the diffusely scattered intensity obtained for a volume disordered medium bounded by a planar interface and the diffusely scattered intensity obtained from the rough surface separating two homogeneous media,  with an effective dielectric constant describing the response of the substrate. The splitting rule, was first demonstrated numerically for a wide range of parameters \cite{Sentenac:02}, and then supported theoretically in a regime where the length scale of the fluctuations is small compared to the wavelength \cite{Guerin:07}. The assumption of independent stochastic processes for the surface roughness and the volume disorder was made in deriving the splitting rule; this may be a necessary condition for the splitting rule to be valid, as intuited by the authors.
 It is known that light scattering by correlated disordered media can exhibit a wide range of phenomena such as structural coloration \cite{Kinoshita:2008}, localization \cite{Riboli:11}, enhanced transparency \cite{Leseur:16}, and absorption \cite{Bigourdan:19}, to name a few. The effect of cross correlation between surface and volume disorders on light scattering has essentially been left unexplored, despite its potential interest for the engineering of correlated photonic materials.

The coherent \cite{Mudaliar:2013,Berginc:2013,Berginc:15} and incoherent \cite{Mudaliar:2013} multiple scattering of electromagnetic waves in combined uncorrelated surface and volume disorders has also been studied by different approaches, starting from the Lippmann-Schwinger equation or from the radiative transfer equation. These studies gave perspectives on the derivation of an effective medium theory for volume disorder, and on the existence of regimes in which the surface or volume scattering can be treated as a perturbation.

In remote sensing, discriminating between surface and volume scattering is a key issue. Polarization measurements have been suggested to discriminate between the two scattering processes \cite{Sorrentini:09,Ghabbach:14,Dupont:14}. The main idea in these studies is that volume scattering depolarizes more efficiently than surface scattering. However, to our knowledge, no systematic study of the regime of multiple scattering, combining measurement of scattering mean free path and polarimetric measurement, has been carried out so far. Comparing the depolarization from rough surfaces and volume disorders having the same scattering strength would clarify their respective contributions to the depolarization process.

The present paper revisits the single-scattering theory for correlated surface and volume disorders. In most studies 
the two stochastic processes were considered to be independent. To our knowledge, the only study including the influence of cross correlations was conducted by Elson \cite{Elson:84}. However, the model turned out to be valid only for processes sharing the same correlation lengths.
Here we start by introducing a general model of correlated surface roughness and volume dielectric fluctuations in Sec.~\ref{sec:scatt:sys}. The model permits arbitrary choices of autocorrelation functions for the surface and volume processes, with different lengths scales, and of cross correlation function (with some constrains).
The single-scattering theory is derived in Sec.~\ref{sec:theory} starting from the volume integral representation of the scattered field. We recover Elson's theory by treating both surface and volume scattering on the same footing, and we derive expressions for the diffusely scattered intensity for correlated disorders in different asymptotic regimes.
These asymptotic regimes are analyzed for uncorrelated disorders, in Sec.~\ref{sec:power}, to map out a diagram of predominance of volume to surface scattering. 
 Then, the effects of cross correlations are studied in different configurations in Sec.~\ref{sec:correlation}. We identify that cross correlations have the strongest impact on scattering when the correlation length along the depth of the layer of dielectric fluctuations is large compared to, or on the order of, the thickness of the heterogeneous medium. The possibility to design specific interference patterns in the diffusely scattered intensity by modulating the cross correlations is also examined. The paper ends with a short discussion on the use of polarization measurements for separating the two scattering contributions in the single-scattering regime in Sec.~\ref{sec:pola}.
 
 The reader primarily interested in the physical understanding rather than the technical theoretical details may skip most of the derivation in Sec.~\ref{sec:theory}, and jump to the end of the section to Eq.~(\ref{eq:avR12:general}) and Table~\ref{tab1}. They summarize the main theoretical results of the present paper and are the starting point for all subsequent discussions. Their interpretation at a more conceptual level is given in the last paragraph of Sec.~\ref{sec:theory}. 

\section{Correlated surface and volume disorders}\label{sec:scatt:sys}

\subsection{Dielectric function and surface profile}

The scattering system that we consider is composed of a semi-infinite heterogeneous medium bounded by a rough interface separating it from a homogeneous medium. The homogeneous medium (medium 1) occupies region $\Omega_1$, and is characterized by a dielectric constant $\varepsilon_1$. The heterogeneous medium (medium 2) occupies region $\Omega_2$, and is characterized by a  dielectric function of the form $\varepsilon_2 + \Delta \varepsilon(\Vie{x}{}{})$, where $\varepsilon_2$ is a constant and $\Vie{x}{}{} = x_1 \, \Vie{\hat{e}}{1}{} +  x_2 \, \Vie{\hat{e}}{2}{} + x_3 \, \Vie{\hat{e}}{3}{} = \Vie{x}{\parallel}{} + x_3 \, \Vie{\hat{e}}{3}{}$
 is a point in space. The spatially dependent dielectric function in the whole space can be written as
\begin{equation}
\varepsilon (\Vie{x}{}{}) = \varepsilon_1 + \ind_{\Omega_2}(\Vie{x}{}{}) \, \big( \varepsilon_2 + \Delta \varepsilon (\Vie{x}{}{}) - \varepsilon_1 \big) \: ,
\label{eq:dielectric}
\end{equation}
where we have defined the indicator function $\ind_{A}$ of a set $A$ as being  equal to 1 if its argument belongs to $A$ and zero otherwise.
%
%
 In the following, we will assume that the interface between the two media can be represented by the equation $x_3 = \zeta(\Vie{x}{\parallel}{})$, where $\zeta$ is the surface profile function. We can therefore write
\begin{equation}
\ind_{\Omega_2}(\Vie{x}{}{}) = \mathrm{H} \Big( \zeta(\Vie{x}{\parallel}{}) - x_3 \Big) \: ,
\end{equation}
where $\mathrm{H}$ is the Heaviside step function. Note that the dielectric fluctuation, $\Delta \varepsilon$, may be defined and may take nonzero values outside of $\Omega_2$, since its contribution in Eq.~(\ref{eq:dielectric}) is cut off by the factor $\ind_{\Omega_2}(\Vie{x}{}{})$.

The definition of the dielectric function above can represent a rich variety of scattering systems. For instance, by setting $\zeta = 0$ and $\Delta \varepsilon = 0$, we describe a system made of two homogeneous semi-infinite media separated by a planar interface. If $\zeta$ is a nontrivial function, the surface becomes rough. It could be chosen to be periodic, or to be a realization of a stochastic process. Similarly, the dielectric fluctuations could be piecewise constant in some subdomains hence representing a homogeneous host medium with inclusions, like particles, which may have arbitrary shape, and relative positions. The scattering system could represent a photonic crystal, or a disordered medium with a continuously randomly fluctuating permittivity.
\begin{figure*}[t]
\begin{center}
\includegraphics[width=0.29\textwidth, trim=0.5cm .5cm 1.35cm .2cm,clip]{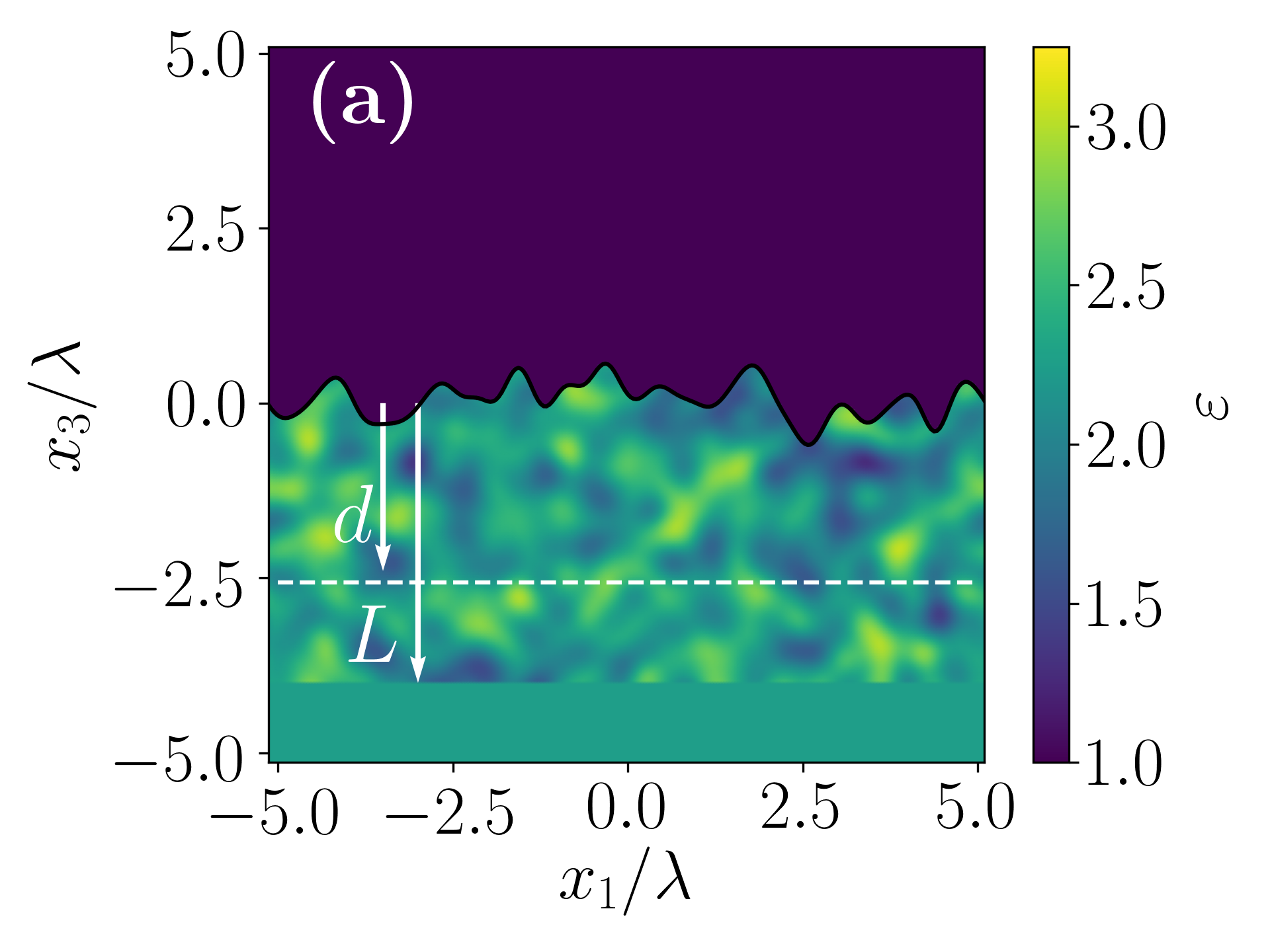}
~
\includegraphics[width=0.29\textwidth, trim=1.35cm .5cm .5cm .2cm,clip]{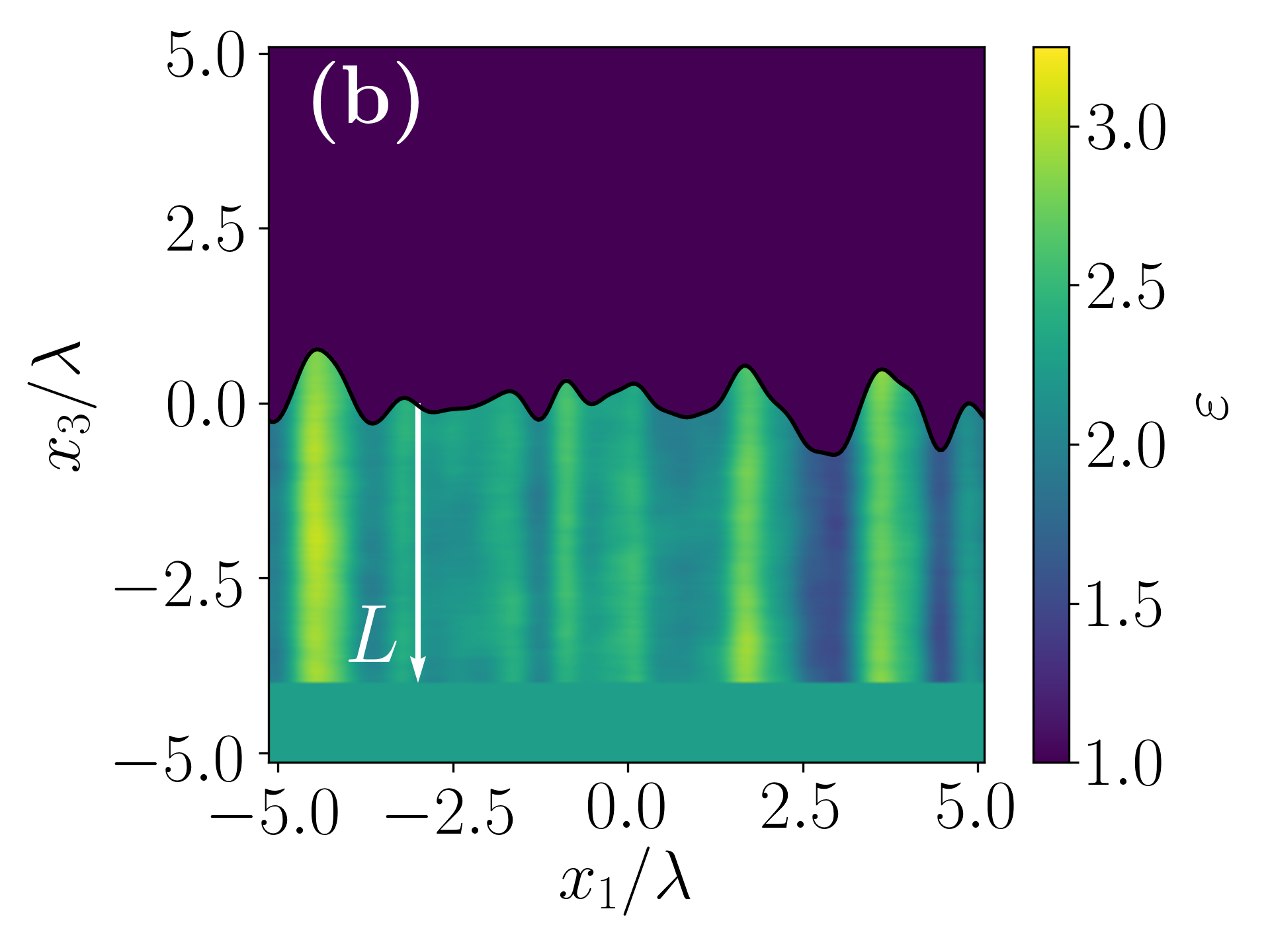}
~
\includegraphics[width=0.38\textwidth, trim=4.5cm 2.5cm 4cm 2.cm,clip]{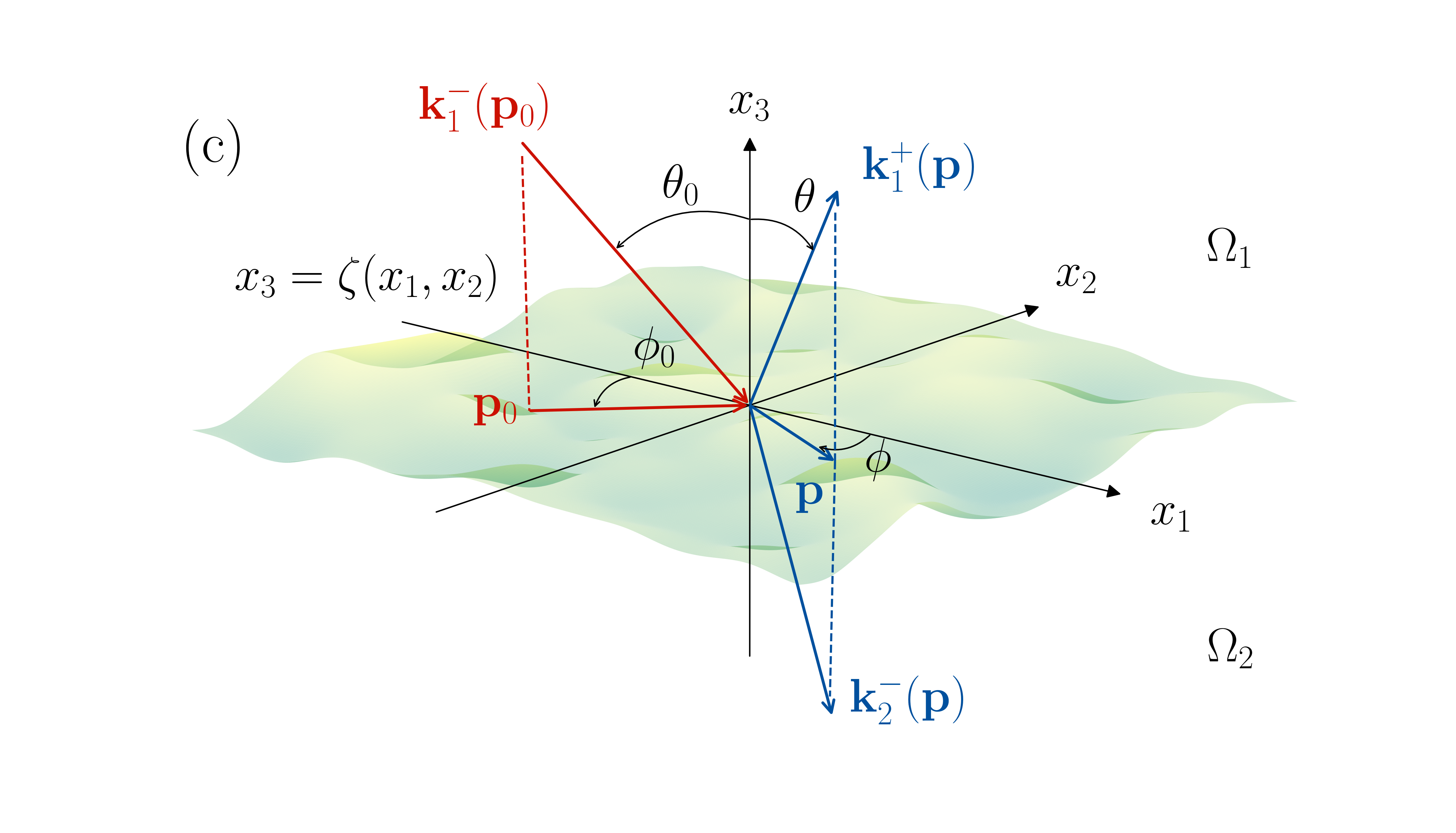}
\caption{Maps of permittivity for typical scattering systems. (a) Genuine volume configuration ($\ell_\varepsilon \ll L$) and (b) surface-like configuration ($\ell_{\varepsilon \perp} > L$) for positively perfectly correlated surface and volume disorder. The white dashed line in (a) indicates the dielectric layer maximally correlated to the surface. (c) Illustration of the definitions of angles and wave vectors.}
\label{fig:volume_vs_surface_like}
\end{center}
\end{figure*}
%

\subsection{Model of correlated processes}

We now introduce a model for a disordered scattering system where the surface profile and the dielectric fluctuations are realizations of stochastic processes with correlations. We start by representing the whole system, boundary and dielectric fluctuations, as a stochastic process the realizations of which are denoted by $\Delta \tilde{\varepsilon} (\Vie{x}{}{})$. It can be written as a function of two sub-processes $\zeta (\Vie{x}{\parallel}{})$ and $\Delta \varepsilon (\Vie{x}{}{})$ in the form
\begin{equation}
\Delta \tilde{\varepsilon} (\Vie{x}{}{}) = \ind_{\Omega_2} (\Vie{x}{}{}) \, \Delta \varepsilon (\Vie{x}{}{}) = \mathrm{H}(\zeta(\Vie{x}{\parallel}{})-x_3) \, \Delta \varepsilon (\Vie{x}{}{}) \: .
\label{eq:delta_tilde}
\end{equation}
%
 Next, we need to define a joint probability density for the two subprocesses $\zeta$ and $\Delta \varepsilon$. For the sake of simplicity, we  define the random vector $\Vie{u}{}{\mathrm{T}} = (\zeta(\Vie{x}{\parallel}{}),\zeta(\Vie{x}{\parallel}{\prime}),\Delta \varepsilon (\Vie{x}{}{}),\Delta \varepsilon (\Vie{x}{}{\prime}))$,  and choose a Gaussian joint probability density 
%
%
\begin{equation}
p \big( \Vie{u}{}{} ; \Vie{x}{}{},\Vie{x}{}{\prime} \big) = \frac{\exp \Big( - \frac{1}{2} \Vie{u}{}{\mathrm{T}} \, \boldsymbol\Sigma ^{-1} (\Vie{x}{}{},\Vie{x}{}{\prime}) \, \Vie{u}{}{} \Big)}{(2 \pi)^{2} \, \mathrm{det} (\boldsymbol \Sigma (\Vie{x}{}{},\Vie{x}{}{\prime}))^{1/2}} \: .
\end{equation}
The covariance matrix $\boldsymbol\Sigma (\Vie{x}{}{},\Vie{x}{}{\prime})$ may depend on $\Vie{x}{}{}$ and $\Vie{x}{}{\prime}$ but must be symmetric and positive definite. We have chosen here for simplicity to have vanishing averages $\langle \zeta (\Vie{x}{\parallel}{}) \rangle = 0$ and $\langle \Delta \varepsilon (\Vie{x}{}{}) \rangle = 0$, independently of the spatial position, where the  brackets $\left\langle \cdot \right\rangle$ denote the ensemble average over realizations of the stochastic process. The covariance matrix $\boldsymbol\Sigma (\Vie{x}{}{},\Vie{x}{}{\prime})$  contains all the information about the possible correlations between the surface profile and the volume dielectric fluctuations.
 Note that the covariance matrix reduces to a $2 \times 2$ block-diagonal matrix for \emph{uncorrelated} surface profile and dielectric fluctuations. In such a case, the joint probability density can be written as the product of two probability densities for $(\zeta(\Vie{x}{\parallel}{}),\zeta(\Vie{x}{\parallel}{\prime}))$ and $(\Delta \varepsilon (\Vie{x}{}{}),\Delta \varepsilon (\Vie{x}{}{\prime}))$, respectively.\\

 We now assume that the rms roughness of the surface is independent of position $\langle \zeta^2 (\Vie{x}{\parallel}{}) \rangle = \sigma^2_{\zeta}$, and that the variance of $\Delta \varepsilon (\Vie{x}{}{})$ depends only on  $x_3$, i.e., $\left\langle \Delta \varepsilon^2 (\Vie{x}{}{}) \right\rangle = f^2(x_3) \, \sigma_\varepsilon^2$ (where $\sigma_\zeta$ and $\sigma_\varepsilon$ are non-negative constants). Indeed, it could be physically realistic to consider that the fluctuations of the dielectric constant are somewhat bounded within a layer with thickness $L$ beneath the average surface. The function $f$ may then be taken to be a smooth sigmoid such that $f(x_3) \to 1$ as $x_3 \to \infty$, and $f(x_3) \to 0$ as $x_3 \to - \infty$ with the transition occurring around a characteristic depth $L$. Alternatively, one could use the step function $f(x_3) = \mathrm{H}(x_3 + L)$.
The latter will be used in the following for the sake of simplicity. Assuming wide-sense stationarity of the stochastic process, i.e., the covariances only depend on the difference between two points, we can see that each $2 \times 2$ block of the covariance matrix is symmetric and depends only on the following covariances:
\begin{subequations}
\begin{align}
\left\langle \zeta(\Vie{x}{\parallel}{}) \, \zeta(\Vie{x}{\parallel}{\prime}) \right\rangle &= \sigma_\zeta^2 \: W_\zeta(\Vie{x}{\parallel}{} - \Vie{x}{\parallel}{\prime}) \: , \\
\left\langle \Delta \varepsilon (\Vie{x}{}{}) \, \Delta \varepsilon (\Vie{x}{}{\prime}) \right\rangle &= \sigma_\varepsilon^2 \: f(x_3) \, f(x_3^\prime) \: W_\varepsilon(\Vie{x}{}{} - \Vie{x}{}{\prime}) \: , \\
\left\langle \zeta (\Vie{x}{\parallel}{}) \, \Delta \varepsilon (\Vie{x}{}{\prime}) \right\rangle &= \sigma_\zeta \, \sigma_\varepsilon \: f(x_3^\prime) \: W_{\zeta \varepsilon}(\Vie{x}{\parallel}{} - \Vie{x}{}{\prime}) \: .
\end{align}
\end{subequations}
Here $W_\zeta$ and $W_\varepsilon$ are the auto-correlation functions of the stochastic processes $\zeta$ and $\Delta \varepsilon$, respectively, and are such that $W_\zeta (\Vie{0}{}{}) = 1$ and $W_\varepsilon (\Vie{0}{}{}) = 1$. The function $W_{\zeta \varepsilon}$ is the cross-correlation function of the processes $\zeta$ and $\Delta \varepsilon$. Note that we do \emph{not} necessarily have  $W_{\zeta \varepsilon} (\Vie{0}{}{}) = 1$ (take, for example, the case where $\zeta$ and $\Delta \varepsilon$ are uncorrelated which gives $W_{\zeta \varepsilon} = 0$ identically).
The positiveness of the covariance matrix imposes bounds on the cross-correlation function. In the following we will assume Gaussian auto-correlation functions given by
\begin{subequations}
\begin{align}
W_\zeta (\Vie{x}{\parallel}{}) &= \exp \left( - \frac{|\Vie{x}{\parallel}{}|^2}{\ell_\zeta^2} \right) \: , \\
W_\varepsilon (\Vie{x}{}{}) &= \exp \left( - \frac{|\Vie{x}{\parallel}{} |^2}{\ell_{\varepsilon \parallel}^2} - \frac{x_3^2}{\ell_{\varepsilon \perp}^2} \right) \: .
\end{align}
\end{subequations}
Here $\ell_\zeta$, $\ell_{\varepsilon \parallel}$ and $\ell_{\varepsilon \perp}$ denote the surface correlation length, the transverse and the perpendicular correlation lengths of the dielectric fluctuations, respectively. The corresponding transverse power spectra, defined as the Fourier transforms of the auto-correlation functions, are thus given by
\begin{subequations}
\begin{align}
\hat{W}_\zeta (\Vie{p}{}{}) &= \pi \ell_\zeta^2 \, \exp \left( - \frac{|\Vie{p}{}{}|^2 \ell_\zeta^2}{4} \right) \: , \\
\hat{W}_\varepsilon (\Vie{p}{}{},x_3) &= \hat{W}_{\varepsilon \parallel}(\Vie{p}{}{}) \: \exp \left( -\frac{x_3^2}{\ell_{\varepsilon \perp}^2} \right) , \label{eq:defWepara}
\end{align}
\end{subequations}
where 
\begin{equation}
\hat{W}_{\varepsilon \parallel}(\Vie{p}{}{}) = \pi \ell_{\varepsilon \parallel}^2  \exp \left( - \frac{|\Vie{p}{}{}|^2 \ell_{\varepsilon \parallel}^2}{4} \right) \: .
\end{equation}
Here and in the following, we denote by
\begin{equation}
\hat{f}(\Vie{p}{}{}) = 
 \int f(\Vie{x}{\parallel}{}) \: e^{-i \Vie{p}{}{} \cdot \Vie{x}{\parallel}{}} \: \mathrm{d}^2x_\parallel \: , 
\end{equation}
the two-dimensional Fourier transform of a function $f$.
We model the cross-correlation function via the power spectra of the auto-correlation functions as
\begin{align}
\hat{W}_{\zeta \varepsilon} (\Vie{p}{}{},x_3) = \: &\gamma(\Vie{p}{}{}) \, \hat{W}_{\zeta}^{1/2}(\Vie{p}{}{}) \, \hat{W}_{\varepsilon \parallel}^{1/2}(\Vie{p}{}{}) \nonumber\\
&\times \exp\Big( - \frac{(x_3+d)^2}{\ell_{\varepsilon \perp}^2} \Big)  \: .
\end{align}
Making use of the expressions for $\hat{W}_\zeta$ and $\hat{W}_{\varepsilon \parallel}$ above, it can be rewritten as
\begin{equation}
\hat{W}_{\zeta \varepsilon} (\Vie{p}{}{},x_3) =  \gamma(\Vie{p}{}{}) \, \pi \ell_\zeta \ell_{\varepsilon \parallel} \, \exp \left( - \frac{|\Vie{p}{}{} |^2 \ell_\parallel^2}{4} - \frac{(x_3+d)^2}{\ell_{\varepsilon \perp}^2} \right)  \: ,
\label{eq:hatWze}
\end{equation}
where the transverse cross-correlation length $\ell_\parallel$ is defined as
\begin{equation}
\ell_\parallel^2 = \frac{1}{2} \left( \ell_\zeta^2 + \ell_{\varepsilon \parallel}^2 \right) \: .
\end{equation}
We have also introduced a distance $d$ such that $0\leq d \leq L$, as an arbitrary offset determining the slice of dielectric fluctuation with which the surface is maximally correlated, namely the slice $\Delta \varepsilon (\Vie{x}{\parallel}{}, -d)$. The factor $\gamma(\Vie{p}{}{})$ is a \emph{spectral correlation modulator} which tunes the cross correlation of different transverse spectral components of $\zeta$ and $\Delta \varepsilon$. It is in general a complex valued function satisfying $|\gamma(\Vie{p}{}{})| \leq 1$ and $\gamma(-\Vie{p}{}{}) = \gamma^*(\Vie{p}{}{})$. In principle, a more exotic dependency of $\hat{W}_{\zeta \varepsilon}$ on $x_3$ may be modeled by letting $\gamma$ be a function of $x_3$. We restrict ourselves to the form given in Eq.~(\ref{eq:hatWze}) for simplicity. The simplest example of a nontrivial spectral correlation modulator would be a constant, $\gamma(\Vie{p}{}{}) = \gamma \in [-1,1]$. In such a case, since $W_\zeta$ and $W_\varepsilon$ are both Gaussian, we can easily obtain $W_{\zeta \varepsilon}$ explicitly by an inverse Fourier transform of Eq.~(\ref{eq:hatWze}), leading to
\begin{equation}
W_{\zeta \varepsilon}(\Vie{x}{}{}) = \gamma \frac{2 \ell_\zeta \ell_{\varepsilon \parallel}}{ \ell_\zeta^2 + \ell_{\varepsilon \parallel}^2} \: \exp \left( - \frac{|\Vie{x}{\parallel}{} |^2}{\ell_{\parallel}^2} - \frac{(x_3+d)^2}{\ell_{\varepsilon \perp}^2} \right) \: .
\label{eq:Wze}
\end{equation}
Another example is $\gamma(\Vie{p}{}{}) = \gamma_0 \, \exp(i \Vie{p}{}{} \cdot \Vie{a}{}{})$ where $\Vie{a}{}{}$ is an arbitrary vector in the $x_1 x_2$ plane. This spectral modulation yields a cross-correlation function $W_{\zeta \varepsilon, \Vie{a}{}{}} = W_{\zeta \varepsilon}(\Vie{x}{}{}-\Vie{a}{}{})$ with $W_{\zeta \varepsilon}$ given by Eq.~(\ref{eq:Wze}).\\

Let us comment on the above construction of the cross-correlation function. In this particular model, the transverse cross-correlation length is such that its square is the average of the squares of the respective surface and transverse permittivity correlation lengths. In other words, we find that the transverse correlation length $\ell_\parallel$ lies between $\ell_\zeta$ and $\ell_{\varepsilon \parallel}$. In the particular case where $\ell_\zeta = \ell_{\varepsilon \parallel}$, one finds $\ell_\parallel = \ell_\zeta = \ell_{\varepsilon \parallel}$, and hence all correlation functions share the same transverse length scale. In addition, in such a case, the prefactor $2 \ell_\zeta \ell_{\varepsilon \parallel} / (\ell_\zeta^2 + \ell_{\varepsilon \parallel}^2)$ becomes unity. This implies that for $\gamma = \pm 1$ the surface profile $\zeta(\Vie{x}{\parallel}{})$ is proportional to the permittivity slice $\Delta \varepsilon (\Vie{x}{\parallel}{},-d)$; more precisely, one has $\zeta(\Vie{x}{\parallel}{}) = \pm \sigma_\zeta \Delta \varepsilon (\Vie{x}{\parallel}{},-d) / \sigma_\varepsilon$.
 In contrast, for $\ell_\zeta \neq \ell_{\varepsilon \parallel}$, the prefactor $2 \ell_\zeta \ell_{\varepsilon \parallel} / (\ell_\zeta^2 + \ell_{\varepsilon \parallel}^2)$ is strictly smaller than unity, which means that even for $|\gamma|=1$ the detuning of the correlation lengths imposes bounds on the maximum correlation between $\zeta(\Vie{x}{\parallel}{})$ and $\Delta \varepsilon (\Vie{x}{\parallel}{},-d)$, namely, the prefactor $2 \ell_\zeta \ell_{\varepsilon \parallel} / (\ell_\zeta^2 + \ell_{\varepsilon \parallel}^2)$ is exactly this bound. When one of the correlation lengths dominates, say $\ell_\zeta \gg \ell_{\varepsilon \parallel}$, the prefactor becomes $2 \ell_\zeta \ell_{\varepsilon \parallel} / (\ell_\zeta^2 + \ell_{\varepsilon \parallel}^2) \sim 2 \ell_{\varepsilon \parallel} / \ell_\zeta \ll 1$, which essentially makes the cross correlation negligible. The intuitive understanding of this result is that one cannot get $\zeta(\Vie{x}{\parallel}{})$ and $\Delta \varepsilon (\Vie{x}{\parallel}{},-d)$ arbitrarily correlated if each process satisfies wide-sense stationarity (i.e., statistical invariance by translation) with different correlation lengths.

\section{Scattering model}\label{sec:theory}

\subsection{Volume integral representation}

Consider the scattering system defined in Sec.~\ref{sec:scatt:sys}, with the dielectric function given by Eq.~(\ref{eq:dielectric}).
 The total electric field $\Vie{E}{}{}$ resulting from the interaction of an incident harmonic field $\Vie{E}{0}{}$ with angular frequency $\omega$ with the scattering system satisfies the Lippmann-Schwinger integral equation (see, e.g., Ref.~\onlinecite{calvo-Perez1999})
\begin{align}
\Vie{E}{}{} (\Vie{x}{}{}) = \: &\Vie{E}{}{(0)} (\Vie{x}{}{}) \label{eq:int:equation:1}\\
&+ k_0^2 \: \int \Vie{G}{}{} (\Vie{x}{}{},\Vie{x}{}{\prime}) \, \big[ \varepsilon(\Vie{x}{}{\prime}) - \varepsilon_\mathrm{ref}(\Vie{x}{}{\prime}) \big] \, \Vie{E}{}{} (\Vie{x}{}{\prime}) \: \mathrm{d}^3x^\prime \: , \nonumber
\end{align}
where $k_0 = \omega / c = 2 \pi / \lambda$, $c$ being the speed of light in vacuum.
Here $\Vie{E}{}{(0)}$ is the total electric field solution of the scattering problem for a planar interface between media 1 and 2, i.e, of a reference system with dielectric function
\begin{equation}
\varepsilon_\mathrm{ref} (\Vie{x}{}{}) = \varepsilon_1 + \mathrm{H}(-x_3) \, (\varepsilon_2 -\varepsilon_1) \: .
\end{equation}
The tensor Green's function $\Vie{G}{}{}$ is the solution to
\begin{equation}
\nabla \times \nabla \times \Vie{G}{}{} (\Vie{x}{}{},\Vie{x}{}{\prime}) - \varepsilon_\mathrm{ref} (\Vie{x}{}{}) \, k_0^2 \Vie{G}{}{} (\Vie{x}{}{},\Vie{x}{}{\prime}) = \delta (\Vie{x}{}{}-\Vie{x}{}{\prime}) \Vie{I}{}{} \: ,
\end{equation}
with outgoing wave conditions at infinity (radiation condition). Note that translational invariance along the $x_1 x_2$ plane allows us to write $\Vie{G}{}{}(\Vie{x}{}{},\Vie{x}{}{\prime}) = \Vie{G}{}{}(\Vie{x}{\parallel}{}-\Vie{x}{\parallel}{\prime},x_3,x_3^\prime)$ whenever this seems adequate. By expanding the dielectric function, we can recast the integral in Eq.~(\ref{eq:int:equation:1}) as the sum of two terms,
\begin{align}
&\Vie{E}{}{} (\Vie{x}{}{}) = \Vie{E}{}{(0)} (\Vie{x}{}{}) \nonumber\\
+ &k_0^2 \: \int \Vie{G}{}{} (\Vie{x}{}{},\Vie{x}{}{\prime}) \, (\varepsilon_2 - \varepsilon_1) \, h(\Vie{x}{}{\prime}) \, \Vie{E}{}{} (\Vie{x}{}{\prime}) \: \mathrm{d}^3x^\prime \label{eq:integral:representation} \\ 
+ &k_0^2 \: \int \Vie{G}{}{} (\Vie{x}{}{}, \Vie{x}{}{\prime}) \, \Delta \varepsilon (\Vie{x}{}{\prime}) \, \mathrm{H}(\zeta(\Vie{x}{\parallel}{\prime}) - x_3^\prime) \, \Vie{E}{}{} (\Vie{x}{}{\prime}) \: \mathrm{d}^3x^\prime \: , \nonumber
\end{align}
with
\begin{equation}
h(\Vie{x}{}{}) =  \mathrm{H}(\zeta(\Vie{x}{\parallel}{}) - x_3) - \mathrm{H}(-x_3) \: .
\end{equation}
%
The first integral term on the right-hand side of Eq.~(\ref{eq:integral:representation}) corresponds to surface scattering. Indeed, the field is scattered by a dielectric fluctuation in the selvedge region, induced by the surface profile, which is piece-wise constant and takes values zero, $\varepsilon_2 - \varepsilon_1$, or $\varepsilon_1 - \varepsilon_2$ depending on the  position $\Vie{x}{}{\prime}$ with respect to the interface and the reference plane $x_3 = 0$.
%
  The second integral term corresponds to scattering by the volume dielectric fluctuations  $\Delta \varepsilon$ located below the interface. 
   Note that the presence of the surface profile in the second term must not be understood as surface scattering. Its role is merely to delimit the volume in which the dielectric fluctuation $\Delta \varepsilon$ contributes.
%
    Nevertheless, this indicates that even for a system for which the stochastic processes $\zeta$ and $\Delta \varepsilon$ are uncorrelated, having dielectric fluctuations bounded by the rough interface induces a correlation between the field described by the two integral terms. This correlation effect is a second-order contribution in the product of $\zeta$ and $\Delta \varepsilon$ and will be neglected in the following.

\subsection{Single-scattering regime}

By writing the total field in the form $\Vie{E}{}{} = \Vie{E}{}{(0)} + \Vie{E}{}{(s)}$ with $\Vie{E}{}{(0)}$ the field in the reference system, and $\Vie{E}{}{(s)}$ the scattered field, 
and by assuming that $\Vie{E}{}{} \approx \Vie{E}{}{(0)}$ in the integrals in Eq.~(\ref{eq:integral:representation}), we obtain the (first) Born approximation for the scattered field  given by
%
\begin{align}
\Vie{E}{}{(1)} (\Vie{x}{}{}) = \: & k_0^2 \, \int \Vie{G}{}{} (\Vie{x}{}{},\Vie{x}{}{\prime}) \, (\varepsilon_2 - \varepsilon_1) \,  h(\Vie{x}{}{\prime}) \, \Vie{E}{}{(0)} (\Vie{x}{}{\prime}) \: \mathrm{d}^3x^\prime \nonumber\\
+ & k_0^2 \, \int \Vie{G}{}{} (\Vie{x}{}{}, \Vie{x}{}{\prime}) \, \Delta \tilde{\varepsilon} (\Vie{x}{}{\prime}) \, \Vie{E}{}{(0)} (\Vie{x}{}{\prime}) \: \mathrm{d}^3x^\prime \: .
\label{eq:single:scattering}
\end{align}
%
%
%
The Born approximation corresponds to single-scattering either at the surface or in the volume.\\

\emph{Zeroth-order field} --- Equation~(\ref{eq:single:scattering}) requires the zeroth-order field, $\Vie{E}{}{(0)}$, solution of the scattering problem for the reference system. In the case of an incident monochromatic plane wave, the zeroth-order field can be written as the sum of the incident plane wave and a reflected plane wave in medium 1, and as a transmitted plane wave in medium 2. The expression of the zeroth-order field is given in Appendix~\ref{App:zero:order}.\\

\emph{Volume contribution} --- The second term on the right-hand side of Eq.~(\ref{eq:single:scattering}) can be approximated by
\begin{align}
&\Vie{E}{\varepsilon}{(1)}(\Vie{x}{}{}) = \label{eq:single:scattering:volume}\\
& k_0^2 \int_{-L}^0 \, \Vie{G}{}{}(\cdot,x_3,x_3^\prime) * \Big[ \Delta \varepsilon (\cdot,x_3^\prime) \, \Vie{E}{}{(0)}(\cdot,x_3^\prime) \Big] (\Vie{x}{\parallel}{}) \, \mathrm{d}x_3^\prime \: , \nonumber
\end{align}
where $*$ denotes the two-dimensional convolution product. Here we have approximated the upper bound in the integral over $x_3^\prime$ by zero, i.e., $\zeta(\Vie{x}{\parallel}{\prime}) \approx 0$. The small amplitude approximation can be considered to be valid in the regime where the typical amplitude of the surface profile is small compared to the wavelength, i.e., $\sigma_\zeta k_1 \ll 1$, $\sigma_\zeta k_2 \ll 1$ with $k_1 = \sqrt{\varepsilon_1} k_0$, and $k_2 = \sqrt{\varepsilon_2} k_0$. This approximation allows us to interchange the order of integration and to obtain the convolution product in the $(x_1,x_2)$ variables as shown in Eq.~(\ref{eq:single:scattering:volume}). Taking the Fourier transform of $\Vie{E}{\varepsilon}{(1)}$ with respect to $\Vie{x}{\parallel}{}$ yields
\begin{align}
&\Vie{\hat{E}}{\varepsilon}{(1)}(\Vie{p}{}{},x_3) = \label{eq:Eeps:fourier:planewave}\\
& k_0^2 \int_{-L}^0 \, \Vie{\hat{G}}{}{}(\Vie{p}{}{},x_3,x_3^\prime)  \Delta \hat{\varepsilon} (\Vie{p}{}{}-\Vie{p}{0}{},x_3^\prime) \, \Vie{\hat{E}}{2}{(0)}(\Vie{p}{0}{},x_3^\prime) \, \mathrm{d}x_3^\prime \: ,\nonumber
\end{align}
where we have used the convolution theorem and the fact that the reference field for $x_3^\prime < 0$ is a plane wave. The factor $\Vie{\hat{E}}{2}{(0)}(\Vie{p}{0}{},x_3^\prime)$ is the Fourier-Weyl amplitude of the transmitted zeroth-order field [see Eq.~(\ref{eq:fresnel:fourier})].\\

\emph{Surface contribution} --- The surface contribution given by the first term on the right-hand side of Eq.~(\ref{eq:single:scattering}) can be treated in a similar fashion, although some care is required. The surface term as written in Eq.~(\ref{eq:single:scattering}) reads
\begin{widetext}
\begin{equation}
\Vie{E}{\zeta}{(1)} (\Vie{x}{}{}) = k_0^2 (\varepsilon_2 - \varepsilon_1) \: \int_{\mathbb{R}^2} \int_0^{\zeta(\Vie{x}{\parallel}{\prime})} \Vie{G}{}{}(\Vie{x}{\parallel}{}-\Vie{x}{\parallel}{\prime},x_3, x_3^\prime) \Vie{E}{}{(0)}(\Vie{x}{\parallel}{\prime},x_3^\prime) \: \mathrm{d}x_3^\prime \: \mathrm{d}^2x_\parallel^\prime \: .
\label{eq:born1}
\end{equation}
However, the Born approximation as written above by approximating the total field $\Vie{E}{}{}$ by $\Vie{E}{}{(0)}$ is a poor choice in this case. The following choice will prove to be more accurate:
\begin{equation}
\Vie{E}{\zeta}{(1)} (\Vie{x}{}{}) = k_0^2 (\varepsilon_2 - \varepsilon_1) \: \int_{\mathbb{R}^2} \int_0^{\zeta(\Vie{x}{\parallel}{\prime})} \Vie{G}{}{}(\Vie{x}{\parallel}{}-\Vie{x}{\parallel}{\prime},x_3, x_3^\prime) \Vie{\tilde{E}}{}{(0)}(\Vie{x}{\parallel}{\prime},x_3^\prime) \: \mathrm{d}x_3^\prime \: \mathrm{d}^2x_\parallel^\prime \: .
\label{eq:born2}
\end{equation}
Here the field $\Vie{\tilde{E}}{}{(0)}(\Vie{x}{\parallel}{\prime},x_3^\prime)$ is the continuation of the reference field $\Vie{E}{}{(0)}(\Vie{x}{\parallel}{\prime},x_3^\prime)$ inside the grooves of the interface. More explicitly, and using the notation from Appendix~\ref{App:zero:order}, $\Vie{\tilde{E}}{}{(0)}(\Vie{x}{\parallel}{},x_3)$ is given by
\begin{equation}
\Vie{\tilde{E}}{}{(0)} (\Vie{x}{}{}) = \begin{cases}
   \Vie{E}{0}{} (\Vie{x}{}{}) +  \left[ r_{21}^{(p)} (\Vie{p}{0}{}) \Cie{E}{0,p}{} \, \Vie{\hat{e}}{1,p}{+} (\Vie{p}{0}{}) +  r_{21}^{(s)} (\Vie{p}{0}{}) \Cie{E}{0,s}{} \, \Vie{\hat{e}}{s}{} (\Vie{p}{0}{}) \right] \: \exp \left( i \Vie{k}{1}{+}(\Vie{p}{0}{}) \cdot \Vie{x}{}{} \right)       & \quad \text{if } x_3 > \zeta(\Vie{x}{\parallel}{}) \\
    \left[ t_{21}^{(p)} (\Vie{p}{0}{}) \Cie{E}{0,p}{} \, \Vie{\hat{e}}{2,p}{-} (\Vie{p}{0}{}) +  t_{21}^{(s)} (\Vie{p}{0}{}) \Cie{E}{0,s}{} \, \Vie{\hat{e}}{s}{} (\Vie{p}{0}{}) \right] \: \exp \left( i \Vie{k}{2}{-}(\Vie{p}{0}{}) \cdot \Vie{x}{}{} \right)  		& \quad \text{if } x_3 < \zeta(\Vie{x}{\parallel}{})
  \end{cases}
 \: .
 \label{eq:fresnel:continuation}
\end{equation}
\end{widetext}
Here $\Cie{E}{0,p}{}$ and $\Cie{E}{0,s}{}$ are the known field amplitudes of the $p$ and $s$ polarization components of the incident plane wave $\Vie{E}{0}{}$ [see Eq.~(\ref{eq:inc:field})], and $\Vie{\hat{e}}{j,p}{\pm}$ and $\Vie{\hat{e}}{s}{}$ are unit polarization vectors defined in Eqs.~(\ref{eq:wave_vector}). The factors $r_{21}^{(p)}$, $r_{21}^{(s)}$, and $t_{21}^{(p)}$, $t_{21}^{(s)}$ are Fresnel reflection and transmission factors [see Eq.~(\ref{eq:fresnel:rt})].
The physical reason for the choice above can be understood as follows. Picture a point $\Vie{x}{\parallel}{} + x_3 \Vie{\hat{e}}{3}{}$ in the vicinity of the surface such that $\zeta(\Vie{x}{\parallel}{}) < x_3 < 0$, i.e., inside a groove and just above the surface. The approximation given by Eq.~(\ref{eq:born1}) would assume the total field at that point to be $\Vie{E}{}{}(\Vie{x}{\parallel}{},x_3) \approx \Vie{E}{}{(0)}(\Vie{x}{\parallel}{},x_3) = \Vie{E}{2}{(0)} (\Vie{x}{\parallel}{},x_3)$, i.e., the zeroth-order field transmitted in medium 2. However, for a smooth perturbation of the surface profile, the total field just above the interface is expected to be close to the reference field in medium 1 rather than that in medium 2 (and conversely for a point just below the surface). A more mathematically oriented justification may also be given. The perturbation in the dielectric function at a given point, $\Vie{x}{\parallel}{} + x_3 \Vie{\hat{e}}{3}{}$, with say $x_3 < 0$, induced by the surface profile will exhibit a jump from $\varepsilon_2$ to $\varepsilon_1$ as the amplitude of the profile is continuously deformed from say $\zeta(\Vie{x}{\parallel}{}) = 0$ to $\zeta(\Vie{x}{\parallel}{}) < x_3$. Thus no matter how small $|x_3|$ is, the perturbation of the dielectric function induced by the surface will lead to a jump for sufficiently large values of $\sigma_\zeta$. Hence, even though the perturbation of the profile is continuous, the induced dielectric perturbation is not. This justifies the use of the continuation of the reference field to points belonging to the same medium in order to compensate for the discontinuous perturbation of the dielectric function.\\

We now apply the small amplitude approximation to the lowest nonvanishing order in Eq.~(\ref{eq:born2}), i.e., we assume $\int_0^\zeta f(x_3^\prime) \mathrm{d}x_3^\prime \approx f(0^\mathrm{sgn(\zeta)}) \zeta$ where $f(0^\pm)$ denotes the limit of $f$ when $x_3^\prime$ goes to zero from above or below. This leads to
\begin{widetext}
\begin{align}
\Vie{E}{\zeta}{(1)} (\Vie{x}{}{}) &= k_0^2 (\varepsilon_2 - \varepsilon_1) \: \sum_\pm \Vie{G}{}{}(\cdot,x_3, 0^\pm) * \Big[ \zeta^\pm (\cdot) \Vie{\tilde{E}}{}{(0)}(\cdot,0^\pm) \Big] (\Vie{x}{\parallel}{}) \nonumber\\
&=  k_0^2 (\varepsilon_2 - \varepsilon_1) \: \Bigg[ \Vie{G}{}{}(\cdot,x_3, 0^+) * \Big[ \zeta^+ (\cdot) \Vie{E}{}{(0)}(\cdot,0^-) \Big] (\Vie{x}{\parallel}{}) + \Vie{G}{}{}(\cdot,x_3, 0^-) * \Big[ \zeta^- (\cdot) \Vie{E}{}{(0)}(\cdot,0^+) \Big] (\Vie{x}{\parallel}{}) \Bigg] \: ,
\label{eq:Ezeta}
\end{align}
%
where $\zeta^+ = \max(\zeta,0)$ and $\zeta^- = \min(\zeta,0)$. This approximation is expected to be accurate for small surface roughness, i.e., $\sigma_\zeta  k_1 \ll 1$ and $\sigma_\zeta k_2 \ll 1$. 
By taking the Fourier transform of Eq.~(\ref{eq:Ezeta}) with respect to $\Vie{x}{\parallel}{}$ and using the convolution theorem, we obtain
%
\begin{equation}
\Vie{\hat{E}}{\zeta}{(1)} (\Vie{p}{}{},x_3) 
= k_0^2 (\varepsilon_2 -\varepsilon_1) \Big[ \Vie{\hat{G}}{}{} \big(\Vie{p}{}{},x_3, 0^{+} \big) \, \hat{\zeta}^+ (\Vie{p}{}{}-\Vie{p}{0}{}) \, \Vie{\hat{E}}{2}{(0)} \! \big(\Vie{p}{0}{},0\big) + \Vie{\hat{G}}{}{}\big(\Vie{p}{}{},x_3, 0^{-} \big)  \, \hat{\zeta}^- (\Vie{p}{}{}-\Vie{p}{0}{}) \, \Vie{\hat{E}}{1}{(0)} \! \big(\Vie{p}{0}{},0\big) \Big] \: ,
\label{eq:Ezeta:fourier:planewave:pm}
\end{equation}
where $\hat{\zeta}^\pm$ denotes the Fourier transform of $\zeta^\pm$, and the amplitudes $\Vie{\hat{E}}{1}{(0)}$ and $\Vie{\hat{E}}{2}{(0)}$ are defined in Appendix~\ref{App:zero:order} [see Eqs.~(\ref{eq:fresnel:fourier}) and (\ref{eq:E0})].
Next we use the following identity proven in Appendix~\ref{App:GE:identity}:
\begin{equation}
\Vie{\hat{G}}{}{} \big(\Vie{p}{}{},x_3, 0^{+} \big) \Vie{\hat{E}}{2}{(0)} \! \big(\Vie{p}{0}{},0\big) = \Vie{\hat{G}}{}{} \big(\Vie{p}{}{},x_3, 0^{-} \big) \Vie{\hat{E}}{1}{(0)} \! \big(\Vie{p}{0}{},0\big) \: .
\label{eq:GE:id}
\end{equation}
%
 This result allows us, for instance, to factorize $\Vie{\hat{G}}{}{} \big(\Vie{p}{}{},x_3, 0^{-} \big) \Vie{\hat{E}}{1}{(0)} \big(\Vie{p}{0}{},0\big)$ in Eq.~(\ref{eq:Ezeta:fourier:planewave:pm}). Making use of $\hat{\zeta}^+ + \hat{\zeta}^- = \hat{\zeta}$, we finally obtain
\begin{equation}
\Vie{\hat{E}}{\zeta}{(1)} (\Vie{p}{}{},x_3) 
= k_0^2 (\varepsilon_2 -\varepsilon_1) \, \hat{\zeta} (\Vie{p}{}{}-\Vie{p}{0}{}) \, \Vie{\hat{G}}{}{}\big(\Vie{p}{}{},x_3, 0^{-} \big) \, \Vie{\hat{E}}{1}{(0)} \! \big(\Vie{p}{0}{},0\big) \: . \label{eq:Ezeta:fourier}
\end{equation}

\subsection{Scattering amplitudes and mean differential scattering coefficients}

\emph{Reflection amplitudes} --- The Weyl representation of the Green's function as given in Ref.~\cite{Sipe:87} is recalled in Appendix~\ref{App:GE:identity}. Substituting the expressions for the Green's function Eqs.~(\ref{eq:green:pp}) and (\ref{eq:green:pm}), and for the reference field Eqs.~(\ref{eq:E01}) and (\ref{eq:E02}) into  Eqs.~(\ref{eq:Eeps:fourier:planewave}) and (\ref{eq:Ezeta:fourier}) yields
\begin{subequations}
\begin{align}
\Vie{\hat{E}}{\varepsilon}{(1)} (\Vie{p}{}{},x_3) &= \sum_{\mu = p,s} \Vie{\hat{e}}{1,\mu}{+} (\Vie{p}{}{}) \sum_{\nu = p,s} R_{\varepsilon, \mu \nu}^{(1)} (\Vie{p}{}{},\Vie{p}{0}{}) \, \Cie{E}{0,\nu}{} \: \exp \Big( i \alpha_1(\Vie{p}{}{}) \, x_3 \Big) \\
\Vie{\hat{E}}{\zeta}{(1)} (\Vie{p}{}{},x_3) &= \sum_{\mu = p,s} \Vie{\hat{e}}{1,\mu}{+} (\Vie{p}{}{}) \sum_{\nu = p,s} R_{\zeta, \mu \nu}^{(1)}(\Vie{p}{}{},\Vie{p}{0}{}) \, \Cie{E}{0,\nu}{} \: \exp \Big( i \alpha_1(\Vie{p}{}{}) \, x_3 \Big) \: ,
\end{align}
\end{subequations}
for $x_3  > 0$. The first-order volume and surface reflection amplitudes are given by
\begin{subequations}
\begin{align}
R_{\varepsilon, \mu \nu}^{(1)} (\Vie{p}{}{},\Vie{p}{0}{}) &= \frac{i k_0^2}{2\alpha_2(\Vie{p}{}{})} \psi^{+}(\Vie{p}{}{},\Vie{p}{0}{}) \: \rho_{\varepsilon, \mu \nu}(\Vie{p}{}{},\Vie{p}{0}{}) \label{eq:Reps}  \\
R_{\zeta, \mu \nu}^{(1)}(\Vie{p}{}{},\Vie{p}{0}{}) &= \frac{i k_0^2}{2 \alpha_2(\Vie{p}{}{})} \, (\varepsilon_2 -\varepsilon_1) \hat{\zeta}(\Vie{p}{}{}-\Vie{p}{0}{}) \: \rho_{\zeta, \mu \nu}(\Vie{p}{}{},\Vie{p}{0}{})\label{eq:Rzeta}
\: .
\end{align}
\end{subequations}
In writing Eq.~(\ref{eq:Reps}), we have introduced the quantity
\begin{equation}
\psi^{\pm} (\Vie{p}{}{},\Vie{p}{0}{}) = \int_{-L}^0 \Delta \hat{\varepsilon} (\Vie{p}{}{} - \Vie{p}{0}{},x_3^\prime) \, \exp \Big[ - i \big( \pm \alpha_{2}(\Vie{p}{}{}) + \alpha_2(\Vie{p}{0}{}) \big) \, x_3^\prime \Big] \: \mathrm{d}x_3^\prime \: .
\label{eq:psi}
\end{equation}
The polarization coupling amplitudes $\rho_{\varepsilon, \mu \nu}$ and $\rho_{\zeta, \mu \nu}$ for the polarization states $\mu, \nu \in \{p,s\}$ are defined by
\begin{subequations}
\begin{align}
\rho_{\varepsilon, \mu \nu} (\Vie{p}{}{},\Vie{p}{0}{}) &= t_{12}^{(\mu)}(\Vie{p}{}{}) \, \Vie{\hat{e}}{2,\mu}{+}(\Vie{p}{}{}) \cdot \Vie{\hat{e}}{2,\nu}{-}(\Vie{p}{0}{}) \, t_{21}^{(\nu)}(\Vie{p}{0}{}) \\
\rho_{\zeta, \mu \nu} (\Vie{p}{}{},\Vie{p}{0}{}) &= t_{12}^{(\mu)}(\Vie{p}{}{}) \, \Vie{\hat{e}}{2,\mu}{+}(\Vie{p}{}{}) \cdot \Big[ \Vie{\hat{e}}{1,\nu}{-}(\Vie{p}{0}{}) + r_{21}^{(\nu)}(\Vie{p}{0}{}) \Vie{\hat{e}}{1,\nu}{+}(\Vie{p}{0}{}) \Big] \: .
\end{align}
\end{subequations}
The total scattered field for $x_3 > 0$, including the surface and volume contributions, can thus be written as
\begin{equation}
\Vie{\hat{E}}{}{(1)} (\Vie{p}{}{},x_3) = \sum_{\mu = p,s} \Vie{\hat{e}}{1,\mu}{+} (\Vie{p}{}{}) \sum_{\nu = p,s} R_{\mu \nu}^{(1)}(\Vie{p}{}{},\Vie{p}{0}{}) \, \Cie{E}{0,\nu}{} \: \exp \Big( i \alpha_1(\Vie{p}{}{}) \, x_3 \Big) \: ,
\end{equation}
where we have identified the first-order (total) reflection amplitude $R_{\mu \nu}^{(1)}$ as
\begin{equation}
R_{\mu \nu}^{(1)}(\Vie{p}{}{},\Vie{p}{0}{}) = R_{\zeta, \mu \nu}^{(1)} (\Vie{p}{}{},\Vie{p}{0}{}) + R_{\varepsilon, \mu \nu}^{(1)}(\Vie{p}{}{},\Vie{p}{0}{}) = \frac{i k_0^2}{2\alpha_2(\Vie{p}{}{})} \Big[ (\varepsilon_2 -\varepsilon_1) \hat{\zeta}(\Vie{p}{}{}-\Vie{p}{0}{}) \: \rho_{\zeta, \mu \nu}(\Vie{p}{}{},\Vie{p}{0}{}) + \psi^{+}(\Vie{p}{}{},\Vie{p}{0}{}) \: \rho_{\varepsilon, \mu \nu}(\Vie{p}{}{},\Vie{p}{0}{}) \Big] \: .
\label{eq:R1}
\end{equation}
A similar expression for $x_3 < -L$ can be derived for the transmission amplitude, and is detailed in Appendix~\ref{App:transmission:amplitude} [see Eq.~(\ref{eq:T1})].
\end{widetext}

\emph{Physical interpretation of the scattering amplitudes} --- The reflection amplitude $R_{\mu \nu} (\Vie{p}{}{},\Vie{p}{0}{})$ is the probability amplitude for an incident plane wave with incident in-plane wave vector $\Vie{p}{0}{}$ and polarization state $\nu$ to be scattered in reflection in the direction defined by the in-plane wave vector $\Vie{p}{}{}$ with polarization state $\mu$ (see Fig.~\ref{fig:volume_vs_surface_like}(c) for a schematic representation of the incident and scattering wave vectors). The superscript $(1)$ indicates that it is the first-order correction to the reflection amplitude in a power expansion of the disorder, the zeroth-order being given by the Fresnel reflection factor times a Dirac mass $\delta (\Vie{p}{}{} - \Vie{p}{0}{})$ [see Eq.~(\ref{eq:fresnel:fourier})].
 Equation~(\ref{eq:R1}) shows that the first-order reflection amplitude $R_{\mu \nu}^{(1)}$ can be decomposed as the sum of a contribution originating from surface scattering and a contribution from volume scattering.
  The volume scattering contribution, $R_{\varepsilon, \mu \nu}^{(1)}$, is the product of a factor $i k_0^2 \psi^+ / (2 \alpha_2)$, independent of polarization, and which sums the contribution of all single-scattering paths issued from the dielectric fluctuations in the layer $-L < x_3 < 0$, and a factor $\rho_{\varepsilon, \mu \nu}$, proportional to $\Vie{\hat{e}}{2,\mu}{+}(\Vie{p}{}{}) \cdot \Vie{\hat{e}}{2,\nu}{-}(\Vie{p}{0}{})$, which encodes the polarization coupling [Eq.~(\ref{eq:Reps})]. The factor $\psi^+$ hence encodes the speckle field, i.e., the interference of the scattering paths, and depends on the specific realization of the disorder [presence of $\Delta \varepsilon$ in Eq.~(\ref{eq:psi})]. The volume polarization coupling factor, $\rho_{\varepsilon, \mu \nu}$, is \emph{independent} of the specific realization of the disorder, and corresponds to the polarization response of a dipole source below the reference interface. This factor can be interpreted as follows. The reference field in medium 2, proportional to $t_{21}^{(\nu)}(\Vie{p}{0}{})\Vie{\hat{e}}{2,\nu}{-}(\Vie{p}{0}{})$, is projected along the polarization vector $\Vie{\hat{e}}{2,\mu}{+}(\Vie{p}{}{})$ which is the \emph{Snell-conjugate} polarization vector of the measured wave in medium 1, $\Vie{\hat{e}}{1,\mu}{+}(\Vie{p}{}{})$, and the transmission Fresnel factor $t_{12}^{(\mu)}(\Vie{p}{}{})$ accounts for  transmission of the scattered path from medium 2 to medium 1. The concept of Snell-conjugate waves was introduced in Ref.~\onlinecite{Banon:2018} for light scattering by a weakly rough interface. It was shown to be a useful tool for the physical interpretation of the perturbative solution of the reduced Rayleigh equations to first order in the surface profile function, and in particular in giving an explanation of the Yoneda and Brewster scattering phenomena.
  
The surface contribution to the reflection amplitude can be interpreted in a similar fashion. It is written as the product of a factor $i k_0^2 (\varepsilon_2 -\varepsilon_1) \hat{\zeta} / (2 \alpha_2)$ which encodes the speckle field (and depends on the realization of the disorder), and a polarization coupling factor $\rho_{\zeta, \mu \nu}$. In fact, the factor $(\varepsilon_2 -\varepsilon_1) \hat{\zeta}$ can be thought of as a particular case of $\psi^+$ for an infinitesimal layer of dielectric fluctuations with nonvanishing integral [a Dirac layer $\Delta \hat{\varepsilon}(\Vie{p}{}{}-\Vie{p}{0}{},x_3^\prime) = (\varepsilon_2 - \varepsilon_1) \hat{\zeta}(\Vie{p}{}{}-\Vie{p}{0}{}) \, \delta(x_3^\prime)$]. This results from the small amplitude approximation.
 There is, however, an important distinction between the polarization coupling factors in the surface and volume contributions.
  Scattering from the dielectric fluctuations results from dipole sources excited in medium 2 by the reference field, while scattering from the surface results from dipole sources located near the interface, in either medium 1 or 2, but excited by the continuation of the reference field in the vicinity of the surface. In other words, the dipole sources in the selvedge region oscillate in phase with the elementary dipoles in the rest of the medium in which they lie. It is interesting to note that the contributions from the induced dipoles above and below the reference plane share the same polarization coupling in virtue of the identity Eq.~(\ref{eq:GE:id}). The result we have obtained here for the surface contribution to the reflection amplitude is in agreement with similar perturbation theories derived from the extinction theorem, or the reduced Rayleigh equations (see, e.g., Refs.~\cite{Agarwal1977,Elson:84,Banon:2018}). In particular, our derivation based on a volume integral representation, provides a complementary physical interpretation to that given recently in Ref.~\onlinecite{Banon:2018} based on the reduced Rayleigh equations in terms of Snell-conjugate waves.

 The observation that the two sources of disorder have different polarization responses is not only of fundamental interest. In practice, it can be used to decompose the contribution of the surface and of the volume to the total measured diffusely scattered intensity. Indeed, experimentally, one only measures the total scattered intensity, and estimating the relative surface and volume contributions is a delicate task. Our result shows that this decomposition can in principle be done using polarimetric measurements. This observation was already made by Elson in Ref.~\onlinecite{Elson:84}. In fact, the aim of Elson's work was to explain experimental measurements for which the ratio of scattered intensities for $p$ and $s$ polarizations varied from sample to sample of rough heterogeneous silver surfaces. We will elaborate on Elson's idea and suggest a method for decomposing the diffusely scattered intensity for uncorrelated disorder in Sec.~\ref{sec:pola}.

For scalar waves, the scattering amplitudes can be obtained following a similar derivation as the one presented for polarized electromagnetic waves. 
These expressions can be useful as simplified expressions when polarization effects can be neglected, or for the scattering of other kinds of waves. 
For a scalar wave subjected to the continuity of the field and its normal derivative across the interface~\footnote{This would apply for quantum matter waves for example but not for acoustic waves.}, it suffices to replace all the scalar products between polarization vectors in the reflection amplitudes by unity and the Fresnel amplitudes by the corresponding amplitudes for scalar waves. Explicitly, we find
\begin{widetext}
\begin{equation}
R^{(1)}(\Vie{p}{}{},\Vie{p}{0}{}) = \frac{i k_0^2}{2\alpha_2(\Vie{p}{}{})} \Big[ (\varepsilon_2 -\varepsilon_1) \hat{\zeta}(\Vie{p}{}{}-\Vie{p}{0}{})  + \psi^{+}(\Vie{p}{}{},\Vie{p}{0}{}) \Big] t_{12}(\Vie{p}{}{}) \, t_{21} (\Vie{p}{0}{}) \: .
\label{eq:R1:scalar}
\end{equation}
The transmission amplitude for scalar waves is given in Appendix~\ref{App:transmission:amplitude}.
\end{widetext}

\emph{Mean differential scattering coefficients} --- Let us now examine how the electromagnetic fields scattered by the surface and the volume interfere, and analyze the role played by the cross correlation between the surface and the volume disorder. To this end, we compute the diffusely scattered intensity. To first order in the disorder amplitudes, the diffuse component of the mean differential reflection coefficient (MDRC) is obtained from the relation \cite{Banon:2018}
\begin{widetext}
\begin{equation}
\left\langle \frac{\partial R_{\mu \nu}}{\partial \Omega} (\Vie{p}{}{},\Vie{p}{0}{}) \right\rangle_\mathrm{diff} = \lim_{S\to \infty} \frac{ \varepsilon_1^{1/2} k_0 \, \Re \big( \alpha_1(\Vie{p}{}{}) \big)^2}{S (2 \pi)^2 \alpha_1(\Vie{p}{0}{})} \: \left\langle |R_{\mu \nu}^{(1)} (\Vie{p}{}{},\Vie{p}{0}{})|^2 \right\rangle \: . \label{eq:MDRC}
\end{equation}
 In this expression, $S$ is the area of the mean surface in the $x_1 x_2$ plane (meaning that the disorder is supported by a volume $S \times L$). By substituting Eq.~(\ref{eq:R1}) into Eq.~(\ref{eq:MDRC}), a straightforward but tedious calculation, reported in Appendix~\ref{App:MDXC}, yields
%
\begin{align}
&\left\langle \frac{\partial R_{\mu \nu}}{\partial \Omega} (\Vie{p}{}{},\Vie{p}{0}{}) \right\rangle_\mathrm{diff} = \: C^{(r)} (\Vie{p}{}{},\Vie{p}{0}{}) \, k_0^4 \: \Bigg[ (\varepsilon_2 - \varepsilon_1)^2 \sigma_\zeta^2 \, \hat{W}_\zeta(\Vie{p}{}{}-\Vie{p}{0}{}) \, |\rho_{\zeta, \mu \nu} (\Vie{p}{}{},\Vie{p}{0}{})|^2 \nonumber\\
&+ 2 (\varepsilon_2 - \varepsilon_1) \sigma_\zeta \sigma_\varepsilon \, \hat{W}^{1/2}_{\zeta}(\Vie{p}{}{}-\Vie{p}{0}{}) \, \hat{W}^{1/2}_{\varepsilon \parallel} (\Vie{p}{}{}-\Vie{p}{0}{}) \, \Re \Big( \gamma(\Vie{p}{}{}-\Vie{p}{0}{}) \, J \big(\ell_{\varepsilon \perp},L,d,\alpha^+ (\Vie{p}{}{},\Vie{p}{0}{}) \big) \: \rho_{\zeta, \mu \nu} (\Vie{p}{}{},\Vie{p}{0}{}) \, \rho_{\varepsilon, \mu \nu}^* (\Vie{p}{}{},\Vie{p}{0}{}) \Big)  \nonumber \\
&+ \sigma_\varepsilon^2 \, \hat{W}_{\varepsilon \parallel}(\Vie{p}{}{}-\Vie{p}{0}{}) \, I \big(\ell_{\varepsilon \perp},L,\alpha^+(\Vie{p}{}{},\Vie{p}{0}{}), \alpha^+(\Vie{p}{}{},\Vie{p}{0}{})\big) \: |\rho_{\varepsilon, \mu \nu} (\Vie{p}{}{},\Vie{p}{0}{})|^2 \Bigg] \label{eq:avR12:general} \: .
\end{align}
Here we have used the shorthand notation $\alpha^{\pm}(\Vie{p}{}{},\Vie{p}{0}{}) = \pm \alpha_2(\Vie{p}{}{}) + \alpha_2(\Vie{p}{0}{})$, and the dimensionless factor $C^{(r)} (\Vie{p}{}{},\Vie{p}{0}{})$ is given in Appendix~\ref{App:MDXC}.
A similar, expression is found for the diffuse component of the mean differential transmission coefficient (MDTC see Appendix~\ref{App:MDXC}). For the discussion of the results in Secs.~\ref{sec:power} and \ref{sec:correlation}, it will be  convenient to use the scalar wave approximation which is deduced from Eq.~(\ref{eq:avR12:general}) by replacing all the $\rho_{\mu \nu}$ factors by $t_{12}(\Vie{p}{}{}) t_{21}(\Vie{p}{0}{})$ which then all factorize as $|t_{12}(\Vie{p}{}{}) t_{21}(\Vie{p}{0}{})|^2$ outside of the square bracket.
The functions $I$ and $J$ appearing in Eq.~(\ref{eq:avR12:general}) are defined as
\begin{subequations}
\begin{align}
I (\ell_{\varepsilon \perp}, L, \alpha, \beta) &= \int_{-L}^0 \int_{-L}^0 \exp \left[ - \frac{(x_3-x_3^\prime)^2}{\ell_{\varepsilon \perp}^2} \right] \: \exp \Big[ - i \alpha \, x_3 + i \beta^* \, x_3^{\prime} \Big] \, \mathrm{d}x_3 \, \mathrm{d}x_3^\prime \\
J (\ell_{\varepsilon \perp}, L, d, \alpha) &= \int_{-L}^0 \exp \left[ - \frac{(x_3+d)^2}{\ell_{\varepsilon \perp}^2} \right] \: \exp \Big[ i \alpha \, x_3 \Big] \, \mathrm{d}x_3 \: .
\end{align}
\label{eq:IJ:integral:def}
\end{subequations}
\end{widetext}
In general, the above integrals have to be evaluated numerically. There are, however, asymptotic expressions that can be derived analytically  which correspond to particular configurations of the dielectric fluctuations: the genuine volume configuration for which $\ell_{\varepsilon \perp} \ll L$ and the surfacelike configuration for which $\ell_{\varepsilon \perp} \gg L$. The genuine volume and surface-like configurations correspond to Figs.~\ref{fig:volume_vs_surface_like}(a) and \ref{fig:volume_vs_surface_like}(b), respectively. Note that in the configuration $\ell_{\varepsilon \perp} \gg L$, the depth of the maximally correlated slice $d$ does not play any role since any slice $\Delta \varepsilon (\Vie{x}{_\parallel}{},x_3)$ is essentially equally correlated to the surface profile. In the genuine volume configuration, we consider the situations  $d = 0$ and $\ell_{\varepsilon \perp} \ll \min( d, L-d, L)$. Moreover, one may assume a sub-wavelength regime for the dielectric fluctuation in the $x_3$ direction, i.e., $k_0 \ell_{\varepsilon \perp} \ll 1$, which simplifies further the asymptotics.
With these assumptions, we obtain different  asymptotic regimes, that are identified in Table~\ref{tab1} (the derivation is given in Appendix~\ref{App:asymptotics}).

\begin{table*}[t]
\begin{center}
\caption{Asymptotics of the $I$ and $J$ integrals [Eq.~(\ref{eq:IJ:integral:def})] in different regimes. The - in the last column denotes either an irrelevant regime or an asymptotics which is not easily obtained. $^\mathrm{a}$An additional assumption is made: $\ell_{\varepsilon \perp} \ll \min(d, L-d, L)$.}
\begin{tabular}{c p{3.9cm} p{3.9cm} c}
\hline
\hline
Asymptotic regime & \centering$I (\ell_{\varepsilon \perp}, L, \alpha, \alpha)$ & \centering$J (\ell_{\varepsilon \perp}, L, 0, \alpha)$ & $J (\ell_{\varepsilon \perp}, L, d, \alpha)^\mathrm{a}$ \\[.1cm]
\hline
$\ell_{\varepsilon \perp} \ll L$ and $k_0 \ell_{\varepsilon \perp} \ll 1$ (regime 1) & \centering$\displaystyle \sqrt{\pi} \, L \, \ell_{\varepsilon \perp}$ & \centering$\displaystyle \sqrt{\pi} \, \ell_{\varepsilon \perp} / 2$ & -\\[.3cm]
$\ell_{\varepsilon \perp} \gg L$ and $k_0 \ell_{\varepsilon \perp} \ll 1$ (regime 2) & \centering$\displaystyle L^2$ & \centering$\displaystyle L$ &  -\\[.3cm]
$\ell_{\varepsilon \perp} \ll L$ (regime 3) & \centering$\displaystyle \sqrt{\pi} \, L \, \ell_{\varepsilon \perp} \, \exp \left( - \frac{\alpha^2 \ell_{\varepsilon \perp}^2}{4} \right)$ & \centering$\displaystyle \frac{\sqrt{\pi}}{2} \, \ell_{\varepsilon \perp} \, \exp \left( - \frac{\alpha^2 \ell_{\varepsilon \perp}^2}{4} \right)$ & $\displaystyle \sqrt{\pi} \, \ell_{\varepsilon \perp} \, \exp \left( - \frac{\alpha^2 \ell_{\varepsilon \perp}^2}{4} \right) \, \cos (\alpha d)$\\[.3cm]
$\ell_{\varepsilon \perp} \gg L$ (regime 4) & \centering$\displaystyle \frac{4 \sin^2(\alpha L / 2)}{\alpha^2}$ & \centering$\displaystyle \frac{\sin (\alpha L) }{\alpha}$ & -\\[.3cm]
\hline
\hline
\end{tabular}
\label{tab1}
\end{center}
\end{table*}

In summary, the theoretical results presented in Eq.~(\ref{eq:avR12:general}) can be read, at a more conceptual level, as a classical interference formula for the intensity resulting from two types of paths:
\begin{equation}
\left\langle I_\mathrm{tot} \right\rangle = \left\langle I_\zeta \right\rangle + \left\langle I_\mathrm{corr} \right\rangle +\left\langle I_\varepsilon \right\rangle \: .
\label{eq:concept}
\end{equation}
Here $I_\zeta$ and $I_\varepsilon$ are, respectively, the intensities for paths scattered from the surface or volume dielectric fluctuations only. The term $I_\mathrm{corr}$ corresponds to the interference between the two types of paths which survives the averaging in the presence of surface-volume cross correlation. Each of the terms in Eq.~(\ref{eq:concept}) scales differently with different parameters of the surface and volume disorders, as shown by the asymptotics and scalings in Table~\ref{tab1}. 

\section{Regimes of predominance for uncorrelated surface and volume disorder}\label{sec:power}

In this section we study the relative weight of the surface and volume contributions to the scattered intensity as a function of the parameters defining the disordered system.
To this end, we consider the diffuse reflectance, which, for an incident $\nu$-polarized electromagnetic plane wave, is defined as
\begin{equation}
\mathcal{R}_{\nu,\mathrm{diff}} (\Vie{p}{0}{}) = \sum_{\mu = p, s} \int  \left\langle \frac{\partial R_{\mu \nu}}{\partial \Omega} (\Vie{p}{}{},\Vie{p}{0}{}) \right\rangle_\mathrm{diff} \: \mathrm{d}\Omega \: .
\end{equation}
For unpolarized light, the diffuse reflectance is  given by $\mathcal{R}_{\mathrm{diff}} = (\mathcal{R}_{p,\mathrm{diff}} + \mathcal{R}_{s,\mathrm{diff}})/2$. Next, we define the volume to surface diffuse reflectance ratio
\begin{equation}
\eta = \mathcal{R_{\mathrm{diff},\varepsilon}} / \mathcal{R_{\mathrm{diff},\zeta}} \: ,
\end{equation}
where $\mathcal{R_{\mathrm{diff},\varepsilon}}$ ($\mathcal{R_{\mathrm{diff},\zeta}}$) corresponds to the diffuse reflectance when only the volume (surface) disorder is present. The parameter $\eta$ thus gives the regimes for which (i) volume scattering is negligible compared to surface scattering ($\eta \ll 1$), (ii) volume scattering dominates over surface scattering ($\eta \gg 1$), or (iii) volume scattering is of the same order as surface scattering ($\eta \approx 1$).\\

 In all the illustrative examples that we will consider below, we will assume that $\varepsilon_1 = 1$ and $\varepsilon_2 = 2.25$. In addition, $\sigma_\varepsilon$ will be chosen in such a way that the scattering mean free path for the volume disorder, $\ell_s$, estimated from Eq.~(\ref{eq:ls}) (see Appendix~\ref{App:mean_free_path}), yields an optical thickness $L/\ell_s = 0.5$ independently of the configuration, thus ensuring the validity of the single-scattering approximation. In the present section, the two sources of disorder are taken to be uncorrelated ($\gamma = 0$). The total diffusely scattered intensity can thus be written as the sum of the intensity of the subsystems for which either only the rough surface or the volume disorder contributes. 
The results presented in the figures will be obtained based on the polarized expressions [Eq.~(\ref{eq:avR12:general})] in the case of incident unpolarized light for normal incidence. However, the \emph{scalar wave approximation} [scalar version of Eq.~(\ref{eq:avR12:general})] of the form
\begin{equation}
\left\langle \frac{\partial R}{\partial \Omega} \right\rangle_\mathrm{diff} \propto (\varepsilon_2 -\varepsilon_1)^2 k_0^4 \, \sigma_\zeta^2 \hat{W}_\zeta + \sigma_\varepsilon^2 \, k_0^4 \, \hat{W}_{\varepsilon \parallel} \: I \: ,
\label{eq:Rdiff:scalar}
\end{equation}
for $\gamma = 0$, will be sufficient to understand the phenomena of interest, and will be used in the discussion for the sake of simplicity. The different regimes are analyzed by plugging the asymptotics of $I$ given in Table~\ref{tab1} in Eq.~(\ref{eq:Rdiff:scalar}).\\
 
\emph{Regime 1} --- By inspection of Eq.~(\ref{eq:Rdiff:scalar}) and Table~\ref{tab1} in regime 1, and up to a common prefactor, the surface contribution to the diffuse component of the MDRC scales as $(\varepsilon_2 - \varepsilon_1)^2 \, k_0^4 \, \sigma_\zeta^2 \ell_\zeta^2 $, and the volume contribution scales as $\pi^{1/2}  \sigma_\varepsilon^2 \, k_0^4 \, \ell_{\varepsilon \parallel}^2 \ell_{\varepsilon \perp} L $. For broad power spectral densities, i.e., correlation lengths small compared to the wavelength, the diffuse reflectance will also scale proportionally to the square of the transverse correlation lengths. For power spectral densities which are relatively well confined within the propagation domain $|\Vie{p}{}{}|^2 < \varepsilon_1 k_0^2$, i.e., for transverse correlation lengths not too small compared to the wavelength, we can assume $ \int \hat{W}(\Vie{p}{}{}) \, \mathrm{d}^2p = (2 \pi)^{2}$ for $\hat{W} = \hat{W}_\zeta$ or $\hat{W}_{\varepsilon \parallel}$, and that the remaining $\Vie{p}{}{}$ dependence in the diffuse component of the MDRC is smooth, to obtain that the surface and volume contributions to the diffuse reflectance become essentially independent of the transverse correlation lengths, and scale respectively as $(\varepsilon_2 - \varepsilon_1)^2 \, k_0^2 \, \sigma_\zeta^2$, while the volume contribution scales as $\pi^{1/2}  \sigma_\varepsilon^2 \, k_0^2 \, \ell_{\varepsilon \perp} L $. To summarize, in regime 1, the volume to surface diffuse reflectance ratio can be estimated to be
\begin{equation}
\eta_1 = \frac{\pi^{1/2} \sigma_\varepsilon^2 \ell_{\varepsilon \perp} L}{(\varepsilon_2 -\varepsilon_1)^2 \sigma_\zeta^2} \: ,
\end{equation}
for narrow power spectral densities. If the volume disorder has a broad transverse spectral density ($k_0 \ell_{\varepsilon \parallel} \ll 1$)
the above expression must be corrected by a factor $\ell_{\varepsilon \parallel}^2$. Similarly, if the surface disorder has a broad spectral density ($k_0 \ell_{\zeta} \ll 1$)
the above expression must be corrected by a factor $1/\ell_{\zeta}^2$. This remark being made, we will only consider narrow spectral densities from here on.\\

\emph{Regime 2} --- In this regime, a similar analysis shows that the volume contribution to the diffuse component of the MDRC scales as $ \sigma_\varepsilon^2 k_0^4 \, \ell_{\varepsilon \parallel}^2 L^2$ (see asymptotics of $I$ in Table~\ref{tab1}). Consequently, the volume to surface diffuse reflectance ratio is estimated to be
\begin{equation}
\eta_2 = \frac{\sigma_\varepsilon^2 L^2}{(\varepsilon_2 -\varepsilon_1)^2 \sigma_\zeta^2} \: .
\end{equation}
\begin{figure*}[t]
\begin{center}
\includegraphics[width=0.33\textwidth,trim = 0cm 0cm 0cm 1cm,clip]{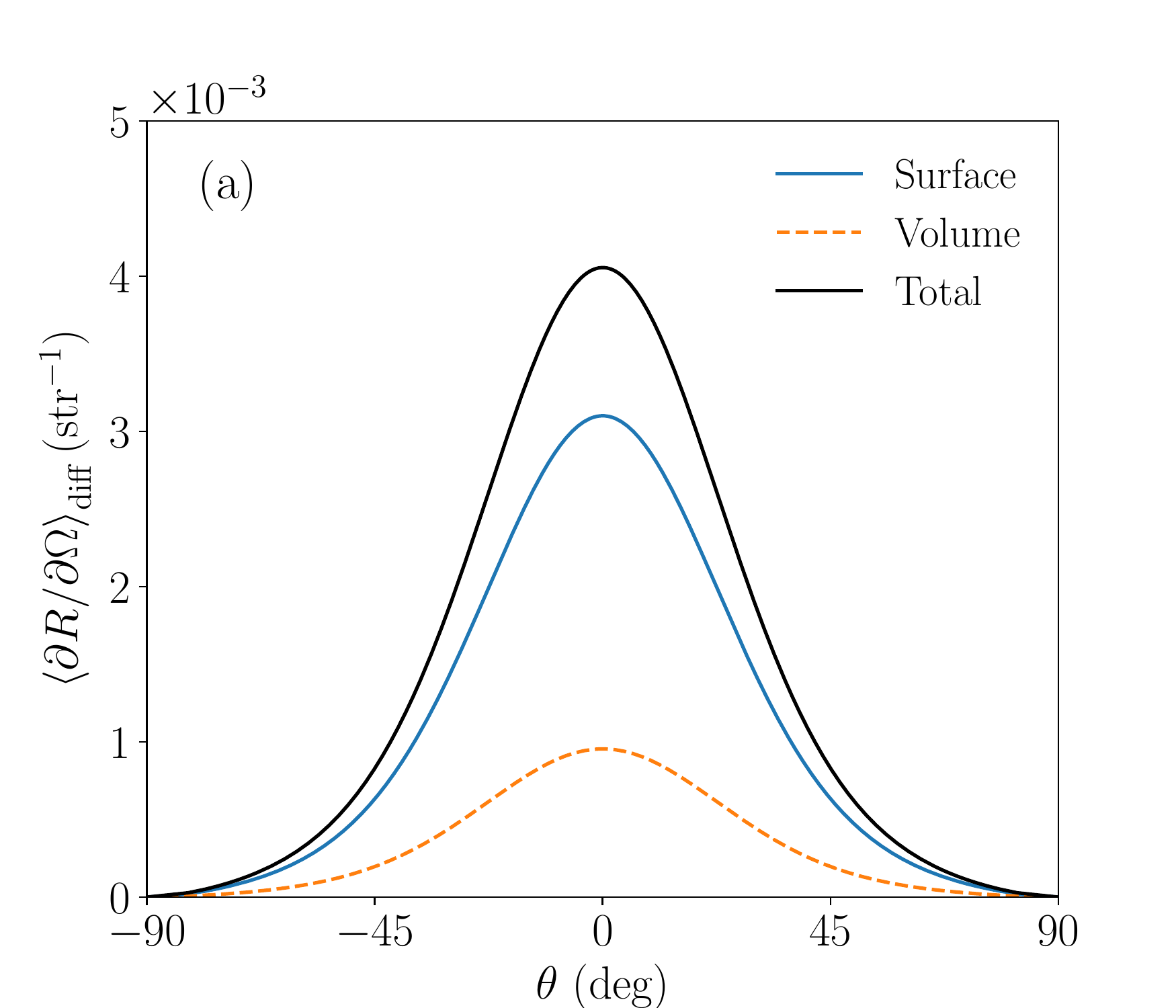}
~
\includegraphics[width=0.33\textwidth,trim = 0cm 0cm 0cm 1cm,clip]{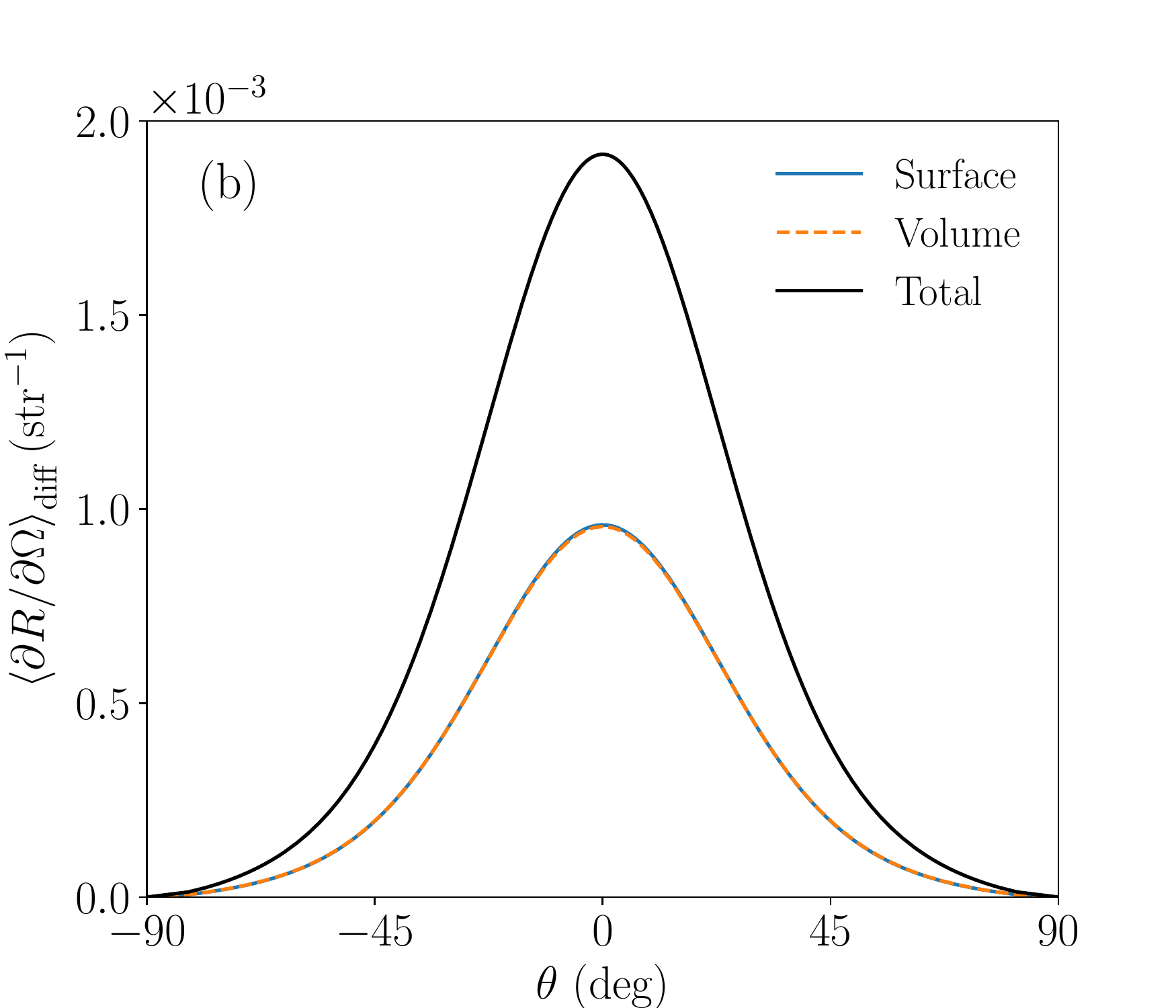}
~
\includegraphics[width=0.33\textwidth,trim = 0cm 0cm 0cm 1cm,clip]{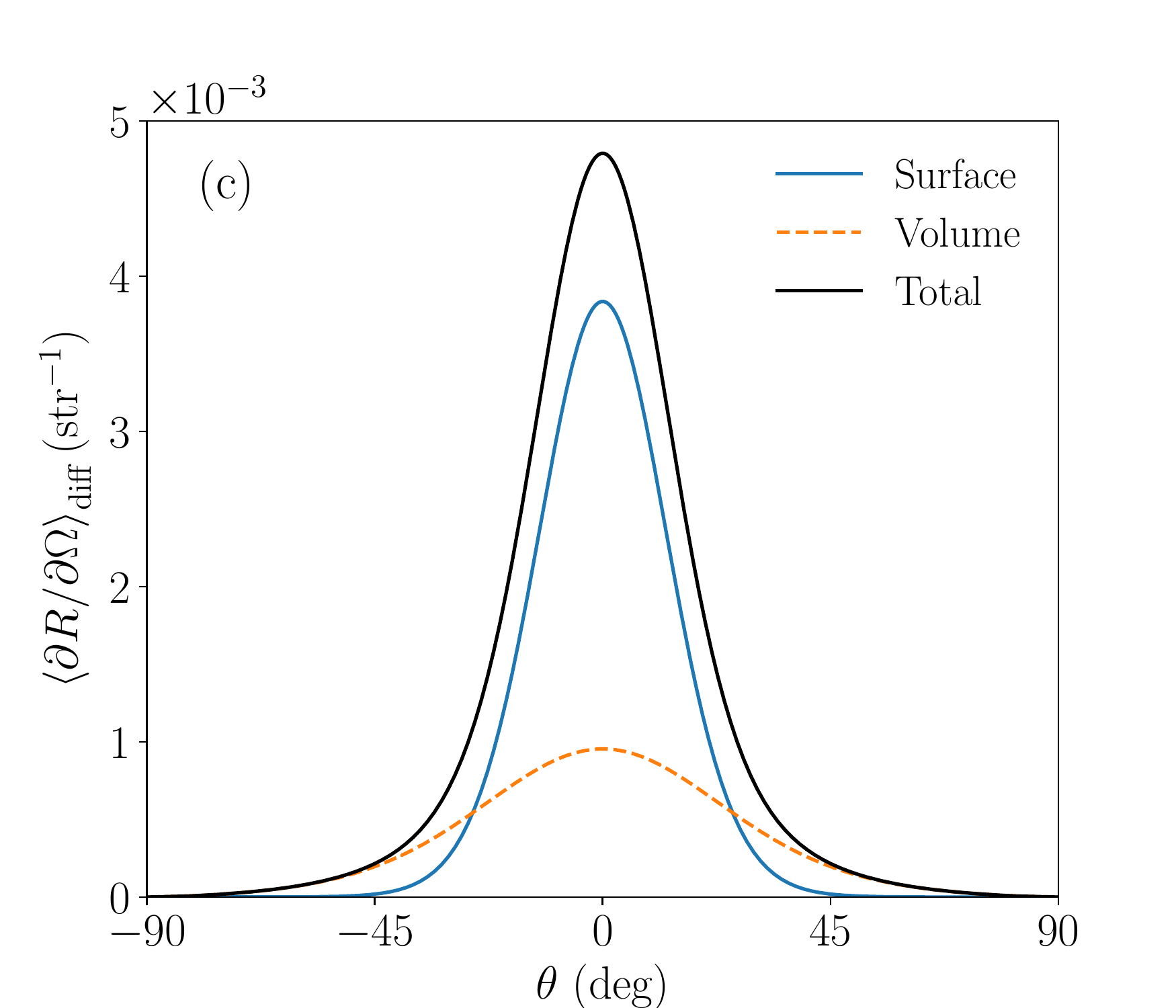}
~
\includegraphics[width=0.33\textwidth,trim = 0cm 0cm 0cm 1cm,clip]{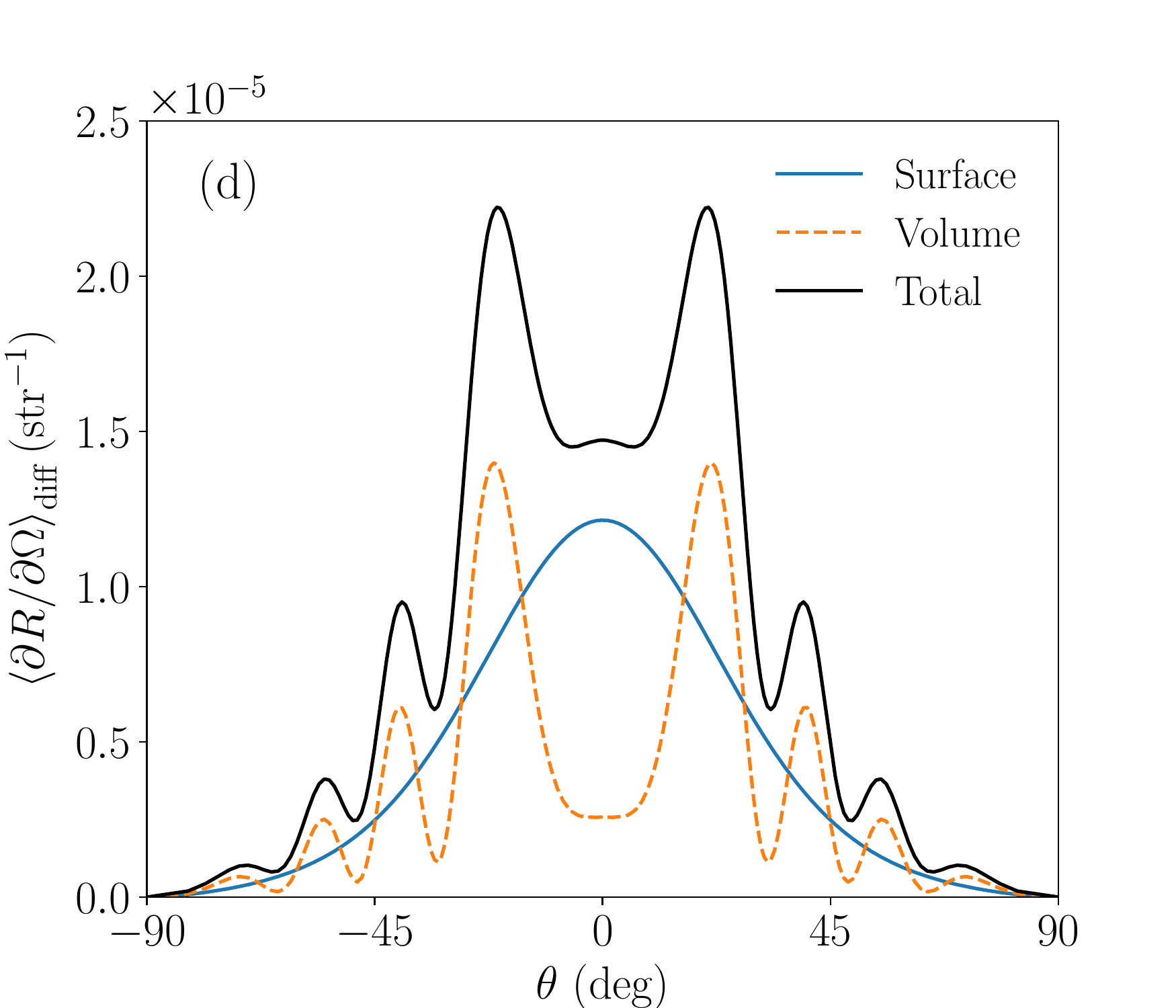}
\caption{Diffuse component of the unpolarized mean differential reflection coefficient for scattering in the plane of incidence as a function of the scattering angle. The surface-like configuration is considered in all cases ($L \ll \ell_{\varepsilon \perp}$), in the regime $L \ll \lambda$ (a, b, c), and in the regime $L > \lambda$ (d).  (a) Surface scattering dominates over volume scattering. (b) Equal contribution from surface and volume scattering. (c) Equal integrated contribution from surface and volume disorder but $\ell_\zeta > \ell_{\varepsilon \parallel}$. (d) Equal contribution and $\ell_\zeta = \ell_{\varepsilon \parallel}$ but in the interference regime. In all cases, the surface profile and dielectric fluctuations are uncorrelated and the optical thickness associated with volume scattering, evaluated following Eq.~(\ref{eq:ls}), is fixed to $L / \ell_s = 0.5$. The parameters assumed were: $\ell_{\varepsilon \parallel} = \lambda / 2$, $\ell_{\varepsilon \perp} = 20 \lambda$, $\ell_\zeta = \lambda / 2$ (a,b,d) or $\ell_\zeta = \lambda$ (c); $L = \lambda / 20$ (a-c) or $L = 10 \lambda$ (d); 
$\sigma_\varepsilon = 0.36$ (a-c) or $\sigma_\varepsilon = 0.026$ (d); and $\sigma_\zeta = \lambda / 40$ (a), $\sigma_\zeta = 14 \times 10^{-3} \lambda$ (b,c), or $\sigma_\zeta = 1.56 \times 10^{-3} \lambda$ (d).}
\label{fig:2}
\end{center}
\end{figure*}
\emph{Comparison of regime 1 and 2} --- We can appreciate the similarity between the surface and volume contributions, in regime 2.
 Indeed, the volume term, proportional to $\sigma_\varepsilon^2 k_0^4 \ell_{\varepsilon \parallel}^2 L^2$, is similar to the surface term, proportional to $(\varepsilon_2 -\varepsilon_1)^2 k_0^4 \ell_{\zeta}^2 \sigma_\zeta^2$, in Eq.~(\ref{eq:Rdiff:scalar}). The role of the dielectric jump $\varepsilon_2 - \varepsilon_1$ is played by the rms of the dielectric fluctuation $\sigma_\varepsilon$; the role of the rms surface roughness $\sigma_\zeta$ is played by the depth $L$; and the role of the in-plane correlation length $\ell_\zeta$ is played by $\ell_{\varepsilon \parallel}$. The denomination of \emph{surfacelike} configuration thus takes its full meaning. Conversely, if we adopt a volume scattering point of view, we can also consider that scattering by a rough surface is equivalent to scattering by a volume with dielectric fluctuations invariant along $x_3$ where $\sigma_\zeta$ is identified with $L$. The genuine volume configuration (regime 1) differs from the surfacelike configuration (regime 2) essentially by the factor $L^2$ which becomes $\ell_{\varepsilon \perp} L$. A first explanation for this difference would be that the power scattered by the volume is always proportional to the depth $L$ and to the correlation length $\ell_{\varepsilon \perp}$ independently of the configuration. However, in the surfacelike configuration, since $\ell_{\varepsilon \perp} \gg L$, the effective out-of-plane correlation length is in fact $L$ because of the depth cut-off. Thus $\ell_{\varepsilon \perp}$ is replaced by $L$ in the surfacelike configuration. A second interpretation of the scattering strength in regime 1 is obtained by estimating the scattering mean free path $\ell_s$ for a system with dielectric fluctuations in an otherwise homogeneous medium. Considering isotropic dielectric fluctuations for simplicity, i.e., $\ell_{\varepsilon \parallel} = \ell_{\varepsilon \perp} = \ell_\varepsilon$, the weight of the volume scattering term in Eq.~(\ref{eq:avR12:general}) becomes
\begin{equation}
\pi^{3/2} k_0^4 \sigma_\varepsilon^2 \ell_\varepsilon^3 L = \frac{4 \pi}{\varepsilon_2^{2}} \, \ell_s^{-1} L \: .
\label{eq:optical:thickness}
\end{equation}
The scattering mean free path $\ell_s$ for an infinite medium with average dielectric function $\varepsilon_2$ (and wave number $k_2 = \sqrt{\varepsilon_2} k_0$) and isotropic dielectric fluctuations with Gaussian statistics is (see Appendix~\ref{App:mean_free_path})
\begin{equation}
\ell_s^{-1} = \frac{\pi^{1/2}}{4} \,  \sigma_\varepsilon^2 \, k_2^4 \, \ell_\varepsilon^3 \: .
\end{equation}
Equation~(\ref{eq:optical:thickness}) thus states that, in the single-scattering regime, the strength of the volume scattering term is controlled by the optical thickness of the layer $L / \ell_s$, i.e., by the average number of scattering events inside the layer.\\
\begin{figure*}[t]
\begin{center}
\includegraphics[width=0.34\textwidth]{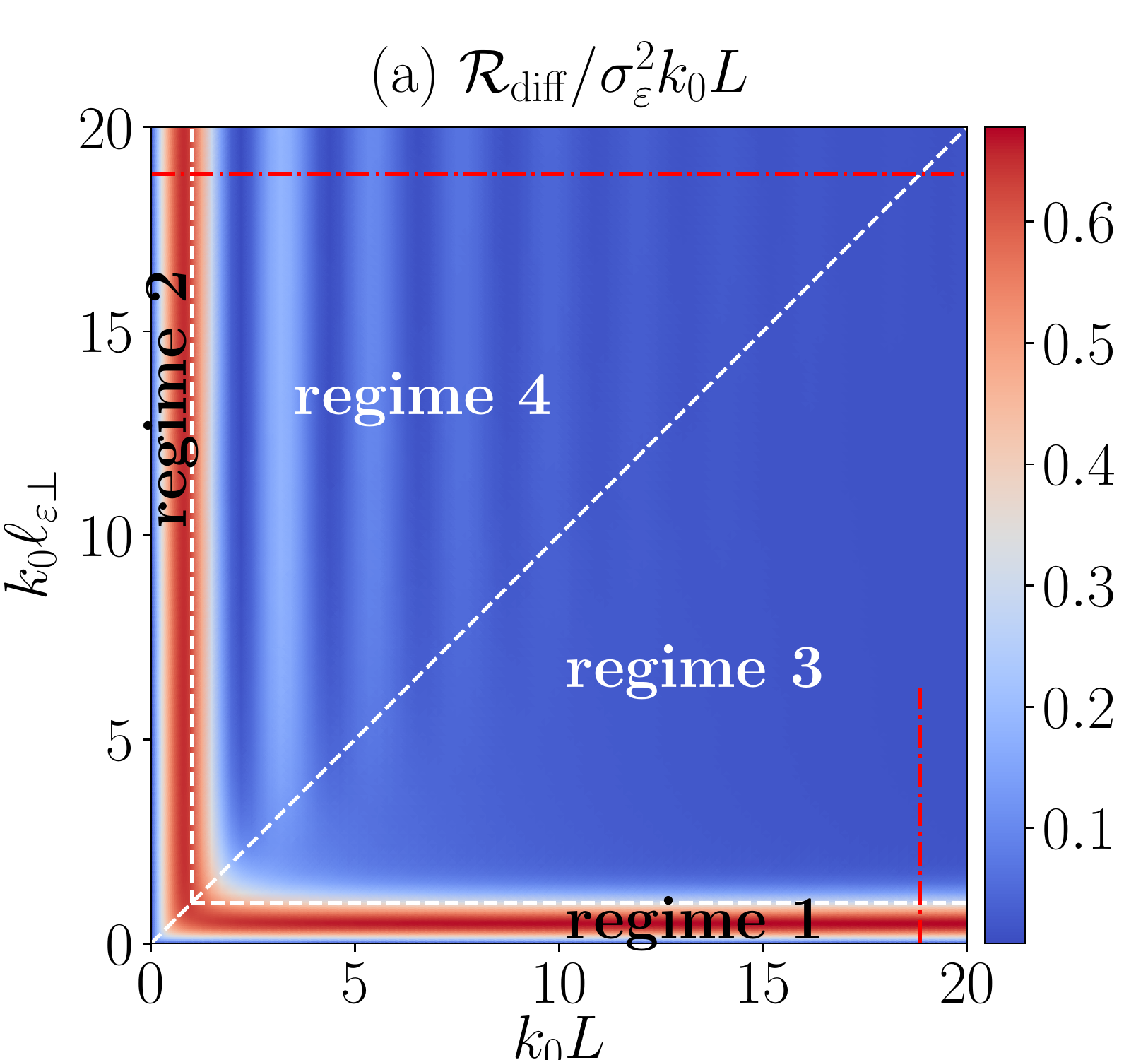}
~
\includegraphics[width=0.31\textwidth]{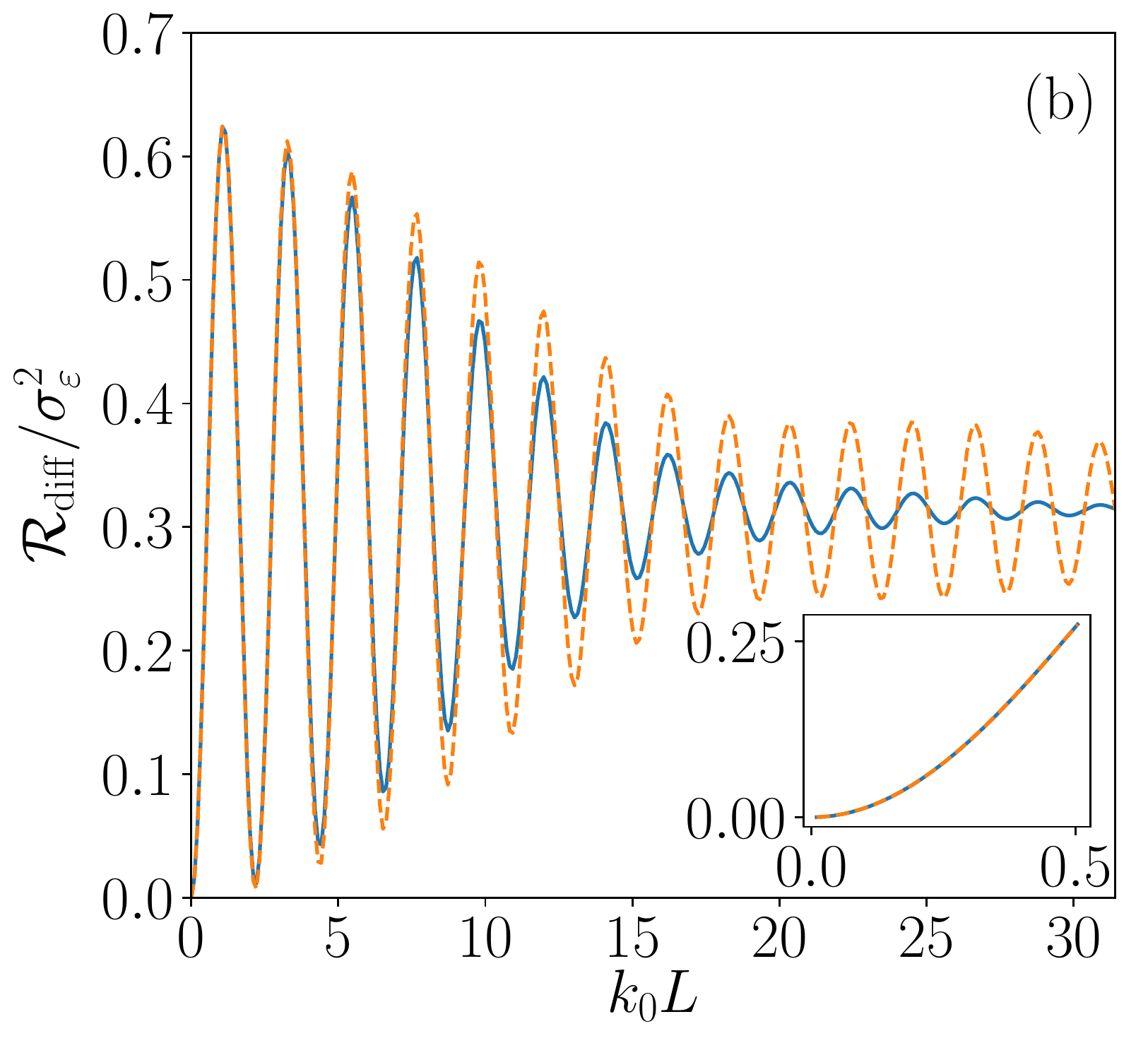}
~
\includegraphics[width=0.31\textwidth]{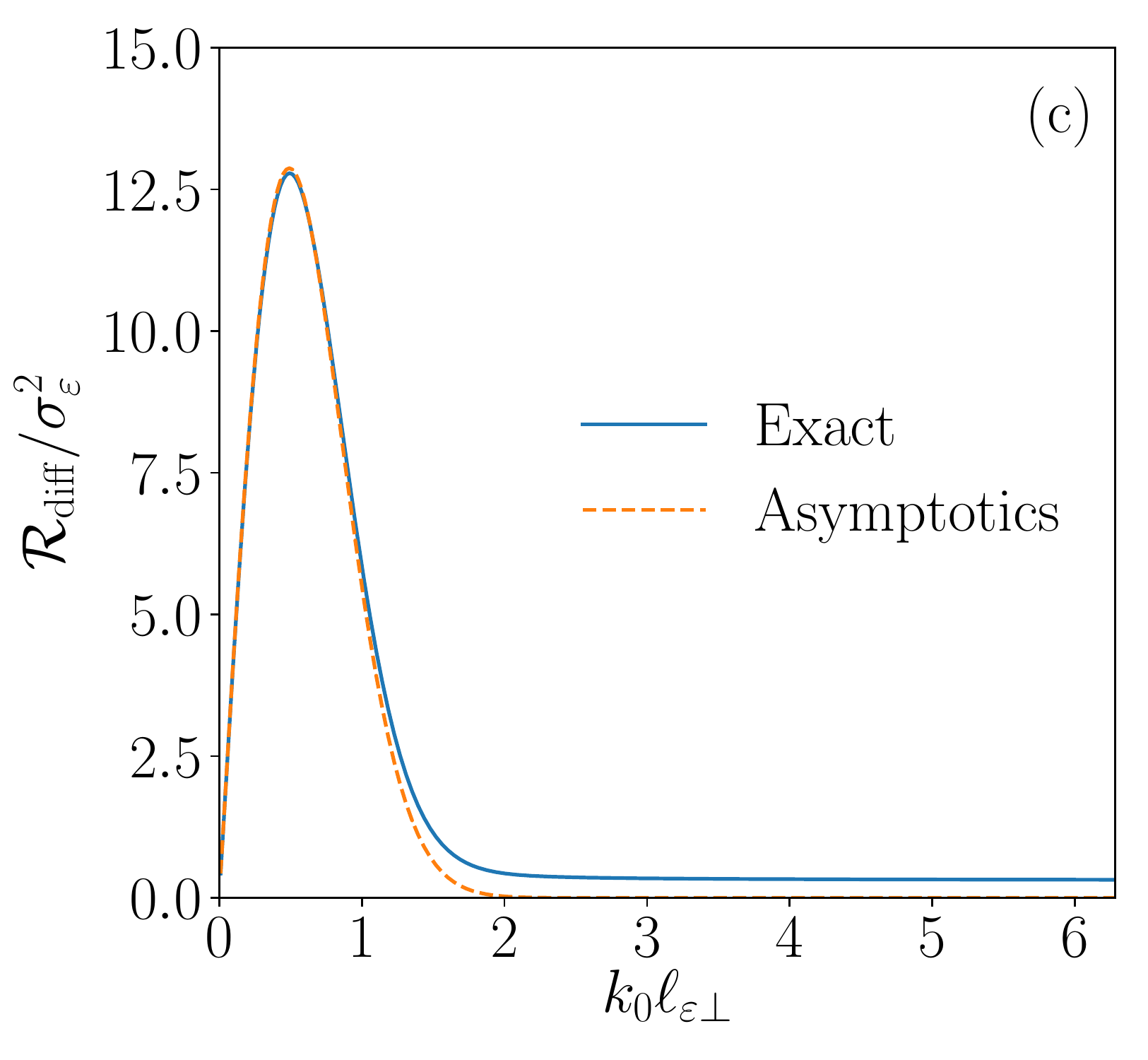}
\caption{(a) Volume diffuse reflectance for unpolarized light $\mathcal{R}_\mathrm{diff} = \mathcal{R}_{\mathrm{diff},\varepsilon}$ in the $(k_0 L, k_0 \ell_{\varepsilon \perp})$ plane. The white dashed lines delimit the different regimes. (b) Cross section showing $\mathcal{R}_\mathrm{diff}$ as a function of $k_0 L$ for $k_0 \ell_{\varepsilon \perp} = 6 \pi$ [horizontal red dash-dotted line in (a)]. (c) Cross section showing $\mathcal{R}_\mathrm{diff}$ as a function of $k_0 \ell_{\varepsilon \perp}$ for $k_0 L = 6 \pi$ [vertical red dash-dotted line in (a)]. The surface was planar, the transverse correlation length was set to $\ell_{\varepsilon \parallel} = \lambda / 4$, and the angle of incidence was $\theta_0 = \ang{0}$. We have normalized the volume diffuse reflectance by $\sigma_\varepsilon^2$ since it is always proportional to $\sigma_\varepsilon^2$ independently of the regime. This allows us to compare the different regimes without taking care of tuning $\sigma_\varepsilon$ to stay within the single-scattering regime as $\ell_{\varepsilon \perp}$ and $L$ vary. In (a), the diffuse reflectance is further normalized by $k_0 L$ in order to compensate for the linear increase of $\mathcal{R}_\mathrm{diff}$ with $L$ in regime 1 (which would otherwise dominate the color map at large $k_0L$).} 
\label{fig:3}
\end{center}
\end{figure*}

 So far we have compared the diffuse reflectance for surface and volume scattering in order to determine their respective regimes of predominance. Note that if we compare their contributions in an elementary solid angle, scattering can be dominated by either surface or volume disorder depending on the scattering angle. Figures~\ref{fig:2}~(a)--(c) present the diffuse component of the MDRC for normally incident and unpolarized light for different cases of uncorrelated surface and volume disorders in regime 2. Figures~\ref{fig:2}(a) and \ref{fig:2}(b) illustrate the cases for which the surface and volume transverse correlation lengths are equal $\ell_\zeta = \ell_{\varepsilon \parallel}$. We observe that either the power diffusely reflected by the surface dominates over the power reflected by the bulk [Fig.~\ref{fig:2}(a)], or both disorders contribute equally to the diffusely reflected power [Fig.~\ref{fig:2}(b)]. In contrast, Fig.~\ref{fig:2}(c) illustrates the case for which both contributions to the integrated reflected power are equal, but the transverse correlation lengths are different [$\ell_\zeta \neq \ell_{\varepsilon \parallel}$]. This results in surface scattering and bulk scattering each having their angular regions of predominance.\\

\emph{Regime 3} --- We now analyze the reflectance beyond the regime $k_0 \ell_{\varepsilon \perp} \ll 1$. Note that the surface contribution still remains in the sub-wavelength limit ($k_j \sigma_\zeta \ll 1$) since this assumption has been made in the first place in the derivation. In regime 3,  the volume term in Eq.~(\ref{eq:Rdiff:scalar}) scales as $ \pi^{1/2} \sigma_\varepsilon^2 k_0^4 \, \ell_{\varepsilon \perp} \, L \, \ell_{\varepsilon \parallel}^2 \, \exp \left( - \alpha^{+2}(\Vie{p}{}{},\Vie{p}{0}{}) \ell_{\varepsilon \perp}^2 / 4 \right)$, i.e., it chiefly decays exponentially with increasing $k_0^2 \ell_{\varepsilon \perp}^2$.
 The exponential decay comes from the specific form assumed for the $x_3$-dependency of the correlation function $W_\varepsilon$. Other forms of the correlation function would lead to a different decaying function. Nevertheless, the diffusely reflected intensity decreases with decreasing wavelength or alternatively increasing correlation length $\ell_{\varepsilon \perp}$. The physical reason for this decay can be understood in terms of the anisotropy factor for the volume disorder. In scattering by a particle, it is known that as one increases the size of the particle compared to the wavelength, the scattering becomes peaked in the forward direction. In the case of continuous dielectric fluctuations, the size of the particle is played by the correlation length. Hence for increasing correlation length $\ell_{\varepsilon \perp}$ beyond the wavelength, scattering by the dielectric fluctuations increases in the forward direction. Thus the reflected scattered light intensity decreases (and, although not shown here, scattering increases in transmission). Beyond the sub-wavelength regime, the genuine volume configuration  yields a wavelength dependent parameter $\eta$ which reads
\begin{equation}
\eta_3 = \eta_1 \, \exp \left( - k_2^2 \, \ell_{\varepsilon \perp}^2 \right) \: .
\end{equation}

\emph{Regime 4} --- In regime 4, the contribution from the volume disorder to the diffusely reflected intensity behaves as $4 k_0^4 \sigma_\varepsilon^2 \hat{W}_{\varepsilon \parallel} (\Vie{p}{}{}-\Vie{p}{0}{}) \sin^2 \big[ \alpha^+(\Vie{p}{}{},\Vie{p}{0}{}) L / 2\big] / \alpha^{+2}(\Vie{p}{}{},\Vie{p}{0}{})$. It exhibits oscillations, hence generating rings in the diffusely reflected intensity, the frequency of which increases with the depth $L$ [see Fig.~\ref{fig:2}(d)]. This is a clear interference phenomenon which survives the averaging. Furthermore, this contribution is bounded by $4 k_0^4 \sigma_\varepsilon^2 \hat{W}_{\varepsilon \parallel} (\Vie{p}{}{}-\Vie{p}{0}{}) / \alpha^{+2}(\Vie{p}{}{},\Vie{p}{0}{})$ and the diffusely reflected power thus scales as $4 \sigma_\varepsilon^2 k_0^2 \, \ell_{\varepsilon \parallel}^2 / \varepsilon_2$. This is a radically different scaling from that observed in the subwavelength regime (regime 2). In particular, the scaling in regime 4 is proportional to $k_0^2$, which differs from the $k_0^4$ scaling in regime 2, and becomes independent of the depth $L$. The depth only controls the angular positions of the interference rings. The parameter $\eta$ of the volume to surface power ratio thus reads
\begin{equation}
\eta_4 = \frac{ 4 \sigma_\varepsilon^2}{\varepsilon_2 \, (\varepsilon_2 - \varepsilon_1)^2 k_0^2 \, \sigma_\zeta^2 } \: .
\end{equation}
The behavior in this regime contrasts with the behavior in the sub-wavelength regime since the volume to surface power ratio $\eta$ depends on the wavelength. This means that the system may undergo a transition from a surface dominated regime to a volume dominated regime as the wavelength is varied.\\

Figure 3 illustrates the different regimes of volume scattering in more details. Figure~\ref{fig:3}(a) presents a contour map of the normalized diffuse reflectance $\mathcal{R}_{\mathrm{diff}} / (\sigma_\varepsilon^2 k_0 L)$ in the $(k_0 L, k_0 \ell_{\varepsilon \perp})$ plane and Figs.~\ref{fig:3}(b) and \ref{fig:3}(c) are cross sections of $\mathcal{R}_{\mathrm{diff}} / \sigma_\varepsilon^2$  for fixed values of $k_0 \ell_{\varepsilon \perp}$ and $k_0 L$, respectively. The two curves in theses figures labeled as "Exact" were obtained by numerical evaluation of the $I$ and $J$ integrals [Eq.~(\ref{eq:IJ:integral:def})] instead of the asymptotic expressions (Table~\ref{tab1}). The different aforementioned regimes are readily observed in Fig.~\ref{fig:3}(a) from the features of the diffuse reflectance. The subwavelength regimes in the genuine volume configuration (regime 1) and in the surfacelike configuration (regime 2), respectively, are bounded by local maxima ridges in the $(k_0 L, k_0 \ell_{\varepsilon \perp})$ plane. Indeed, we recognize on the cross sections a quadratic increase of the reflectance with $k_0 \ell_{\varepsilon \perp}$ [see inset in Fig.~\ref{fig:3}(b)] and a linear increase with $k_0 L$ [Fig.~\ref{fig:3}(c)] in the sub-wavelength limit. In regime 4, we observe that the reflectance oscillates with $k_0 L$, as interference rings appear in the MDRC (this is also seen in Fig.~\ref{fig:5}(a) that will be discussed below). The oscillations are damped and stabilize around a constant value as $k_0 L$ increases. This is due to the fact that as more rings appear in the MDRC, the integration of the MDRC becomes less sensitive to the apparition of new rings [see Figs.~\ref{fig:3}(a, b)]. In regime 3, we initially observe an exponential decay with $k_0^2 \ell_{\varepsilon \perp}^2$ which saturates to an almost constant value for large $k_0 \ell_{\varepsilon \perp}$ (the asymptotic expression becomes inaccurate), matching the value one would obtain coming from regime 4 by increasing $k_0 L$ as the oscillations dampen [i.e., coming from either sides of the diagonal in Fig.~\ref{fig:3}(a)].
\begin{figure*}[t]
\begin{center}
\includegraphics[width=0.32\textwidth]{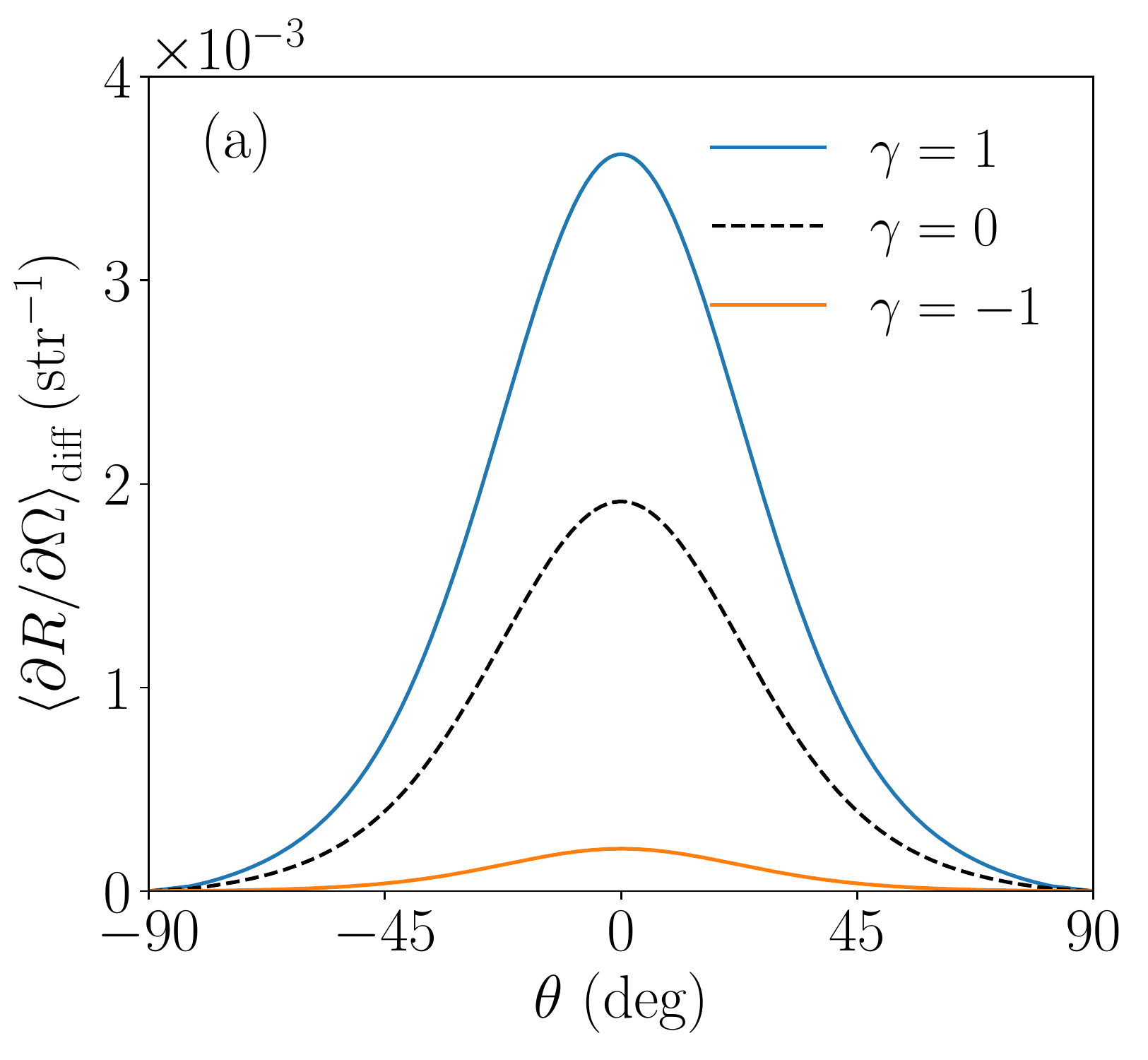}
~
\includegraphics[width=0.32\textwidth]{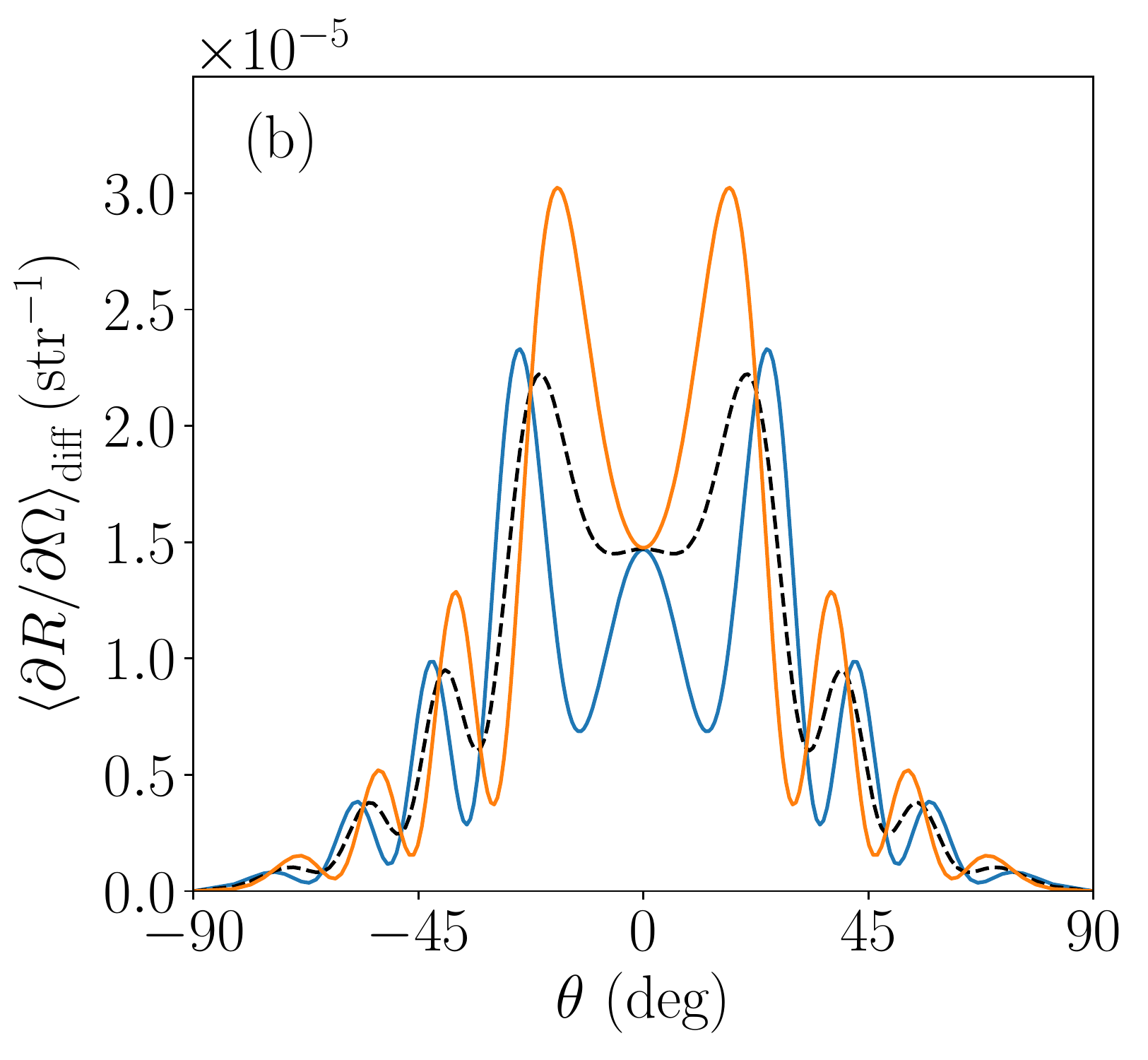}
~
\includegraphics[width=0.32\textwidth]{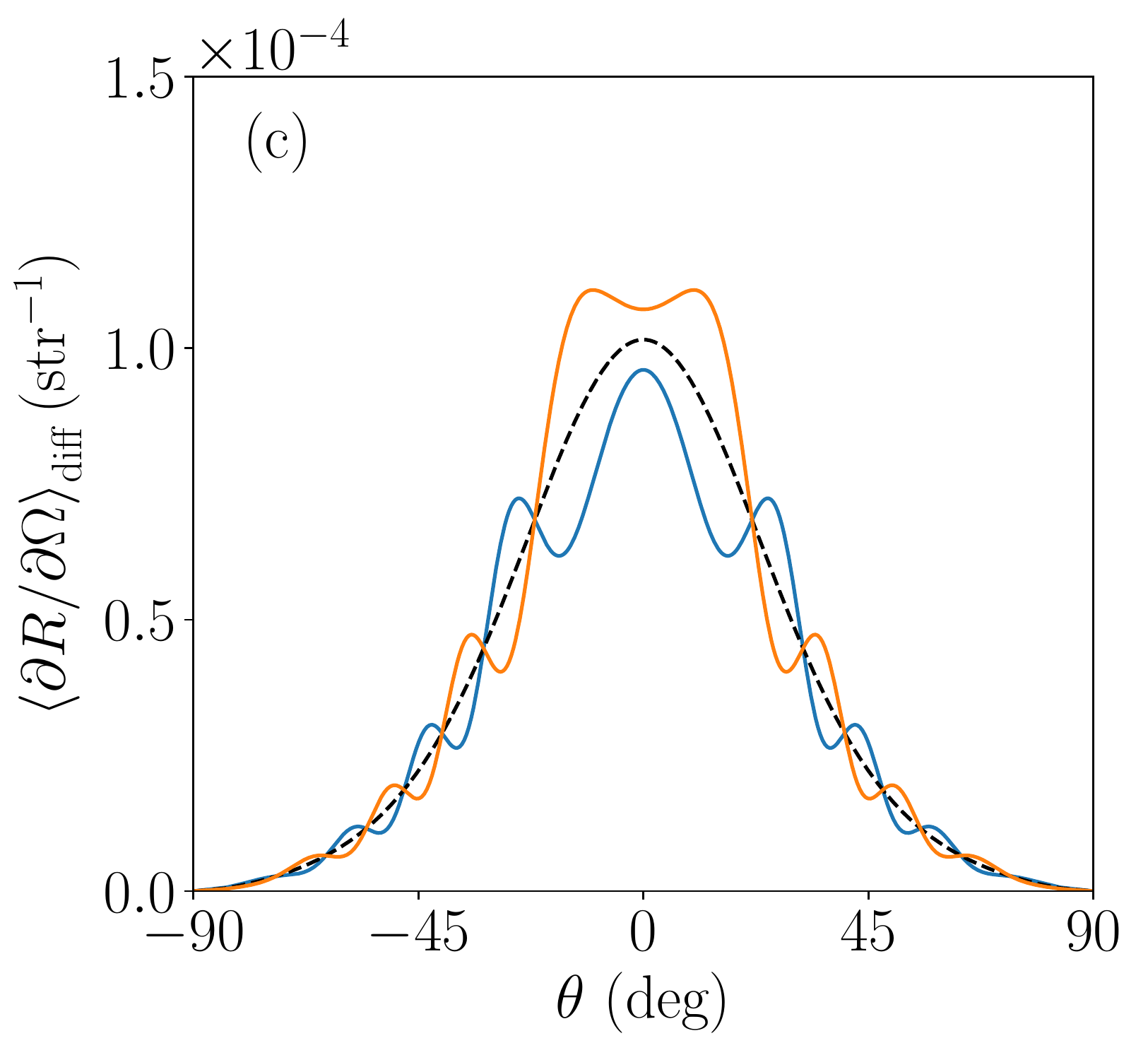}
\caption{The diffuse component of the MDRC for a normally incident and unpolarized plane wave in the surfacelike configuration for (a) $k_0 L \ll 1$, (b) $k_0 L > 1$, and (c) in the genuine volume regime. For (a) and (b) the parameters are identical to those of Fig.~\ref{fig:2}(b). For (c) the parameters are $L = 10 \lambda$, $\ell_\zeta = \ell_{\varepsilon \parallel} = \ell_{\varepsilon \perp} = \lambda /2$, $d= L - \ell_{\varepsilon \perp}$, $\sigma_\zeta = 3.2 \times 10^{-3} \lambda$, and $\sigma_\varepsilon = 0.051$. The dash black line corresponds to the response of a system for which $\eta \approx 1$ and uncorrelated surface and volume disorder $\gamma = 0$. The blue (resp. orange) solid line corresponds to a correlated surface and volume disorder with $\gamma = 1$ (resp. $\gamma = -1$).}
\label{fig:4}
\end{center}
\end{figure*}
\begin{figure*}[t]
\begin{center}
\includegraphics[width=0.32\textwidth,trim = 0.cm 1cm 0cm 0.cm,clip]{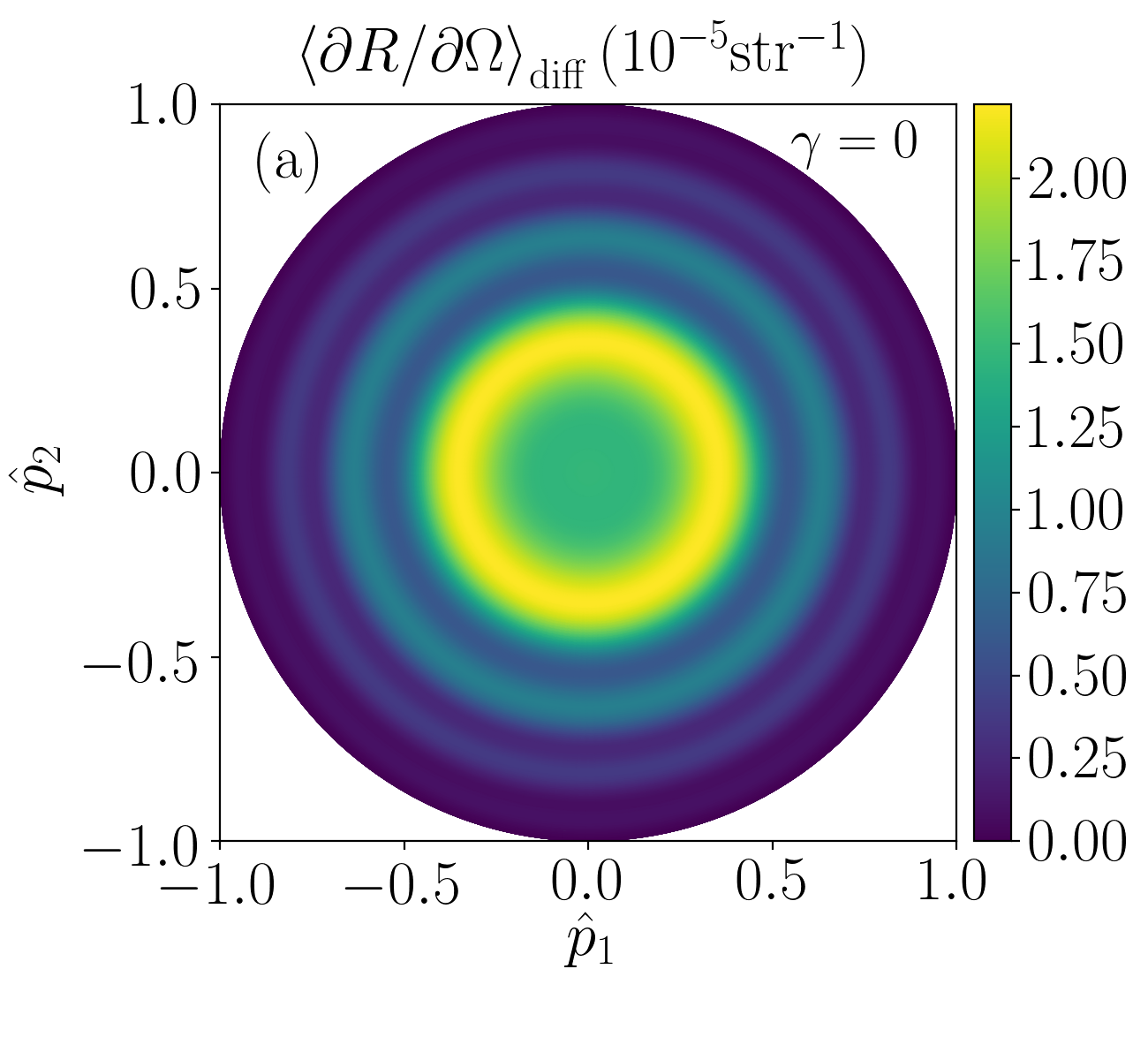}
~
\includegraphics[width=0.32\textwidth,trim = 0.cm 1cm 0cm 0.cm,clip]{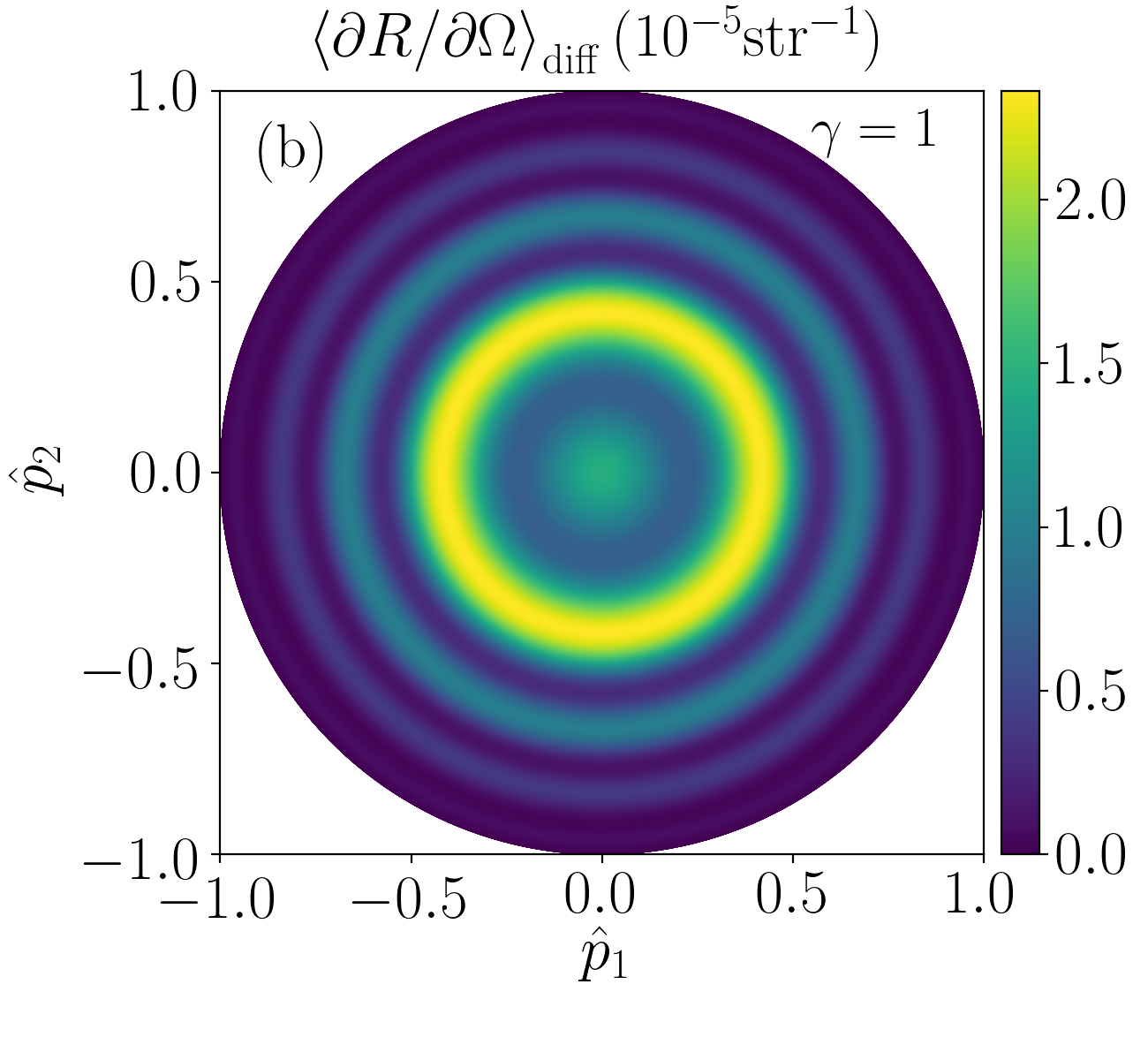}
~
\includegraphics[width=0.32\textwidth,trim = 0.cm 1cm 0cm 0.cm,clip]{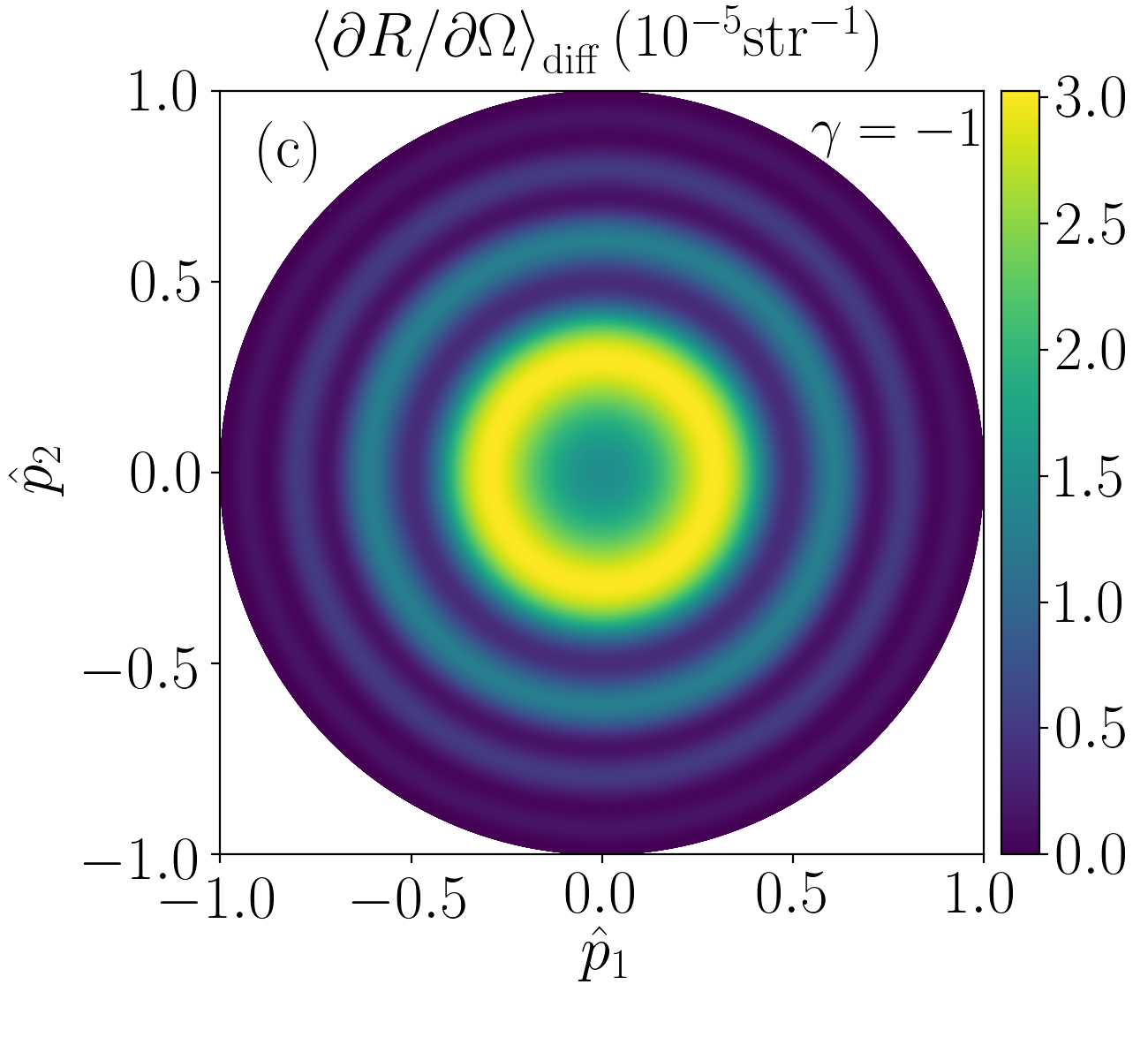}
\caption{The diffuse component of the MDRC in the $\Vie{p}{}{}$ plane (axis normalized as $\Vie{p}{}{} = \sqrt{\varepsilon_1} k_0 \Vie{\hat{p}}{}{}$) for a normally incident and unpolarized plane wave. The system is in the surface-like configuration in the regime $L  > \lambda$. (a) Uncorrelated ($\gamma = 0$), (b) positively correlated ($\gamma = 1$), and (c) negatively correlated ($\gamma = -1$) surface and permittivity fluctuations. The remaining parameters were those assumed in producing the results of Fig.~\ref{fig:2}(d).}
\label{fig:5}
\end{center}
\end{figure*}
\begin{figure*}[t]
\begin{center}
\includegraphics[width=0.32\textwidth]{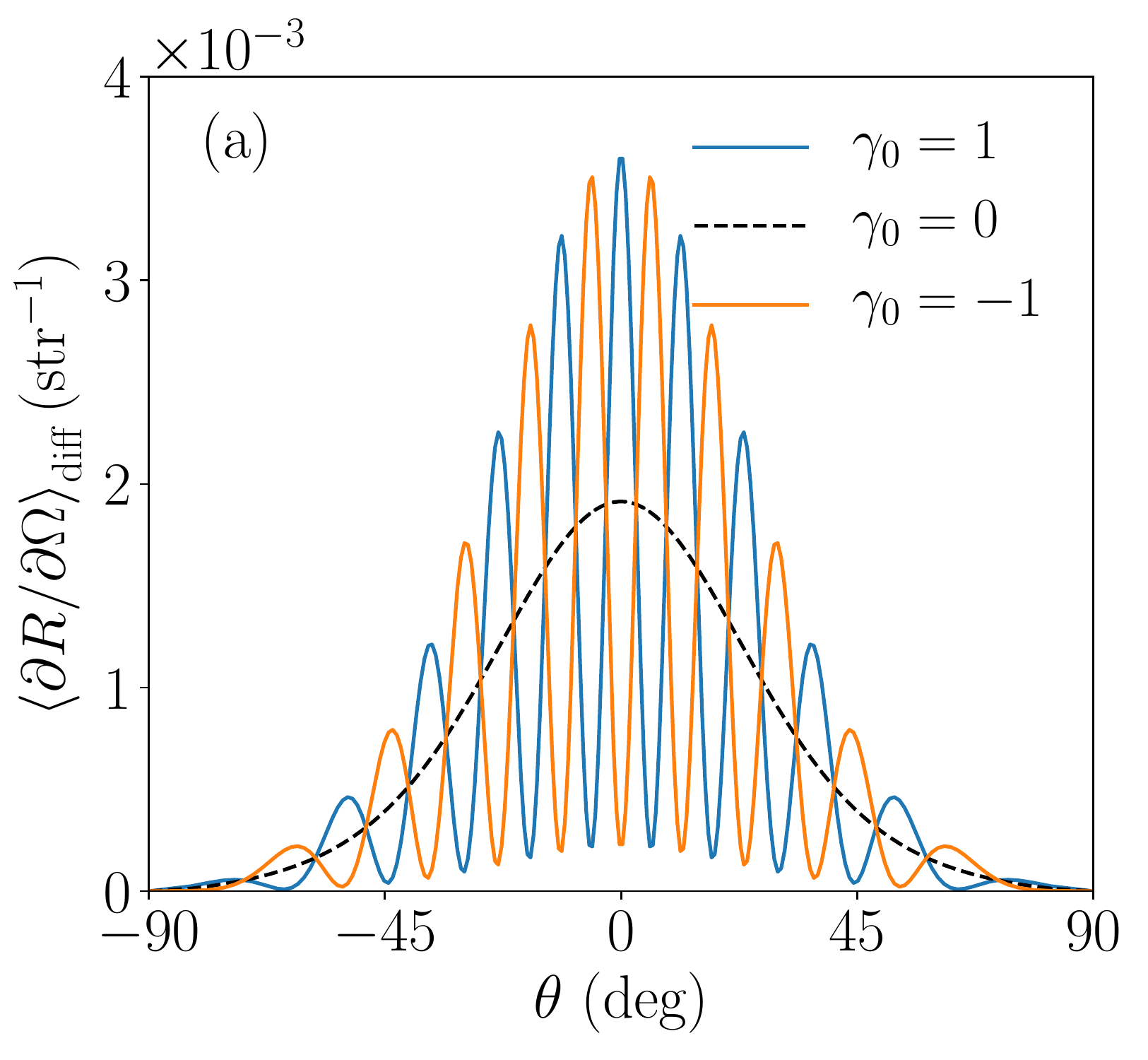}
~
\includegraphics[width=0.32\textwidth]{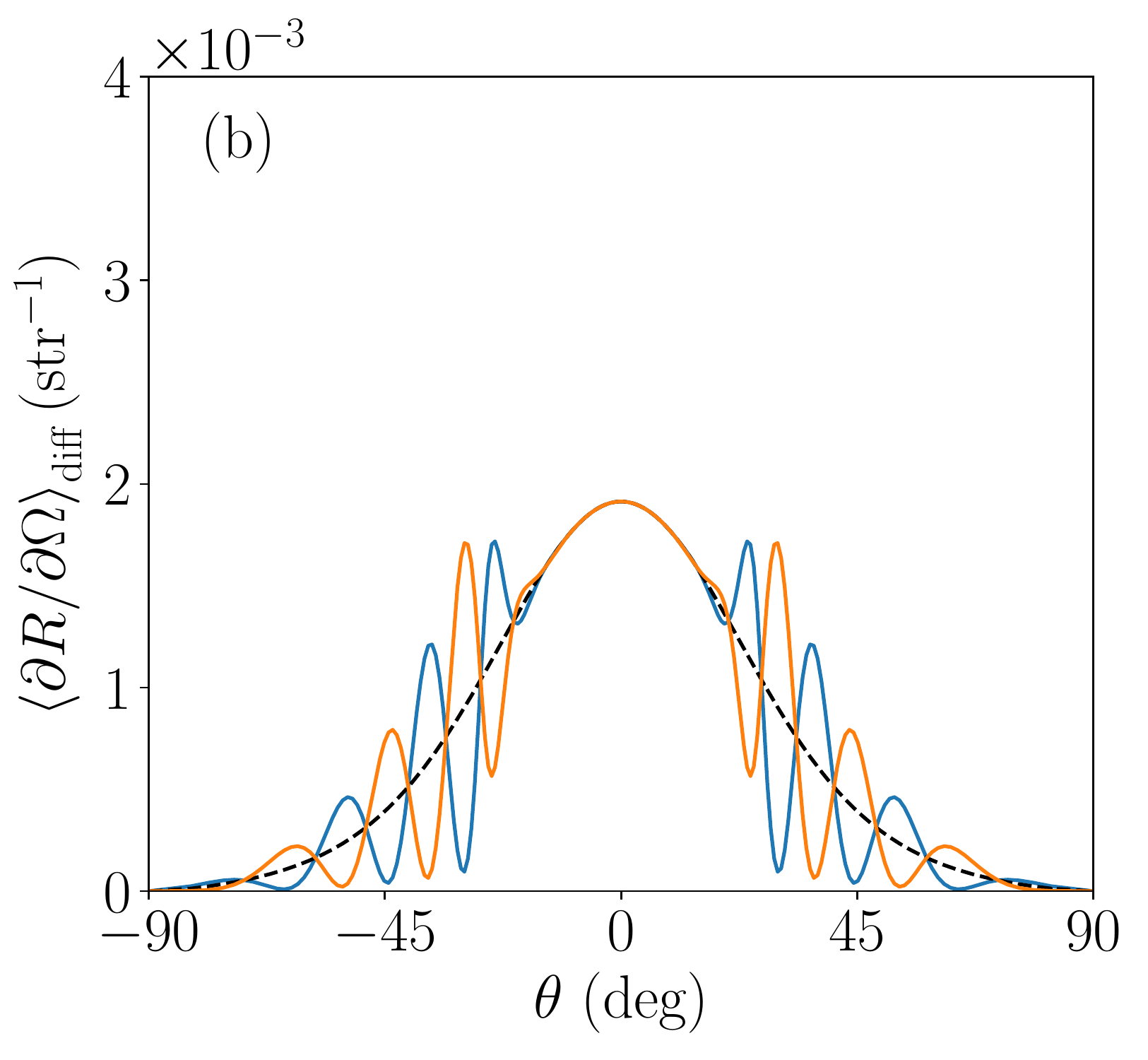}
~
\includegraphics[width=0.32\textwidth]{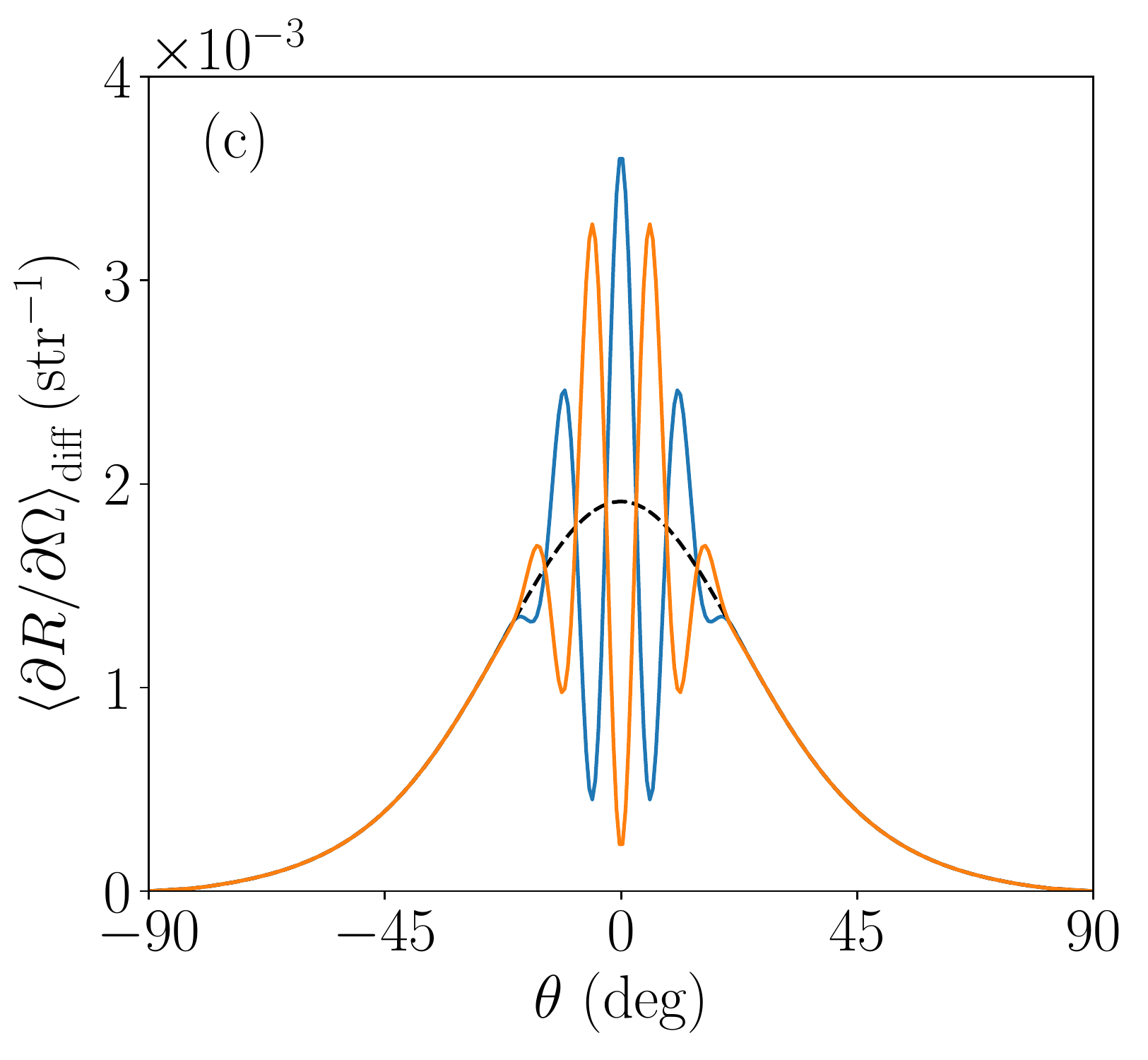}
\includegraphics[width=0.32\textwidth,trim = 0cm 1cm 0cm 0cm,clip]{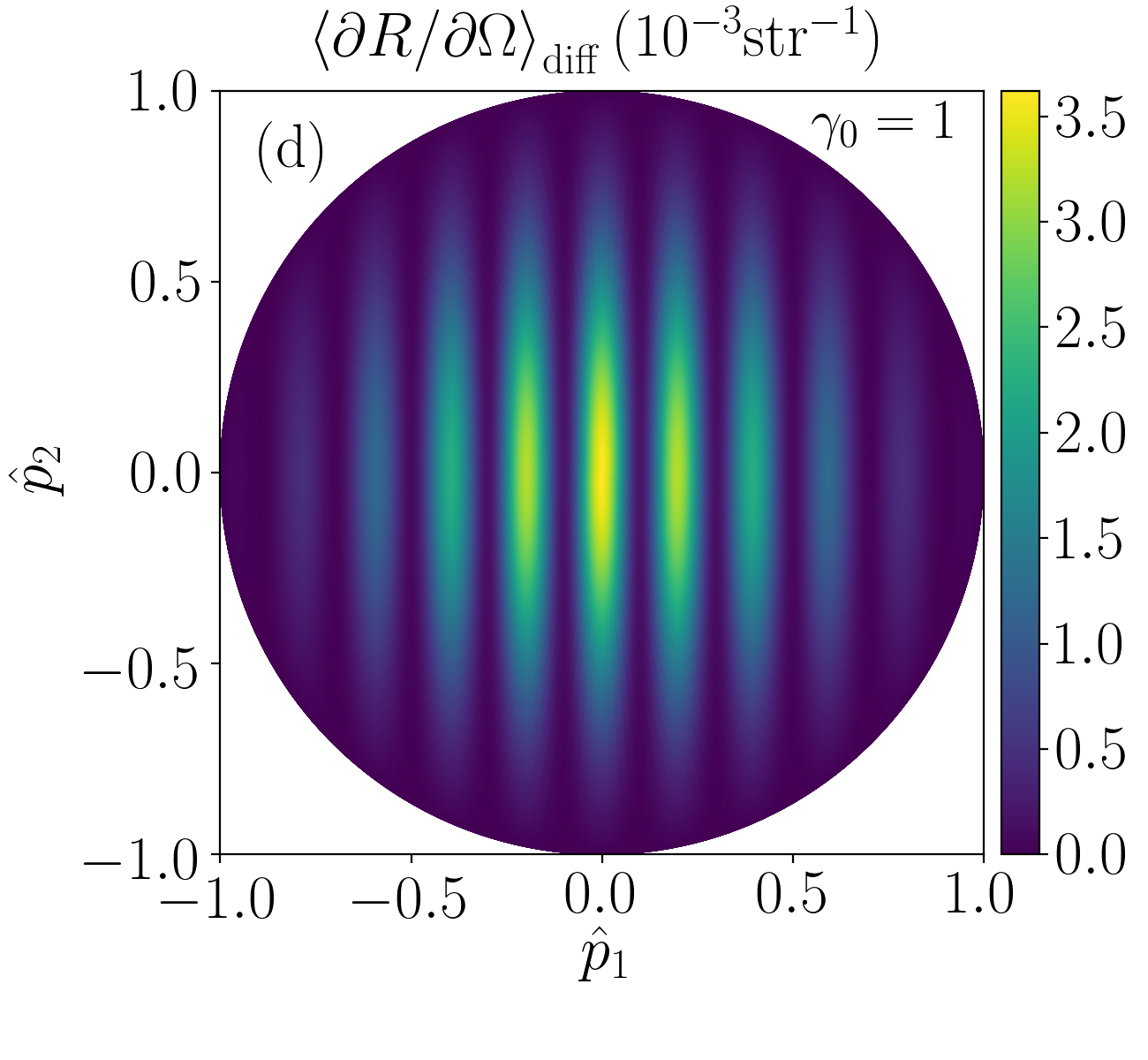}
~
\includegraphics[width=0.32\textwidth,trim = 0cm 1cm 0cm 0cm,clip]{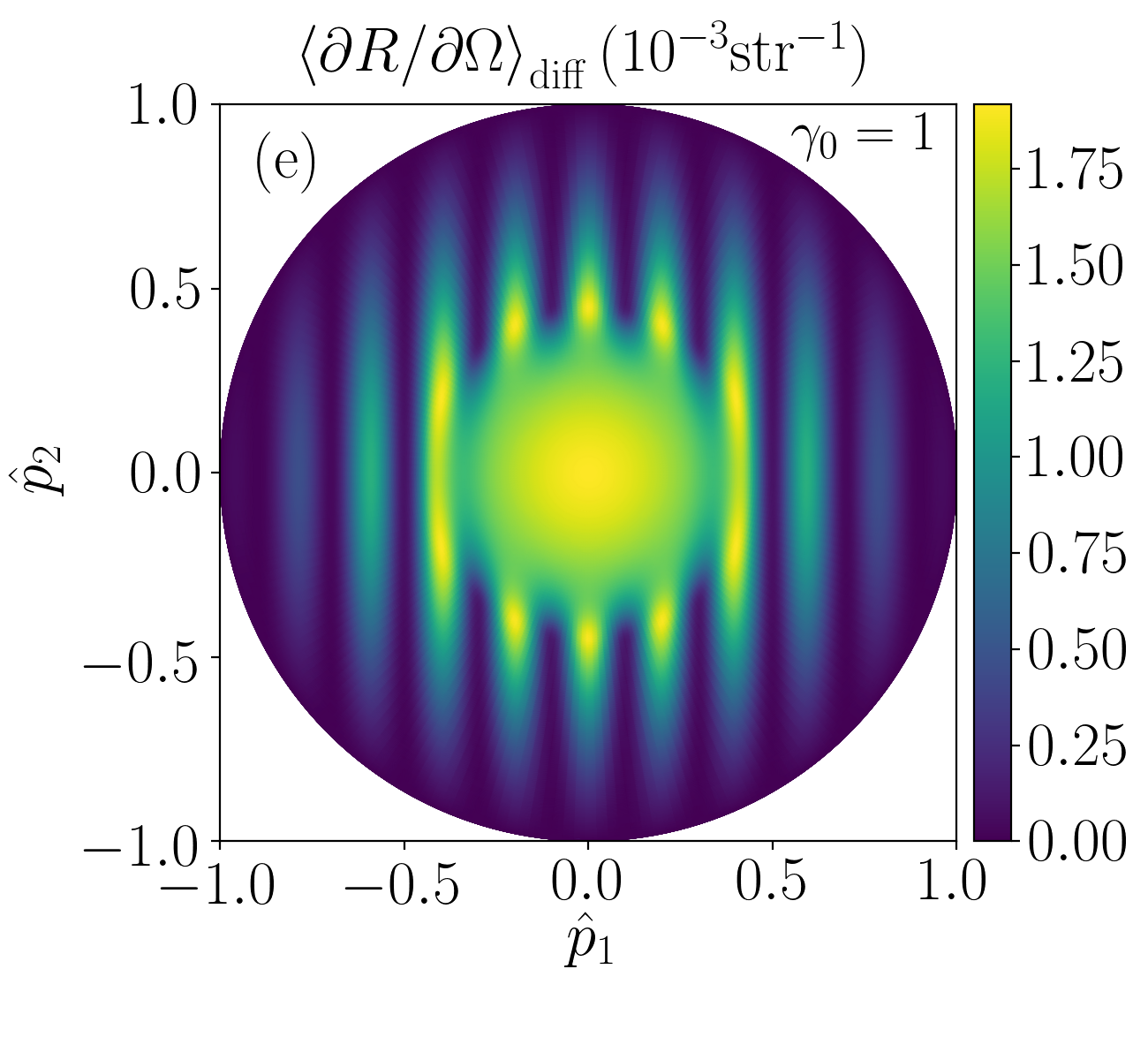}
~
\includegraphics[width=0.32\textwidth,trim = 0cm 1cm 0cm 0cm,clip]{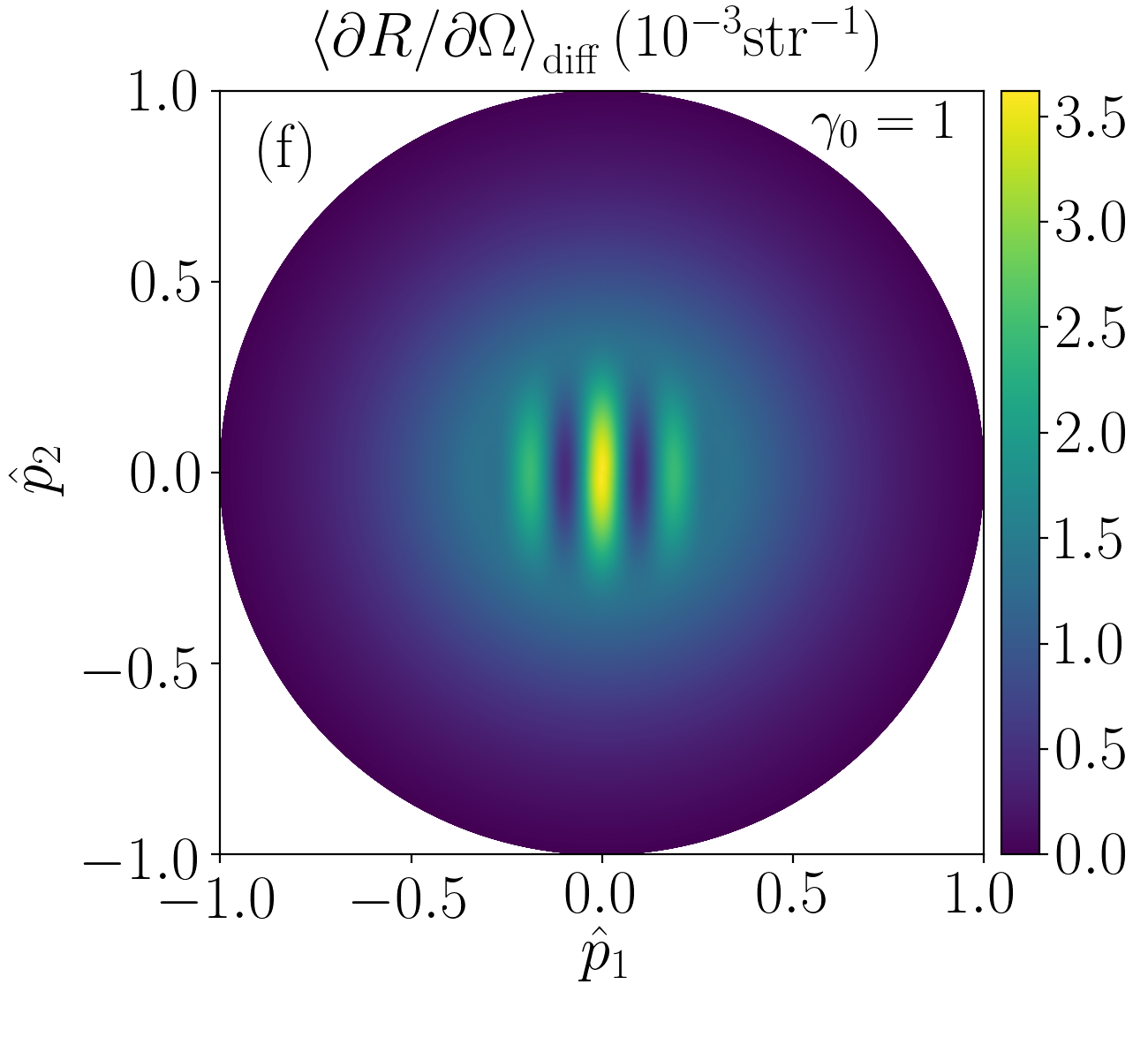}
\caption{The diffuse component of the MDRC as a function of the angle of scattering for in-plane scattering (a-c) and in the $\Vie{p}{}{}$ plane (d-f). All parameters are identical to those assumed in producing the results in Fig.~\ref{fig:2}(b) except for the spectral correlation modulator $\gamma$. (a, d) Shift modulation: $\gamma(\Vie{p}{}{})= \gamma_0 \, \exp(i \Vie{p}{}{} \cdot \Vie{a}{}{})$ with $\Vie{a}{}{} = 5 \lambda \Vie{\hat{e}}{1}{}$. (b, e) Shift and forbidden correlation in the central region: $\gamma(\Vie{p}{}{})= \gamma_0 \, \exp(i \Vie{p}{}{} \cdot \Vie{a}{}{}) \, \big[1- \varphi(2 |\Vie{p}{}{}| / k_1) \big]$. (c, f) Shift and forbidden correlation in the outer region: $\gamma(\Vie{p}{}{})= \gamma_0 \, \exp(i \Vie{p}{}{} \cdot \Vie{a}{}{}) \, \varphi(2 |\Vie{p}{}{}| / k_1)$. The function $\varphi$ is a smooth function with compact support $[-1,1]$ taking values between 0 and 1: $\varphi(x) = \mathrm{H}(1-x^2) \, \exp(4+4/(x^2 -1))$.}
\label{fig:6}
\end{center}
\end{figure*}

\section{Correlated surface and volume disorder}\label{sec:correlation}

\subsection{Surfacelike configuration}

\emph{Uniform spectral correlation} --- We now turn to the situation of correlated surface and volume disorder. In the surfacelike configuration, the depth $d$ of the maximally correlated slice plays a negligible role. We will first take $\gamma(\Vie{p}{}{}) = \gamma \in [-1,1]$ to be a real constant. This corresponds to the case of the surface profile being  correlated with any slice $\Delta \varepsilon (\cdot, x_3)$ without specific tuning of the spectral correlations. In regime 2, the expression in the square brackets in the scalar approximation of Eq.~(\ref{eq:avR12:general}) reads
\begin{align}
&\left\langle \frac{\partial R}{\partial \Omega} \right\rangle_\mathrm{diff} \propto (\varepsilon_2 -\varepsilon_1)^2 \sigma_\zeta^2 \hat{W}_\zeta + \sigma_\varepsilon^2 L^2 \hat{W}_{\varepsilon \parallel} \nonumber \\
&+ 2 (\varepsilon_2 -\varepsilon_1) \sigma_\zeta \sigma_\varepsilon \gamma \, L \hat{W}_\zeta^{1/2} \hat{W}_{\varepsilon \parallel}^{1/2}  \: ,
\label{eq:Rdiff:surflike:qs}
\end{align}
where we have dropped the arguments in the functions for clarity. Note that the positivity of the intensity is ensured by the stochastic model itself. Indeed, it suffices to apply the well-known inequality $2 |a b| \leq a^2+b^2$ with $a = (\varepsilon_2 -\varepsilon_1) \sigma_\zeta \hat{W}_\zeta^{1/2}$ and $b = \sigma_\varepsilon L \hat{W}_{\varepsilon \parallel}^{1/2}$, and to notice that $|\gamma| \leq 1$ to check that the right-hand side of Eq.~(\ref{eq:Rdiff:surflike:qs}) is positive. From this simple remark, it also follows that in order to maximize the effect of the cross correlations on the intensity one must have $|\gamma| = 1$ and equal transverse correlation lengths $\ell_\zeta = \ell_{\varepsilon \parallel}$. In this case, we may assume $\gamma = \pm 1$, which yields
\begin{equation}
\left\langle \frac{\partial R}{\partial \Omega} \right\rangle_\mathrm{diff} \propto \Big[ (\varepsilon_2 -\varepsilon_1) \sigma_\zeta  \pm \sigma_\varepsilon L \Big]^2 \hat{W}_\zeta \: .
\end{equation}
Consequently, for $(\varepsilon_2 -\varepsilon_1) \sigma_\zeta = \sigma_\varepsilon L$, i.e., for equal contribution from the surface and volume disorder to the scattering ($\eta_2 = 1$), the resulting diffusely scattered intensity may completely vanish ($\gamma = -1$) or may double ($\gamma = 1$) compared to the uncorrelated case. Such situations are illustrated in Fig.~\ref{fig:4}(a) where the diffuse component of the MDRC for in-plane scattering is presented for uncorrelated, positively correlated and negatively correlated surface and volume disorder in regime 2. 
 The reason why the MDRC does not perfectly vanish for $\gamma = -1$ (nor is it exactly doubled for $\gamma = 1$) is that we used the numerical evaluation of $I$ and $J$ (which can be considered as exact) rather than their asymptotic expressions. As $L \to 0$, the asymptotic expressions would become more accurate and the signal would indeed vanish for $\gamma = -1$. It is important to note that we have constructed a first example for which the splitting rule~\cite{Sentenac:02,Guerin:07} for the intensity does not apply, even in the single-scattering regime. The splitting rule fails here due to the constructive or destructive interference induced by the cross correlation between paths scattered on the surface or in the volume. This result should not come as a surprise though. Indeed, in the chosen regime, the dielectric fluctuations occur only in a thin layer below the surface since we have $L \ll \ell_{\varepsilon \perp} \ll \lambda$. Positively correlating the surface profile and the dielectric fluctuations can be considered as producing an effective surface with larger  dielectric jumps or larger rms roughness, hence enhancing the diffusely scattered power. Conversely, negatively correlating the surface profile and the dielectric fluctuations can be considered as dampening the dielectric jumps for the equivalent surface, hence reducing the scattered power. As an illustrative picture, the reader may refer to the scattering geometry in Fig.~\ref{fig:volume_vs_surface_like}(b) and let $L$ be as small as the rms roughness of the surface profile.
 Note that a similar enhancement or attenuation of scattering due to surface-surface correlation was  observed for randomly rough films \cite{Amra:86,Amra:92,Soriano:19,Banon2018a}.

Can such dramatic effects be observed beyond the sub-wavelength regime? We have seen in Fig.~\ref{fig:2}(d) that in regime 4, the contribution from the permittivity fluctuations to the diffuse component of the MDRC exhibits interference rings, which adds to the broad bell shaped signal coming from the surface. Let us revisit this situation in the presence of correlations between the surface profile and the permittivity fluctuations. In this regime, the diffuse component of the MDRC is proportional to
\begin{align}
&\left\langle \frac{\partial R}{\partial \Omega} \right\rangle_\mathrm{diff} \propto (\varepsilon_2 -\varepsilon_1)^2 \sigma_\zeta^2 \hat{W}_\zeta + \sigma_\varepsilon^2 \hat{W}_{\varepsilon \parallel} \, \frac{4 \sin^2 (\alpha L /2)}{\alpha^2} \nonumber \\
&+ 2 (\varepsilon_2 -\varepsilon_1) \sigma_\zeta \sigma_\varepsilon \gamma \hat{W}_\zeta^{1/2} \hat{W}_{\varepsilon \parallel}^{1/2} \, \frac{\sin(\alpha L)}{\alpha} \: .
\end{align}
Here again it is straightforward to verify that the intensity remains positive. In contrast to the sub-wavelength case, we observe that the oscillations in the coupling term have the same frequency as the oscillations of the volume contribution but phase shifted by $\pi / 2$. This results in a modification of the interference pattern as illustrated in Fig.~\ref{fig:4}(b). 
 Figure~\ref{fig:5} shows the full angular distribution of the diffuse component of the MDRC for the three aforementioned cases ($\gamma = 0, \pm 1$) where the modulation of the interference rings can be appreciated.

\emph{Modulated spectral correlation} --- Let us now explore the additional degree of freedom offered by the spectral correlation modulator, and let $\gamma$ explicitly depend on the in-plane wave vector $\Vie{p}{}{}$. To this end, we reconsider the situation in Fig.~\ref{fig:4}(a) (regime 2) but for different forms for $\gamma$, namely a shift of cross correlation $\gamma(\Vie{p}{}{}) = \gamma_0 \, \exp(i \Vie{p}{}{} \cdot \Vie{a}{}{})$, and a shift combined with a spectral forbidden region of correlation  localized in either the domain $|\Vie{p}{}{}|< k_1/2$ or the domain $|\Vie{p}{}{}| > k_1 / 2$ (see the caption of Fig.~\ref{fig:6} for details). Figure~\ref{fig:6} presents the diffuse component of the MDRC for these three cases. First, we observe in Figs.~\ref{fig:6}(a) and \ref{fig:6}(d) that the correlation shift induces interference fringes the frequency of which in the $\Vie{p}{}{}$ plane and orientation are determined by the shift vector $\Vie{a}{}{}$. We also note that the resulting MDRC is bounded by the MDRCs obtained in Fig.~\ref{fig:4}(a) for uniform spectral correlations, and thus they determine the envelop of the oscillating MDRC in Fig.~\ref{fig:6}(a). Note that the maxima and minima of the MDRC can be exchanged by choosing $\gamma_0$ to be equal to either $-1$ or $1$. The two other forms of $\gamma$ that we consider in Figs.~\ref{fig:6}(b, c, e, f) exhibit fringes as in the previous case but only in the determined allowed regions, either outside [Figs.~\ref{fig:6}(b, e)] or inside [Figs.~\ref{fig:6}(c, f)] of the domain $|\Vie{p}{}{}|<k_1/2$. 

These examples of engineering of the cross correlation illustrate a very general concept. The average interference pattern results from constructive and destructive interference between correlated optical paths. In the single-scattering regime, an average interference pattern results either from the design of the power spectral density of a single stochastic process (e.g. the surface profile), or from the design of the cross-spectral power density between two stochastic processes. Well known examples, in the case of a single stochastic process, are the surface scattering of band-limited uniform diffusers \cite{Leskova:98} and the volume scattering of hyperuniform lattices of scatterers \cite{Torquato:03,Leseur:16}. The originality here, is that we have assumed Gaussian forms of the auto-power spectral densities for the surface and volume disorder, which \emph{independently} diffuse broadly, but which can exhibit exotic interference patterns when correlated.

Beyond the subwavelength regime, the interference pattern observed in regime 2 remains but is combined with the interference rings discussed previously in Figs.~\ref{fig:4} and \ref{fig:5}. We would like to stress the  physical origin of these two types of interference. The rings observed in the regime $k_0 L > 1$ result from the constructive and destructive interferences between optical paths scattered along a line $\Vie{x}{\parallel}{} = $ constant, for which $\Delta \varepsilon$ is constant (keep in mind that we consider the surfacelike regime here). On the other hand, the interference pattern originating from the cross correlation is chiefly an effect resulting from correlated paths involving scattering centers located at different positions in the $x_1 x_2$ plane. A clear example is the shift of correlation, which even in the subwavelength regime, i.e., when the phase shifts due to propagation along $x_3$ can be neglected, produces interference fringes which are entirely determined by the in-plane shift vector $\Vie{a}{}{}$.

\subsection{Genuine volume configuration}

In regimes 1 and 3, the physics discussed in the surfacelike configuration (regimes 2 and 4) remains valid but the interference effect induced by the cross correlation is weaker. Indeed, in regime 1 for example, the expression for the MDRC in the scalar wave approximation reads
\begin{align}
&\left\langle \frac{\partial R}{\partial \Omega} \right\rangle_\mathrm{diff} \propto (\varepsilon_2 -\varepsilon_1)^2 \sigma_\zeta^2 \hat{W}_\zeta + \sqrt{\pi} L \ell_{\varepsilon \perp} \, \sigma_\varepsilon^2 \hat{W}_{\varepsilon \parallel} \nonumber \\
&+ 2 (\varepsilon_2 -\varepsilon_1) \sigma_\zeta \sigma_\varepsilon \, \frac{\sqrt{\pi}}{2} \ell_{\varepsilon \perp} \, \gamma \hat{W}_\zeta^{1/2} \hat{W}_{\varepsilon \parallel}^{1/2} \: .
\end{align}
We see that while the volume term scales as $L \ell_{\varepsilon \perp}$, the cross term scales as $\ell_{\varepsilon \perp}$, which is the length scale for the range of the cross correlation. This is in contrast with regime 2 where the cross term scales as $L$. Consequently, the cross term is small compared to the volume term  (and the surface term for $\eta = 1$) in the regime $\ell_{\varepsilon \perp} \ll L$. This situation is illustrated in Fig.~\ref{fig:4}(c) where the depth of  the maximally correlated dielectric layer was chosen to be taken one correlation length away from the bottom edge of the fluctuating domain, $d = L - \ell_{\varepsilon \perp}$. We observe that the diffuse component of the MDRC for the correlated systems oscillates weakly around that of the uncorrelated system. The physical origin of these oscillations is similar to that of the Sel\'{e}nyi rings occurring in rough dielectric films \cite{Selenyi1911,Lu1998,Banon2018a}. The reason for the less pronounced effect is that only a thin layer, of thickness $\ell_{\varepsilon \perp}$,  contributes to the average interference effect, on top of the background signal coming from the thick layer of thickness $L-\ell_{\varepsilon \perp}$ the dielectric fluctuations of which are not correlated to the surface. The amplitude of the oscillations with respect to the uncorrelated signal thus scales roughly as $\ell_{\varepsilon \perp} / L$.

\section{Polarization response and surface-volume decomposition}\label{sec:pola}

 Before concluding, it is interesting to discuss how the polarization response can be used to decompose the total scattered intensity into its surface and volume contributions in the case of uncorrelated surface and volume disorder. The starting point of this discussion will be based on  Eq.~(\ref{eq:avR12:general}) for $\gamma = 0$. The diffuse component of the MDRC in this case is of the form
\begin{align}
\left\langle \frac{\partial R_{\mu \nu}}{\partial \Omega} (\Vie{p}{}{},\Vie{p}{0}{}) \right\rangle_\mathrm{diff} = \:  &\mathcal{S}(\Vie{p}{}{},\Vie{p}{0}{}) \, |\rho_{\zeta, \mu \nu} (\Vie{p}{}{},\Vie{p}{0}{})|^2 \nonumber\\
&+ \mathcal{V}(\Vie{p}{}{},\Vie{p}{0}{}) \, |\rho_{\varepsilon, \mu \nu} (\Vie{p}{}{},\Vie{p}{0}{})|^2 \: .
\label{eq:Rmunu}
\end{align}
The functions $\mathcal{S}$ and $\mathcal{V}$ are proportional to $\sigma_\zeta^2 \hat{W}_\zeta$ and $\sigma_\varepsilon^2 \hat{W}_\varepsilon$, respectively. In an experimental setup for optical sample characterization, these functions are unknowns. However, within the single-scattering approximation, the polarization coupling factors are known. To determine $\mathcal{S}$ and $\mathcal{V}$ (and consequently assess the surface and volume statistical property of the sample) one may proceed as follows. Measure $\left\langle \frac{\partial R_{p p}}{\partial \Omega} (\Vie{p}{}{},\Vie{p}{0}{}) \right\rangle_\mathrm{diff}$ and $\left\langle \frac{\partial R_{s s}}{\partial \Omega} (\Vie{p}{}{},\Vie{p}{0}{}) \right\rangle_\mathrm{diff}$ in the plane of incidence for a given oblique angle of incidence (or a set of angles of incidence) for which $\rho_{\zeta, pp}(\Vie{p}{}{},\Vie{p}{0}{}) \neq \rho_{\varepsilon, pp}(\Vie{p}{}{},\Vie{p}{0}{})$. Note that in the plane of incidence $\rho_{\zeta, ss} = \rho_{\varepsilon, ss} $. We thus have for each set of measurements in a direction $\Vie{p}{}{}$, a linear set of two equations with two unknowns, namely Eq.~(\ref{eq:Rmunu}) with $\mu = \nu = p$ or $\mu = \nu = s$, which can be inverted to give
\begin{subequations}
\begin{align}
\mathcal{S} &= \frac{|\rho_{\varepsilon, ss}|^2 \left\langle \frac{\partial R_{pp}}{\partial \Omega} \right\rangle_\mathrm{diff} - |\rho_{\varepsilon, pp}|^2 \left\langle \frac{\partial R_{ss}}{\partial \Omega} \right\rangle_\mathrm{diff} }{ |\rho_{\varepsilon, ss}|^2 |\rho_{\zeta, pp}|^2 -  |\rho_{\varepsilon, pp}|^2 |\rho_{\zeta, ss}|^2} \\
\mathcal{V} &=  \frac{|\rho_{\zeta, ss}|^2 \left\langle \frac{\partial R_{pp}}{\partial \Omega} \right\rangle_\mathrm{diff} - |\rho_{\zeta, pp}|^2 \left\langle \frac{\partial R_{ss}}{\partial \Omega} \right\rangle_\mathrm{diff} }{ |\rho_{\zeta, ss}|^2 |\rho_{\varepsilon, pp}|^2 -  |\rho_{\zeta, pp}|^2 |\rho_{\varepsilon, ss}|^2} \: .
\end{align}
\label{eq:invert_system}
\end{subequations}
A situation of particular interest is that corresponding to the Brewster scattering angle, for which $\rho_{\zeta, pp}$ vanishes. This angle of scattering depends on the angle of incidence and can readily be predicted from the definition of $\rho_{\zeta, pp}$ \citep{Banon:2018}. In principle, one could vary the angles of incidence and of observation so that $\rho_{\zeta, pp} = 0$, and  measure an intensity resulting only from volume scattering. However, it may be simpler to use Eq.~(\ref{eq:invert_system}) which is valid for any set of angles of incidence and scattering provided that $\rho_{\zeta, pp}(\Vie{p}{}{},\Vie{p}{0}{}) \neq \rho_{\varepsilon, pp}(\Vie{p}{}{},\Vie{p}{0}{})$ (in the plane of incidence).\\

\section{Conclusion and perspectives}\label{sec:conclusion}

The single-scattering theory derived in this paper has allowed us to obtain a number of results. First, we have mapped out a diagram of predominance of surface and volume scattering depending on length scales characteristic of the disorder. Second, we have shown how polarimetric measurements can be used to discriminate surface scattering from volume scattering in the total diffusely scattered intensity. Finally, we have explored interference effects induced by surface-volume correlations. This study required the construction of a model of surface-volume correlation, to understand the degrees of freedom involved, and how the latter can be used either for shaping the diffuse interference pattern or for assessing statistical information about the disorder.

New perspectives are now open in different directions. First, it will be of interest to investigate to which extent the presented regimes and interference effects are robust when multiple scattering events are taken into account.
 Second, despite its relative simplicity, the single-scattering theory is of interest for a wide range of applications. A first application is the optical characterization of disordered thin films, where  both fluctuation of refractive index, due to material phase separation, and surface roughness can simultaneously be present and correlated \cite{Aytug:2013,Aytug:2015}.  
 Another application is the detection of label-free single nano-objects (like proteins) in optical interferometric microscopy. The Rayleigh scattering signal from such small objects is often merged in a background of speckles coming from weak, nanometric, surface roughness or density fluctuations in the substrate or cover slip \cite{Piliarik:2014, Taylor:2019}. Including small objects in the framework presented in the paper is straightforward, opening a way to analytical treatments for precise background subtraction by taking advantage of the knowledge of the interference between the background speckle field and the field scattered by the nano-object. Finally, we have shown that the cross correlation function could in principle be designed to create exotic interference patterns in the diffusely scattered light. This is a general single-scattering result for two types of correlated disorders, here a surface and a volume with a fluctuating index of refraction. These results suggest that information could in principle be encoded in a pair of disordered media which would only be decoded by a light scattering experiment from or through both media. This idea is reminiscent of that of optical image encryption with random phase masks introduced by Refregier and Javidi in Ref.~\cite{Refregier:95}.

\section*{Acknowledgment}
This research was supported by the French National Research Agency (Grant No. ANR-15-CHIN-0003) and the LABEX WIFI (Laboratory of Excellence within the French Program "Investment for the Future") under Grants No. ANR-10-LABX-24 and No. ANR-10-IDEX-0001-02-PSL*. The authors are grateful to Romain Pierrat for fruitful discussions.


\appendix

\begin{widetext}

\section{Zeroth-order field} \label{App:zero:order}

Equation~(\ref{eq:single:scattering}) giving the field obtained in the Born approximation requires the zeroth-order field $\Vie{E}{}{(0)}$ solution of the scattering problem for the reference system. If we consider the case of a monochromatic incident plane wave
\begin{equation}
\Vie{E}{0}{} (\Vie{x}{}{}) = \left[ \Cie{E}{0,p}{} \, \Vie{\hat{e}}{1,p}{-} (\Vie{p}{0}{}) +  \Cie{E}{0,s}{} \, \Vie{\hat{e}}{s}{} (\Vie{p}{0}{}) \right] \: \exp \left( i \Vie{k}{1}{-}(\Vie{p}{0}{}) \cdot \Vie{x}{}{} \right) \: ,
\label{eq:inc:field}
\end{equation}
the reference field is given by
\begin{equation}
\Vie{E}{}{(0)} (\Vie{x}{}{}) = \begin{cases}
   \Vie{E}{0}{} (\Vie{x}{}{}) +  \left[ r_{21}^{(p)} (\Vie{p}{0}{}) \Cie{E}{0,p}{} \, \Vie{\hat{e}}{1,p}{+} (\Vie{p}{0}{}) +  r_{21}^{(s)} (\Vie{p}{0}{}) \Cie{E}{0,s}{} \, \Vie{\hat{e}}{s}{} (\Vie{p}{0}{}) \right] \: \exp \left( i \Vie{k}{1}{+}(\Vie{p}{0}{}) \cdot \Vie{x}{}{} \right)  & \quad \text{if } x_3 > 0\\
    \left[ t_{21}^{(p)} (\Vie{p}{0}{}) \Cie{E}{0,p}{} \, \Vie{\hat{e}}{2,p}{-} (\Vie{p}{0}{}) +  t_{21}^{(s)} (\Vie{p}{0}{}) \Cie{E}{0,s}{} \, \Vie{\hat{e}}{s}{} (\Vie{p}{0}{}) \right] \: \exp \left( i \Vie{k}{2}{-}(\Vie{p}{0}{}) \cdot \Vie{x}{}{} \right)  		& \quad \text{if } x_3 < 0
  \end{cases}
 \: .
 \label{eq:ref_field}
\end{equation}
\end{widetext}
Here we have defined the wave vectors and polarization vectors parametrized by $\Vie{p}{}{} = p_1 \Vie{\hat{e}}{1}{} + p_2 \Vie{\hat{e}}{2}{}$ and $j \in \{1, 2\}$ by
\begin{subequations}
\begin{align}
\Vie{k}{j}{\pm} (\Vie{p}{}{}) &= \Vie{p}{}{} \pm \alpha_j(\Vie{p}{}{}) \, \Vie{\hat{e}}{3}{} \\
k_j &= |\Vie{k}{j}{\pm}| = \varepsilon_j^{1/2} k_0 \\
\alpha_j(\Vie{p}{}{}) &= \left( k_j^2 - \Vie{p}{}{2} \right)^{1/2} \:, \: \Re(\alpha_j) \geq 0, \: \Im(\alpha_j) \geq 0 \\
\Vie{\hat{e}}{s}{} (\Vie{p}{}{}) &= \Vie{\hat{e}}{3}{} \times \Vie{\hat{p}}{}{} \\
\Vie{\hat{e}}{j,p}{\pm} (\Vie{p}{}{}) &= \frac{\pm \alpha_j(\Vie{p}{}{}) \Vie{\hat{p}}{}{} - |\Vie{p}{}{}| \, \Vie{\hat{e}}{3}{}}{k_j} \: , \label{eq:wave_vector:ep}
\end{align}
\label{eq:wave_vector}
\end{subequations}

and the Fresnel factors for a plane wave incident from medium $i$ to $j$ are given by
\begin{subequations}
\begin{align}
r_{ji}^{(s)}(\Vie{p}{}{}) &= \frac{\alpha_i(\Vie{p}{}{})-\alpha_j(\Vie{p}{}{})}{\alpha_i(\Vie{p}{}{})+\alpha_j(\Vie{p}{}{})} \\
r_{ji}^{(p)}(\Vie{p}{}{}) &= \frac{\varepsilon_j \alpha_i(\Vie{p}{}{})-\varepsilon_i \alpha_j(\Vie{p}{}{})}{\varepsilon_j \alpha_i(\Vie{p}{}{})+\varepsilon_i \alpha_j(\Vie{p}{}{})} \label{eq:fresnel:rp}\\
t_{ji}^{(s)}(\Vie{p}{}{}) &= \frac{2 \alpha_i(\Vie{p}{}{})}{\alpha_i(\Vie{p}{}{})+\alpha_j(\Vie{p}{}{})} \\
t_{ji}^{(p)}(\Vie{p}{}{}) &= \frac{2\sqrt{\varepsilon_j \varepsilon_i} \alpha_i(\Vie{p}{}{})}{\varepsilon_j \alpha_i(\Vie{p}{}{})+\varepsilon_i \alpha_j(\Vie{p}{}{})} \: . \label{eq:fresnel:tp}
\end{align}
\label{eq:fresnel:rt}
\end{subequations}
\begin{widetext}
The two-dimensional Fourier transform of the zeroth-order field, also known as Weyl expansion, thus reads
\begin{equation}
\Vie{\hat{E}}{}{(0)} (\Vie{p}{}{},x_3) = \begin{cases}
 (2 \pi)^2 \, \delta(\Vie{p}{}{} - \Vie{p}{0}{}) \:  \Vie{\hat{E}}{1}{(0)} (\Vie{p}{0}{},x_3)  & \quad \text{if } x_3 > 0\\
 (2 \pi)^2 \, \delta(\Vie{p}{}{} - \Vie{p}{0}{}) \:  \Vie{\hat{E}}{2}{(0)} (\Vie{p}{0}{},x_3)  & \quad \text{if } x_3 < 0
  \end{cases}
 \: ,
 \label{eq:fresnel:fourier}
\end{equation}
where the zeroth-order fields $\Vie{\hat{E}}{1}{(0)}$ and $\Vie{\hat{E}}{2}{(0)}$ are given by
\begin{subequations}
\begin{align}
\Vie{\hat{E}}{1}{(0)} (\Vie{p}{0}{},x_3) =  &\Big[ \Cie{E}{0,p}{} \, \Vie{\hat{e}}{1,p}{-} (\Vie{p}{0}{}) + \Cie{E}{0,s}{} \, \Vie{\hat{e}}{s}{} (\Vie{p}{0}{}) \Big] \, \exp \left( - i \alpha_1(\Vie{p}{0}{}) \, x_3 \right) \nonumber\\
&+  \Big[ r_{21}^{(p)} (\Vie{p}{0}{}) \, \Cie{E}{0,p}{}  \, \Vie{\hat{e}}{1,p}{+} (\Vie{p}{0}{}) + r_{21}^{(s)} (\Vie{p}{0}{}) \, \Cie{E}{0,s}{}  \, \Vie{\hat{e}}{s}{} (\Vie{p}{0}{}) \Big] \: \exp \left( i \alpha_1(\Vie{p}{0}{}) \, x_3 \right) \label{eq:E01}\\
\Vie{\hat{E}}{2}{(0)} (\Vie{p}{0}{},x_3) = &\Big[  t_{21}^{(p)} (\Vie{p}{0}{})  \, \Cie{E}{0,p}{}  \, \Vie{\hat{e}}{2,p}{-} (\Vie{p}{0}{}) + t_{21}^{(s)} (\Vie{p}{0}{}) \, \Cie{E}{0,s}{}  \, \Vie{\hat{e}}{s}{} (\Vie{p}{0}{})  \Big] \: \exp \left( - i \alpha_2(\Vie{p}{0}{}) \, x_3 \right) \: . \label{eq:E02}
\end{align}
\label{eq:E0}
\end{subequations}

\section{Derivation of identity~(\ref{eq:GE:id})} \label{App:GE:identity}

We prove here the identity given in Eq.~(\ref{eq:GE:id}). To this end, let us recall the expression of the Green's function as given in Ref.~\cite{Sipe:87}.
The Green's function for the reference system expressed for $x_3 > x_3^\prime > 0$ reads
\begin{equation}
\Vie{\hat{G}}{}{}(\Vie{p}{}{},x_3,x_3^\prime) = \Vie{\hat{G}}{1}{(d)}(\Vie{p}{}{},x_3-x_3^\prime) + \Vie{\hat{G}}{}{(r)}(\Vie{p}{}{},x_3,x_3^\prime) \: ,
\label{eq:green:pp}
\end{equation}
with
\begin{align}
\Vie{\hat{G}}{1}{(d)}(\Vie{p}{}{},x_3-x_3^\prime) &= \frac{i}{2 \alpha_1(\Vie{p}{}{})} \, \Big[ \Vie{\hat{e}}{1,p}{+}(\Vie{p}{}{}) \otimes \Vie{\hat{e}}{1,p}{+}(\Vie{p}{}{}) + \Vie{\hat{e}}{s}{}(\Vie{p}{}{}) \otimes \Vie{\hat{e}}{s}{}(\Vie{p}{}{}) \Big] \, \exp \Big( i \alpha_1(\Vie{p}{}{}) (x_3 - x_3^\prime) \Big)\\
\Vie{\hat{G}}{}{(r)}(\Vie{p}{}{},x_3,x_3^\prime)  &= \frac{i}{2 \alpha_1(\Vie{p}{}{})} \, \Big[ r_{21}^{(p)}(\Vie{p}{}{}) \Vie{\hat{e}}{1,p}{+}(\Vie{p}{}{}) \otimes \Vie{\hat{e}}{1,p}{-}(\Vie{p}{}{}) + r_{21}^{(s)}(\Vie{p}{}{}) \Vie{\hat{e}}{s}{}(\Vie{p}{}{}) \otimes \Vie{\hat{e}}{s}{}(\Vie{p}{}{}) \Big] \, \exp \Big( i \alpha_1(\Vie{p}{}{}) (x_3 + x_3^\prime) \Big) \: .
\end{align}
The term $\Vie{\hat{G}}{1}{(d)}$ is the Green's function in the homogeneous space with dielectric constant $\varepsilon_1$ and corresponds to the contribution of the \emph{direct path} from the source point $x_3^\prime$ to the observation point $x_3$. The term $\Vie{\hat{G}}{}{(r)}$ corresponds to the contribution of the \emph{reflected path} on the reference interface ($x_3=0$). The Green's function for the reference system expressed for $x_3 > 0$ and $x_3^\prime < 0$ reads
\begin{equation}
\Vie{\hat{G}}{}{} (\Vie{p}{}{},x_3,x_3^\prime) = \frac{i}{2 \alpha_2(\Vie{p}{}{})} \: \Big[ t_{12}^{(p)} (\Vie{p}{}{})  \, \Vie{\hat{e}}{1,p}{+} (\Vie{p}{}{}) \otimes \Vie{\hat{e}}{2,p}{+} (\Vie{p}{}{}) + t_{12}^{(s)} (\Vie{p}{}{}) \, \Vie{\hat{e}}{s}{} (\Vie{p}{}{}) \otimes \Vie{\hat{e}}{s}{} (\Vie{p}{}{}) \Big] \: \exp \Big( i \alpha_1(\Vie{p}{}{}) \, x_3 - i \alpha_2(\Vie{p}{}{}) \, x_3^\prime  \Big) \: ,
\label{eq:green:pm}
\end{equation}
and corresponds to a \emph{transmitted path} from a source below the reference interface to an observation point above the interface. By using Eqs.~(\ref{eq:green:pp},\ref{eq:green:pm}) and (\ref{eq:E0}) we get on the one hand
\begin{align}
\Vie{\hat{G}}{}{}(\Vie{p}{}{},x_3,0^+) \Vie{\hat{E}}{2}{}(\Vie{p}{0}{},0) = &\frac{i}{2 \alpha_1(\Vie{p}{}{})} \, \Bigg[ \Vie{\hat{e}}{1,p}{+}(\Vie{p}{}{}) \otimes \left( \Vie{\hat{e}}{1,p}{+}(\Vie{p}{}{}) + r_{21}^{(p)} (\Vie{p}{}{})  \Vie{\hat{e}}{1,p}{-}(\Vie{p}{}{}) \right) + \left( 1 + r_{21}^{(s)} (\Vie{p}{}{})  \right) \Vie{\hat{e}}{s}{}(\Vie{p}{}{}) \otimes \Vie{\hat{e}}{s}{}(\Vie{p}{}{})  \Bigg] \nonumber \\
&\times \Big[  t_{21}^{(p)} (\Vie{p}{0}{})  \, \Cie{E}{0,p}{}  \, \Vie{\hat{e}}{2,p}{-} (\Vie{p}{0}{}) + t_{21}^{(s)} (\Vie{p}{0}{}) \, \Cie{E}{0,s}{}  \, \Vie{\hat{e}}{s}{} (\Vie{p}{0}{})  \Big] \exp \Big( i \alpha_1(\Vie{p}{}{}) \, x_3 \Big) \nonumber\\
= & \sum_{\mu = p,s} \Vie{\hat{e}}{1,\mu}{+} (\Vie{p}{}{}) \sum_{\nu =p,s}  \rho_{ \mu \nu}^{}(\Vie{p}{}{},\Vie{p}{0}{}) \, \Cie{E}{0,\nu}{} \: \exp \Big( i \alpha_1(\Vie{p}{}{}) \, x_3 \Big) \: ,
\end{align}
where we have used the convention $\Vie{\hat{e}}{j,s}{\pm} (\Vie{p}{}{}) \equiv \Vie{\hat{e}}{s}{} (\Vie{p}{}{})$ and where $\rho_{ \mu \nu}^{}(\Vie{p}{}{},\Vie{p}{0}{})$ is given by
\begin{equation}
\rho_{ \mu \nu}^{}(\Vie{p}{}{},\Vie{p}{0}{}) = \frac{i}{2 \alpha_1(\Vie{p}{}{})} \, \Big[ \Vie{\hat{e}}{1,\mu}{+} (\Vie{p}{}{}) + r_{21}^{(\mu)}(\Vie{p}{}{}) \Vie{\hat{e}}{1,\mu}{-} (\Vie{p}{}{}) \Big] \cdot t_{21}^{(\nu)}(\Vie{p}{0}{}) \Vie{\hat{e}}{2,\nu}{-} (\Vie{p}{0}{}) \: .
\label{eq:rho}
\end{equation}
On the other hand, we have
\begin{align}
\Vie{\hat{G}}{}{}(\Vie{p}{}{},x_3,0^-) \Vie{\hat{E}}{1}{}(\Vie{p}{0}{},0) = &\frac{i}{2 \alpha_2(\Vie{p}{}{})} \, \Bigg[  t_{12}^{(p)} (\Vie{p}{}{})  \, \Vie{\hat{e}}{1,p}{+} (\Vie{p}{}{}) \otimes \Vie{\hat{e}}{2,p}{+} (\Vie{p}{}{}) + t_{12}^{(s)} (\Vie{p}{}{}) \, \Vie{\hat{e}}{s}{} (\Vie{p}{}{}) \otimes \Vie{\hat{e}}{s}{} (\Vie{p}{}{})  \Bigg] \exp \Big( i \alpha_1(\Vie{p}{}{}) \, x_3 \Big) \nonumber \\
&\times \Big[  \Cie{E}{0,p}{} \, \Big( \Vie{\hat{e}}{1,p}{-} (\Vie{p}{0}{}) + r_{21}^{(p)} (\Vie{p}{0}{}) \, \Vie{\hat{e}}{1,p}{+} (\Vie{p}{0}{}) \Big)  + \Cie{E}{0,s}{} \, \Big( 1 + r_{21}^{(s)} (\Vie{p}{0}{}) \Big)  \, \Vie{\hat{e}}{s}{} (\Vie{p}{0}{})  \Big]  \nonumber\\
= & \sum_{\mu = p,s} \Vie{\hat{e}}{1,\mu}{+} (\Vie{p}{}{}) \sum_{\nu = p,s}  \rho_{ \mu \nu}^{\prime}(\Vie{p}{}{},\Vie{p}{0}{}) \, \Cie{E}{0,\nu}{} \: \exp \Big( i \alpha_1(\Vie{p}{}{}) \, x_3 \Big) \: ,
\end{align}
where $\rho_{ \mu \nu}^{\prime}(\Vie{p}{}{},\Vie{p}{0}{})$ is given by
\begin{equation}
\rho_{ \mu \nu}^{\prime}(\Vie{p}{}{},\Vie{p}{0}{}) = \frac{i}{2 \alpha_2(\Vie{p}{}{})} \,  t_{12}^{(\mu)}(\Vie{p}{}{}) \Vie{\hat{e}}{2,\mu}{+} (\Vie{p}{}{}) \cdot \Big[ \Vie{\hat{e}}{1,\nu}{-} (\Vie{p}{0}{}) + r_{21}^{(\nu)}(\Vie{p}{0}{}) \Vie{\hat{e}}{1,\nu}{+} (\Vie{p}{0}{}) \Big] \: .
\label{eq:rhop}
\end{equation}
Thus showing that $\Vie{\hat{G}}{}{}(\Vie{p}{}{},x_3,0^+) \Vie{\hat{E}}{2}{}(\Vie{p}{0}{},0) = \Vie{\hat{G}}{}{}(\Vie{p}{}{},x_3,0^-) \Vie{\hat{E}}{1}{}(\Vie{p}{0}{},0)$ is equivalent to showing that $\rho_{ \mu \nu}^{}(\Vie{p}{}{},\Vie{p}{0}{}) = \rho_{ \mu \nu}^{\prime}(\Vie{p}{}{},\Vie{p}{0}{})$. Note that this should hold for all $\mu, \nu \in \{p, s\} $ and $\Vie{p}{}{}, \Vie{p}{0}{} \in \mathbb{R}^2$. Tedious but straightforward algebra leads to $\rho_{ \mu \nu}^{}(\Vie{p}{}{},\Vie{p}{0}{}) = \rho_{ \mu \nu}^{\prime}(\Vie{p}{}{},\Vie{p}{0}{})$ by substituting Eqs.~(\ref{eq:wave_vector},\ref{eq:fresnel:rt}) into Eqs.~(\ref{eq:rho},\ref{eq:rhop}). We show here the main steps for the case $\mu = \nu = p$, but the remaining polarization couplings can be treated in a similar manner. Let us first insert Eqs.~(\ref{eq:wave_vector:ep},\ref{eq:fresnel:rp}) and (\ref{eq:fresnel:tp}) into Eq.~(\ref{eq:rho}), and we get
\begin{align}
\rho_{pp}(\Vie{p}{}{},\Vie{p}{0}{}) &= \frac{i \sqrt{\varepsilon_1 \varepsilon_2} \alpha_1(\Vie{p}{0}{})  
\begin{bmatrix}
\Big( \varepsilon_2 \alpha_1(\Vie{p}{}{}) + \varepsilon_1 \alpha_2(\Vie{p}{}{}) \Big)
\Big( -\alpha_1(\Vie{p}{}{}) \alpha_2(\Vie{p}{0}{}) \Vie{\hat{p}}{}{} \cdot \Vie{\hat{p}}{0}{} + |\Vie{p}{}{}| |\Vie{p}{0}{}| \Big) \\
 + \Big( \varepsilon_2 \alpha_1(\Vie{p}{}{}) - \varepsilon_1 \alpha_2(\Vie{p}{}{}) \Big)
\Big( \alpha_1(\Vie{p}{}{}) \alpha_2(\Vie{p}{0}{}) \Vie{\hat{p}}{}{} \cdot \Vie{\hat{p}}{0}{} + |\Vie{p}{}{}| |\Vie{p}{0}{}| \Big)
\end{bmatrix}
}{k_1 k_2 \alpha_1(\Vie{p}{}{}) \Big( \varepsilon_2 \alpha_1(\Vie{p}{}{}) + \varepsilon_1 \alpha_2(\Vie{p}{}{}) \Big) \Big( \varepsilon_2 \alpha_1(\Vie{p}{0}{}) + \varepsilon_1 \alpha_2(\Vie{p}{0}{}) \Big)} \nonumber \\
&= \frac{2 i \sqrt{\varepsilon_1 \varepsilon_2} \alpha_1(\Vie{p}{0}{})  \Big[ \varepsilon_2 |\Vie{p}{}{}| |\Vie{p}{0}{}| - \varepsilon_1 \alpha_2(\Vie{p}{}{})   \alpha_2(\Vie{p}{0}{}) \Vie{\hat{p}}{}{} \cdot \Vie{\hat{p}}{0}{} \Big]
}{k_1 k_2 \Big( \varepsilon_2 \alpha_1(\Vie{p}{}{}) + \varepsilon_1 \alpha_2(\Vie{p}{}{}) \Big) \Big( \varepsilon_2 \alpha_1(\Vie{p}{0}{}) + \varepsilon_1 \alpha_2(\Vie{p}{0}{}) \Big)} \: ,
\end{align}
and similarly, by inserting Eqs.~(\ref{eq:wave_vector:ep},\ref{eq:fresnel:rp}) and (\ref{eq:fresnel:tp}) into Eq.~(\ref{eq:rhop}), we get
\begin{align}
\rho_{pp}^\prime (\Vie{p}{}{},\Vie{p}{0}{}) &= \frac{i \sqrt{\varepsilon_1 \varepsilon_2}
\begin{bmatrix}
\Big( \varepsilon_2 \alpha_1(\Vie{p}{0}{}) + \varepsilon_1 \alpha_2(\Vie{p}{0}{}) \Big)
\Big( -\alpha_2(\Vie{p}{}{}) \alpha_1(\Vie{p}{0}{}) \Vie{\hat{p}}{}{} \cdot \Vie{\hat{p}}{0}{} + |\Vie{p}{}{}| |\Vie{p}{0}{}| \Big) \\
 + \Big( \varepsilon_2 \alpha_1(\Vie{p}{0}{}) - \varepsilon_1 \alpha_2(\Vie{p}{0}{}) \Big)
\Big( \alpha_2(\Vie{p}{}{}) \alpha_1(\Vie{p}{0}{}) \Vie{\hat{p}}{}{} \cdot \Vie{\hat{p}}{0}{} + |\Vie{p}{}{}| |\Vie{p}{0}{}| \Big)
\end{bmatrix}
}{k_1 k_2 \Big( \varepsilon_2 \alpha_1(\Vie{p}{}{}) + \varepsilon_1 \alpha_2(\Vie{p}{}{}) \Big) \Big( \varepsilon_2 \alpha_1(\Vie{p}{0}{}) + \varepsilon_1 \alpha_2(\Vie{p}{0}{}) \Big)} \nonumber \\
&= \frac{2 i \sqrt{\varepsilon_1 \varepsilon_2} \alpha_1(\Vie{p}{0}{})  \Big[ \varepsilon_2 |\Vie{p}{}{}| |\Vie{p}{0}{}| - \varepsilon_1 \alpha_2(\Vie{p}{}{})   \alpha_2(\Vie{p}{0}{}) \Vie{\hat{p}}{}{} \cdot \Vie{\hat{p}}{0}{} \Big]
}{k_1 k_2 \Big( \varepsilon_2 \alpha_1(\Vie{p}{}{}) + \varepsilon_1 \alpha_2(\Vie{p}{}{}) \Big) \Big( \varepsilon_2 \alpha_1(\Vie{p}{0}{}) + \varepsilon_1 \alpha_2(\Vie{p}{0}{}) \Big)} = \rho_{pp} (\Vie{p}{}{},\Vie{p}{0}{}) \: .
\end{align}

\section{Derivation of the transmission amplitude} \label{App:transmission:amplitude}

We derive here the transmission amplitudes. The derivation is similar to that of the reflection amplitudes. To this end, we need the Weyl expansion of the Green's function for observation points $x_3 < -L$. The Green's function for the reference system expressed for $x_3 < -L < x_3^\prime < 0$ reads \cite{Sipe:87}
\begin{equation}
\Vie{\hat{G}}{}{} (\Vie{p}{}{},x_3,x_3^\prime) = \Vie{\hat{G}}{2}{(d)} (\Vie{p}{}{},x_3 - x_3^\prime) + \Vie{\hat{G}}{}{(r)} (\Vie{p}{}{},x_3,x_3^\prime) \: ,
\label{eq:green:mm}
\end{equation}
with
\begin{subequations}
\begin{align}
\Vie{\hat{G}}{2}{(d)} (\Vie{p}{}{},x_3-x_3^\prime) = &\frac{i}{2 \alpha_2(\Vie{p}{}{})} \: \Big[ \Vie{\hat{e}}{2,p}{-} (\Vie{p}{}{}) \otimes \Vie{\hat{e}}{2,p}{-} (\Vie{p}{}{}) + \Vie{\hat{e}}{s}{} (\Vie{p}{}{}) \otimes \Vie{\hat{e}}{s}{} (\Vie{p}{}{}) \Big] \: \exp \Big( - i \alpha_2(\Vie{p}{}{}) \, (x_3 - x_3^\prime)  \Big) \\
\Vie{\hat{G}}{}{(r)} (\Vie{p}{}{},x_3,x_3^\prime)= &\frac{i}{2 \alpha_2(\Vie{p}{}{})} \: \Big[ r_{12}^{(p)} (\Vie{p}{}{})  \, \Vie{\hat{e}}{2,p}{-} (\Vie{p}{}{}) \otimes \Vie{\hat{e}}{2,p}{+} (\Vie{p}{}{}) + r_{12}^{(s)} (\Vie{p}{}{}) \, \Vie{\hat{e}}{s}{} (\Vie{p}{}{}) \otimes \Vie{\hat{e}}{s}{} (\Vie{p}{}{}) \Big] \: \exp \Big( -i \alpha_2(\Vie{p}{}{}) \, (x_3 + x_3^\prime)  \Big) \: .
\end{align}
\label{eq:green2}
\end{subequations}
The two terms in Eq.~(\ref{eq:green:mm}) correspond, respectively, to the Green's function of the infinite homogeneous medium with dielectric constant $\varepsilon_2$, $\Vie{\hat{G}}{2}{(d)}$, which encodes the contribution of a dipole source located at  $x_3^\prime$ to the field measured at point $x_3$ by taking a \emph{direct} path, and a correction due to the presence of the interface with medium 1, $\Vie{\hat{G}}{}{(r)}$, which encodes the contribution of a dipole source located at $x_3^\prime$ to the field measured at $x_3$ by taking a path \emph{reflecting} on the interface $x_3=0$. For $x_3 < 0 < x_3^\prime$ the Green's function reads \cite{Sipe:87}
\begin{equation}
\Vie{\hat{G}}{}{} (\Vie{p}{}{},x_3,x_3^\prime) = \frac{i}{2 \alpha_1(\Vie{p}{}{})} \: \Big[ t_{21}^{(p)} (\Vie{p}{}{})  \, \Vie{\hat{e}}{2,p}{-} (\Vie{p}{}{}) \otimes \Vie{\hat{e}}{1,p}{-} (\Vie{p}{}{}) + t_{21}^{(s)} (\Vie{p}{}{}) \, \Vie{\hat{e}}{s}{} (\Vie{p}{}{}) \otimes \Vie{\hat{e}}{s}{} (\Vie{p}{}{}) \Big] \: \exp \Big( - i \alpha_2(\Vie{p}{}{}) \, x_3 + i \alpha_1(\Vie{p}{}{}) \, x_3^\prime  \Big) \: ,
\label{eq:green:mp}
\end{equation}
and corresponds to a \emph{transmitted path} from a source above the reference interface to an observation point below the interface.
 Inserting the expression for the Green's function Eqs.~(\ref{eq:green:mm},\ref{eq:green:mp}) and for the reference field Eqs.~(\ref{eq:E01},\ref{eq:E02}) into the Eqs.~(\ref{eq:Eeps:fourier:planewave}) and (\ref{eq:Ezeta:fourier}) yields for $x_3 < -L$
\begin{subequations}
\begin{align}
\Vie{\hat{E}}{\varepsilon}{(1)} (\Vie{p}{}{},x_3) &= \sum_{\mu = p,s} \Vie{\hat{e}}{2,\mu}{-} (\Vie{p}{}{}) \sum_{\nu = p,s} \Big[ T_{\varepsilon, \mu \nu}^{(1,d)} (\Vie{p}{}{},\Vie{p}{0}{}) + T_{\varepsilon, \mu \nu}^{(1,r)} (\Vie{p}{}{},\Vie{p}{0}{}) \Big] \, \Cie{E}{0,\nu}{} \: \exp \Big( - i \alpha_2(\Vie{p}{}{}) \, x_3 \Big) \\
\Vie{\hat{E}}{\zeta}{(1)} (\Vie{p}{}{},x_3) &= \sum_{\mu = p,s} \Vie{\hat{e}}{2,\mu}{-} (\Vie{p}{}{}) \sum_{\nu = p,s} T_{\zeta, \mu \nu}^{(1)}(\Vie{p}{}{},\Vie{p}{0}{})  \, \Cie{E}{0,\nu}{} \: \exp \Big( - i \alpha_2(\Vie{p}{}{}) \, x_3 \Big) \: ,
\end{align}
\end{subequations}
with
\end{widetext}
\begin{subequations}
\begin{align}
T_{\varepsilon, \mu \nu}^{(1,d)} (\Vie{p}{}{},\Vie{p}{0}{}) &=  \frac{i k_0^2}{2\alpha_2(\Vie{p}{}{})} \psi^{-}(\Vie{p}{}{},\Vie{p}{0}{}) \: \tau_{\varepsilon, \mu \nu}^{(d)}(\Vie{p}{}{},\Vie{p}{0}{}) \\
T_{\varepsilon, \mu \nu}^{(1,r)} (\Vie{p}{}{},\Vie{p}{0}{}) &= \frac{i k_0^2}{2\alpha_2(\Vie{p}{}{})} \psi^{+}(\Vie{p}{}{},\Vie{p}{0}{}) \: \tau_{\varepsilon, \mu \nu}^{(r)}(\Vie{p}{}{},\Vie{p}{0}{}) \\
T_{\zeta, \mu \nu}^{(1)}(\Vie{p}{}{},\Vie{p}{0}{}) &= \frac{i k_0^2}{2\alpha_1(\Vie{p}{}{})} (\varepsilon_2 - \varepsilon_1) \hat{\zeta}(\Vie{p}{}{}-\Vie{p}{0}{}) \: \tau_{\zeta, \mu \nu}(\Vie{p}{}{},\Vie{p}{0}{}) \: ,
\end{align}
\end{subequations}
and where the polarization coupling factors $\tau^{(d)}_{\varepsilon, \mu \nu}$, $\tau^{(r)}_{\varepsilon, \mu \nu}$, and $\tau_{\zeta, \mu \nu}$ are defined by
\begin{subequations}
\begin{align}
\tau_{\varepsilon, \mu \nu}^{(d)}(\Vie{p}{}{},\Vie{p}{0}{}) &=  \Vie{\hat{e}}{2,\mu}{-} (\Vie{p}{}{}) \cdot \Vie{\hat{e}}{2,\nu}{-} (\Vie{p}{0}{})  \, t_{21}^{(\nu)}(\Vie{p}{0}{}) \\
\tau_{\varepsilon, \mu \nu}^{(r)}(\Vie{p}{}{},\Vie{p}{0}{}) &=  r_{12}^{(\mu)}(\Vie{p}{}{}) \Vie{\hat{e}}{2,\mu}{+} (\Vie{p}{}{}) \cdot \Vie{\hat{e}}{2,\nu}{-} (\Vie{p}{0}{})  \, t_{21}^{(\nu)}(\Vie{p}{0}{}) \\
\tau_{\zeta, \mu \nu}(\Vie{p}{}{},\Vie{p}{0}{}) &=   t_{21}^{(\nu)}(\Vie{p}{}{}) \, \Vie{\hat{e}}{1,\mu}{-} (\Vie{p}{}{}) \cdot \Vie{\hat{e}}{2,\nu}{-} (\Vie{p}{0}{})  \, t_{21}^{(\nu)}(\Vie{p}{0}{})  \: .
\end{align}
\label{eq:taumunu}
\end{subequations}
The definition of $\psi^\pm$ is given in Eq.~(\ref{eq:psi}).
\begin{widetext}
The total scattered field for $x_3 < -L$ is thus given by
\begin{equation}
\Vie{\hat{E}}{}{(1)} (\Vie{p}{}{},x_3) = \sum_{\mu = p,s} \Vie{\hat{e}}{2,\mu}{-} (\Vie{p}{}{}) \sum_{\nu = p,s} T_{\mu \nu}^{(1)}(\Vie{p}{}{},\Vie{p}{0}{}) \, \Cie{E}{0,\nu}{} \: \exp \Big( - i \alpha_2(\Vie{p}{}{}) \, x_3 \Big) \: ,
\end{equation}
where we have identified the first order transmission amplitude $T_{\mu \nu}^{(1)}$ as
\begin{align}
T_{\mu \nu}^{(1)}(\Vie{p}{}{},\Vie{p}{0}{}) &= T_{\zeta, \mu \nu}^{(1)}(\Vie{p}{}{},\Vie{p}{0}{}) + T_{\varepsilon, \mu \nu}^{(1,d)}(\Vie{p}{}{},\Vie{p}{0}{}) + T_{\varepsilon, \mu \nu}^{(1,r)}(\Vie{p}{}{},\Vie{p}{0}{}) \label{eq:T1} \\
&= \frac{i k_0^2}{2\alpha_2(\Vie{p}{}{})} \Bigg[ \frac{\alpha_2(\Vie{p}{}{})}{\alpha_1(\Vie{p}{}{})} (\varepsilon_2 -\varepsilon_1) \hat{\zeta}(\Vie{p}{}{}-\Vie{p}{0}{}) \tau_{\zeta,\mu \nu}(\Vie{p}{}{},\Vie{p}{0}{})  + \psi^- (\Vie{p}{}{},\Vie{p}{0}{}) \tau_{\varepsilon, \mu \nu}^{(d)}(\Vie{p}{}{},\Vie{p}{0}{}) + \psi^+ (\Vie{p}{}{},\Vie{p}{0}{}) \tau_{\varepsilon, \mu \nu}^{(r)}(\Vie{p}{}{},\Vie{p}{0}{}) \Bigg] \: . \nonumber
\end{align}
The scalar wave transmission amplitude is given by
\begin{equation}
T^{(1)}(\Vie{p}{}{},\Vie{p}{0}{}) = \frac{i k_0^2}{2} \Bigg[ \frac{t_{12}(\Vie{p}{}{})}{\alpha_1(\Vie{p}{}{})} (\varepsilon_2 -\varepsilon_1) \hat{\zeta}(\Vie{p}{}{}-\Vie{p}{0}{}) + \frac{1}{\alpha_2(\Vie{p}{}{})} \Big( \psi^-(\Vie{p}{}{},\Vie{p}{0}{}) + r_{12}(\Vie{p}{}{}) \psi^+(\Vie{p}{}{},\Vie{p}{0}{}) \Big) \Bigg] t_{21}(\Vie{p}{0}{}) \: .
\label{eq:T1:scalar}
\end{equation}

\section{Computation of covariances} \label{App:covariance}

In the evaluation of the diffuse component of the MDRC and MDTC, we have to compute various covariances of the transverse Fourier transforms of $\zeta$ and $\Delta \varepsilon$. For example, we have
\begin{align}
S^{-1} \, \left\langle \hat{\zeta} (\Vie{p}{}{} - \Vie{p}{0}{}) \hat{\zeta}^* (\Vie{p}{}{} - \Vie{p}{0}{}) \right\rangle &= S^{-1} \, \int_S \int_S \left\langle \zeta (\Vie{x}{\parallel}{}) \zeta (\Vie{x}{\parallel}{\prime}) \right\rangle \: e^{-i(\Vie{p}{}{}-\Vie{p}{0}{}) \cdot (\Vie{x}{\parallel}{} - \Vie{x}{\parallel}{\prime})} \mathrm{d}^2x_\parallel \, \mathrm{d}^2x_\parallel^\prime \nonumber\\
&= \frac{\sigma_\zeta^2}{S} \: \int_S \int_S W_\zeta(\Vie{x}{\parallel}{} -\Vie{x}{\parallel}{\prime}) \: e^{-i(\Vie{p}{}{}-\Vie{p}{0}{}) \cdot (\Vie{x}{\parallel}{} - \Vie{x}{\parallel}{\prime})} \mathrm{d}^2x_\parallel \, \mathrm{d}^2x_\parallel^\prime \nonumber\\
&= \frac{\sigma_\zeta^2}{S} \: \int_S \int_{S-\Vie{x}{\parallel}{\prime}} W_\zeta(\Vie{u}{}{}) \: e^{-i(\Vie{p}{}{}-\Vie{p}{0}{}) \cdot \Vie{u}{}{}} \mathrm{d}^2u \, \mathrm{d}^2x_\parallel^\prime \nonumber\\
&= \frac{\sigma_\zeta^2}{S} \: \int_S \int_{\mathbb{R}^2} \ind_{S-\Vie{x}{\parallel}{\prime}} (\Vie{u}{}{}) \: W_\zeta(\Vie{u}{}{}) \: e^{-i(\Vie{p}{}{}-\Vie{p}{0}{}) \cdot \Vie{u}{}{}} \mathrm{d}^2u \, \mathrm{d}^2x_\parallel^\prime \: .
\end{align}

In this last equality we recognize the inner integral to be the Fourier transform of the product $\ind_{S-\Vie{x}{\parallel}{\prime}} \: W_\zeta$. Applying the convolution theorem we obtain
\begin{align}
S^{-1} \, \left\langle \hat{\zeta} (\Vie{p}{}{} - \Vie{p}{0}{}) \hat{\zeta}^* (\Vie{p}{}{} - \Vie{p}{0}{}) \right\rangle &= \frac{\sigma_\zeta^2}{S} \: \int_S \int_{\mathbb{R}^2} \hat{\ind}_{S-\Vie{x}{\parallel}{\prime}} (\Vie{q}{}{}) \: \hat{W}_\zeta(\Vie{p}{}{}-\Vie{p}{0}{} -\Vie{q}{}{}) \frac{\mathrm{d}^2q}{(2 \pi)^2} \, \mathrm{d}^2x_\parallel^\prime \nonumber \\
&= \frac{\sigma_\zeta^2}{S} \: \int_S \int_{\mathbb{R}^2} \hat{\ind}_{S} (\Vie{q}{}{}) \: \hat{W}_\zeta(\Vie{p}{}{}-\Vie{p}{0}{} -\Vie{q}{}{}) \: e^{i \Vie{q}{}{} \cdot \Vie{x}{\parallel}{\prime}}\frac{\mathrm{d}^2q}{(2 \pi)^2} \, \mathrm{d}^2x_\parallel^\prime \nonumber \\
&= \frac{\sigma_\zeta^2}{S} \: \int_S \int_{\mathbb{R}^2} \frac{4 \sin(q_1 D/2) \sin(q_2 D/2) }{q_1 q_2} \: \hat{W}_\zeta(\Vie{p}{}{}-\Vie{p}{0}{} -\Vie{q}{}{}) \: e^{i \Vie{q}{}{} \cdot \Vie{x}{\parallel}{\prime}}\frac{\mathrm{d}^2q}{(2 \pi)^2} \, \mathrm{d}^2x_\parallel^\prime \: .
\end{align}
Here we have assumed a square domain of size $S = D \times D$ for which the Fourier transform of the indicator function is well known. By interchanging the order of integration and integrating over $\Vie{x}{\parallel}{\prime}$, we obtain an additional Fourier transform of the indicator of the domain $S$, hence
\begin{align}
S^{-1} \, \left\langle \hat{\zeta} (\Vie{p}{}{} - \Vie{p}{0}{}) \hat{\zeta}^* (\Vie{p}{}{} - \Vie{p}{0}{}) \right\rangle &= \sigma_\zeta^2 \: \int_{\mathbb{R}^2} \frac{4 \sin^2(q_1 D/2) \sin^2(q_2 D/2) }{\pi^2 q_1^2 q_2^2 S} \: \hat{W}_\zeta(\Vie{p}{}{}-\Vie{p}{0}{} -\Vie{q}{}{}) \: \mathrm{d}^2q \: .
\end{align}
Now noticing that in the limit $D \to \infty$, the function $q \mapsto \frac{2 \sin^2(q D/2)}{\pi q^2 D}$ converges in the sense of distributions towards a Dirac mass centered at zero, we obtain
\begin{equation}
\lim_{S \to \infty} S^{-1} \, \left\langle \hat{\zeta} (\Vie{p}{}{} - \Vie{p}{0}{}) \hat{\zeta}^* (\Vie{p}{}{} - \Vie{p}{0}{}) \right\rangle = \sigma_\zeta^2 \: \hat{W}_\zeta(\Vie{p}{}{}-\Vie{p}{0}{}) \: .
\label{eq:Czz}
\end{equation}
The remaining covariances are evaluated in a similar way and we get
\begin{align}
&\lim_{S \to \infty} S^{-1} \, \Big\langle \Delta \hat{\varepsilon} (\Vie{p}{}{} - \Vie{p}{0}{},x_3) \, \Delta \hat{\varepsilon}^* (\Vie{p}{}{} - \Vie{p}{0}{},x_3^{\prime}) \Big\rangle = \sigma_\varepsilon^2 \, \hat{W}_{\varepsilon \parallel} (\Vie{p}{}{}-\Vie{p}{0}{})  \, f(x_3) \, f(x_3^\prime) \, \exp \left[ - \frac{(x_3-x_3^\prime)^2}{\ell_{\varepsilon \perp}^2} \right] \label{eq:Cee}\\ 
&\lim_{S \to \infty} S^{-1} \, \left\langle \hat{\zeta}(\Vie{p}{}{} - \Vie{p}{0}{}) \, \Delta \hat{\varepsilon}^* (\Vie{p}{}{} - \Vie{p}{0}{}, x_3^\prime)  \right\rangle =  \sigma_\zeta \sigma_\varepsilon \, \gamma (\Vie{p}{}{}-\Vie{p}{0}{}) \hat{W}_{\zeta}^{1/2}(\Vie{p}{}{}-\Vie{p}{0}{}) \hat{W}_{\varepsilon \parallel}^{1/2} (\Vie{p}{}{}-\Vie{p}{0}{}) \, f(x_3^\prime) \:\exp \left[ - \frac{(x_3^\prime+d)^2}{\ell_{\varepsilon \perp}^2} \right] \label{eq:Cze} \: .
\end{align}
In addition, from the definition of $\psi^\pm$ in Eq.~(\ref{eq:psi}) and from the above formulas, we have the following covariances for $a, b = \pm$
\begin{align}
\lim_{S \to \infty} S^{-1} \, \Big\langle \psi^a(\Vie{p}{}{},\Vie{p}{0}{}) \psi^{b*}(\Vie{p}{}{},\Vie{p}{0}{}) \Big\rangle = &\int \int \lim_{S \to \infty} S^{-1} \, \Big\langle \Delta \hat{\varepsilon} (\Vie{p}{}{} - \Vie{p}{0}{},x_3) \, \Delta \hat{\varepsilon}^* (\Vie{p}{}{} - \Vie{p}{0}{},x_3^{\prime}) \Big\rangle \nonumber \\
&\times \exp \Big( - i \alpha^a(\Vie{p}{}{},\Vie{p}{0}{}) x_3 + i \alpha^{b*}(\Vie{p}{}{},\Vie{p}{0}{}) x_3^\prime \Big) \: \mathrm{d} x_3 \, \mathrm{d}x_3^\prime \nonumber\\
= & \: \sigma_\varepsilon^2 \hat{W}_{\varepsilon \parallel}(\Vie{p}{}{}-\Vie{p}{0}{}) \, \int_{-L}^0 \int_{-L}^0  \exp \left[ - \frac{(x_3-x_3^\prime)^2}{\ell_{\varepsilon \perp}^2} \right] \nonumber \\
&\times \exp \Big( - i \alpha^a(\Vie{p}{}{},\Vie{p}{0}{}) x_3 + i \alpha^{b*}(\Vie{p}{}{},\Vie{p}{0}{}) x_3^\prime \Big) \: \mathrm{d} x_3 \, \mathrm{d}x_3^\prime \nonumber\\
= & \: \sigma_\varepsilon^2 \hat{W}_{\varepsilon \parallel}(\Vie{p}{}{}-\Vie{p}{0}{}) \, I \big( \ell_{\varepsilon \perp}, L, \alpha^a(\Vie{p}{}{},\Vie{p}{0}{}), \alpha^b(\Vie{p}{}{},\Vie{p}{0}{}) \big) \: , \\
\lim_{S \to \infty} S^{-1} \, \Big\langle \zeta(\Vie{p}{}{}-\Vie{p}{0}{}) \psi^{b*}(\Vie{p}{}{},\Vie{p}{0}{}) \Big\rangle = &\int \lim_{S \to \infty} S^{-1} \, \left\langle \hat{\zeta}(\Vie{p}{}{} - \Vie{p}{0}{}) \, \Delta \hat{\varepsilon}^* (\Vie{p}{}{} - \Vie{p}{0}{}, x_3)  \right\rangle \, \exp \Big( i \alpha^{b*}(\Vie{p}{}{},\Vie{p}{0}{}) x_3 \Big) \: \mathrm{d} x_3  \nonumber\\
= & \: \sigma_\zeta \sigma_\varepsilon \, \gamma (\Vie{p}{}{}-\Vie{p}{0}{}) \hat{W}_{\zeta}^{1/2}(\Vie{p}{}{}-\Vie{p}{0}{}) \hat{W}_{\varepsilon \parallel}^{1/2} (\Vie{p}{}{}-\Vie{p}{0}{}) \nonumber \\
& \times \int_{-L}^0 \exp \left[ - \frac{(x_3+d)^2}{\ell_{\varepsilon \perp}^2} \right] \, \exp \Big( i \alpha^{b*}(\Vie{p}{}{},\Vie{p}{0}{}) x_3 \Big) \: \mathrm{d} x_3  \nonumber \\
= & \:  \sigma_\zeta \sigma_\varepsilon \, \gamma (\Vie{p}{}{}-\Vie{p}{0}{}) \hat{W}_{\zeta}^{1/2}(\Vie{p}{}{}-\Vie{p}{0}{}) \hat{W}_{\varepsilon \parallel}^{1/2} (\Vie{p}{}{}-\Vie{p}{0}{}) \, J \big( \ell_{\varepsilon \perp}, L, d, \alpha^b(\Vie{p}{}{},\Vie{p}{0}{})\big) \: ,
\end{align}
where the functions $I$ and $J$ are defined as
\begin{subequations}
\begin{align}
I (\ell_{\varepsilon \perp}, L, \alpha, \beta) &= \int_{-L}^0 \int_{-L}^0 \exp \left[ - \frac{(x_3-x_3^\prime)^2}{\ell_{\varepsilon \perp}^2} \right] \: \exp \Big[ - i \alpha \, x_3 + i \beta^* \, x_3^{\prime} \Big] \, \mathrm{d}x_3 \, \mathrm{d}x_3^\prime \\
J (\ell_{\varepsilon \perp}, L, d, \alpha) &= \int_{-L}^0 \exp \left[ - \frac{(x_3+d)^2}{\ell_{\varepsilon \perp}^2} \right] \: \exp \Big[ i \alpha \, x_3 \Big] \, \mathrm{d}x_3 \: ,
\end{align}
\end{subequations}
and where we have introduced $\alpha^\pm (\Vie{p}{}{},\Vie{p}{0}{}) = \pm \alpha_2(\Vie{p}{}{}) + \alpha_2(\Vie{p}{0}{})$.
%
%

\section{Derivation of the diffuse component of the MDRC and MDTC} \label{App:MDXC}

To evaluate the diffuse component of the MDRC and MDTC it suffices to substitute Eqs.~(\ref{eq:R1}) into Eq.~(\ref{eq:MDRC}) and to use the covariances from Appendix~\ref{App:covariance}. The diffuse component of the MDRC reads
\begin{align}
\left\langle \frac{\partial R_{\mu \nu}}{\partial \Omega} (\Vie{p}{}{},\Vie{p}{0}{}) \right\rangle_\mathrm{diff} = &\lim_{S\to \infty} \frac{ \varepsilon_1^{1/2} k_0 \, \Re \big( \alpha_1(\Vie{p}{}{}) \big)^2}{S (2 \pi)^2 \alpha_1(\Vie{p}{0}{})} \: \left\langle |R_{\mu \nu}^{(1)} (\Vie{p}{}{},\Vie{p}{0}{})|^2 \right\rangle \nonumber\\
= &\frac{ \varepsilon_1^{1/2} k_0^5 \, \Re \big( \alpha_1(\Vie{p}{}{}) \big)^2}{4 |\alpha_2(\Vie{p}{}{})|^2 (2 \pi)^2 \alpha_1(\Vie{p}{0}{})} \, \lim_{S \to \infty} \Bigg[ (\varepsilon_2 - \varepsilon_1)^2 \frac{ \left\langle |\zeta(\Vie{p}{}{}-\Vie{p}{0}{})|^2\right\rangle }{S} \, |\rho_{\zeta, \mu \nu}(\Vie{p}{}{},\Vie{p}{0}{})|^2 \nonumber\\
&+ 2 \Re \Bigg( (\varepsilon_2 - \varepsilon_1) \frac{\left\langle \zeta(\Vie{p}{}{}-\Vie{p}{0}{}) \psi^{+*} (\Vie{p}{}{},\Vie{p}{0}{}) \right\rangle}{S} \, \rho_{\zeta,\mu \nu}(\Vie{p}{}{},\Vie{p}{0}{}) \rho_{\varepsilon,\mu \nu}^* (\Vie{p}{}{},\Vie{p}{0}{}) \Bigg) \nonumber\\
&+ \frac{\left\langle |\psi^+ (\Vie{p}{}{},\Vie{p}{0}{}) |^2 \right\rangle}{S} \, |\rho_{\varepsilon, \mu \nu} (\Vie{p}{}{},\Vie{p}{0}{})|^2 \Bigg] \nonumber\\
= & \: C^{(r)}(\Vie{p}{}{},\Vie{p}{0}{}) \Bigg[ (\varepsilon_2 - \varepsilon_1)^2 k_0^4 \sigma_\zeta^2 \hat{W}_\zeta(\Vie{p}{}{}-\Vie{p}{0}{}) \, |\rho_{\zeta, \mu \nu}(\Vie{p}{}{},\Vie{p}{0}{})|^2 \nonumber\\
&+ 2 \Re \Bigg( (\varepsilon_2 - \varepsilon_1) k_0^4 \sigma_\zeta \sigma_\varepsilon \, \gamma(\Vie{p}{}{}-\Vie{p}{0}{}) \hat{W}_{\zeta}^{1/2}(\Vie{p}{}{}-\Vie{p}{0}{}) \hat{W}_{\varepsilon \parallel}^{1/2} (\Vie{p}{}{}-\Vie{p}{0}{})  \nonumber\\
&\times \, J \big( \ell_{\varepsilon \perp}, L, d, \alpha^+(\Vie{p}{}{},\Vie{p}{0}{})\big) \rho_{\zeta,\mu \nu}(\Vie{p}{}{},\Vie{p}{0}{}) \rho_{\varepsilon,\mu \nu}^* (\Vie{p}{}{},\Vie{p}{0}{}) \Bigg) \nonumber\\
&+ \sigma_\varepsilon^2 k_0^4 \hat{W}_{\varepsilon \parallel} (\Vie{p}{}{}-\Vie{p}{0}{}) I \big( \ell_{\varepsilon \perp}, L, \alpha^+(\Vie{p}{}{},\Vie{p}{0}{}), \alpha^+(\Vie{p}{}{},\Vie{p}{0}{})) |\rho_{\varepsilon, \mu \nu}(\Vie{p}{}{},\Vie{p}{0}{} \big) |^2 \Bigg] \: .
\end{align}
A similar derivation yields the diffuse component of the MDTC
\begin{align}
\left\langle \frac{\partial T_{\mu \nu}}{\partial \Omega} (\Vie{p}{}{},\Vie{p}{0}{}) \right\rangle_\mathrm{diff} = &\lim_{S\to \infty} \frac{ \varepsilon_2^{1/2} k_0 \, \Re \big( \alpha_2(\Vie{p}{}{}) \big)^2}{S (2 \pi)^2 \alpha_1(\Vie{p}{0}{})} \: \left\langle |T_{\mu \nu}^{(1)} (\Vie{p}{}{},\Vie{p}{0}{})|^2 \right\rangle \nonumber\\
= & \: C^{(t)}(\Vie{p}{}{},\Vie{p}{0}{}) \, \Bigg[ \left| \frac{\alpha_2(\Vie{p}{}{})}{\alpha_1(\Vie{p}{}{})} \right|^2 (\varepsilon_2 - \varepsilon_1)^2 k_0^4 \sigma_\zeta^2 \hat{W}_\zeta(\Vie{p}{}{}-\Vie{p}{0}{}) \, |\tau_{\zeta, \mu \nu}(\Vie{p}{}{},\Vie{p}{0}{})|^2 \nonumber\\
+ & 2 \Re \Bigg( \frac{\alpha_2(\Vie{p}{}{})}{\alpha_1(\Vie{p}{}{})} (\varepsilon_2 - \varepsilon_1) k_0^4 \sigma_\zeta \sigma_\varepsilon \, \gamma(\Vie{p}{}{}-\Vie{p}{0}{}) \hat{W}_{\zeta}^{1/2}(\Vie{p}{}{}-\Vie{p}{0}{}) \hat{W}_{\varepsilon \parallel}^{1/2} (\Vie{p}{}{}-\Vie{p}{0}{}) \nonumber\\
&\times  \tau_{\zeta,\mu \nu}(\Vie{p}{}{},\Vie{p}{0}{}) \Big[ J \big( \ell_{\varepsilon \perp}, L, d, \alpha^-(\Vie{p}{}{},\Vie{p}{0}{})\big) \tau_{\varepsilon,\mu \nu}^{(d) *} (\Vie{p}{}{},\Vie{p}{0}{}) + J \big( \ell_{\varepsilon \perp}, L, d, \alpha^+(\Vie{p}{}{},\Vie{p}{0}{})\big) \tau_{\varepsilon,\mu \nu}^{(r) *} (\Vie{p}{}{},\Vie{p}{0}{}) \Big] \Bigg) \nonumber\\
+ & 2 \Re \Bigg( \sigma_\varepsilon^2 \hat{W}_{\varepsilon \parallel} (\Vie{p}{}{}-\Vie{p}{0}{}) \, I\big(\ell_{\varepsilon \perp}, L, \alpha^-(\Vie{p}{}{},\Vie{p}{0}{}), \alpha^+(\Vie{p}{}{},\Vie{p}{0}{}) \big) \, \tau_{\varepsilon,\mu \nu}^{(d)} (\Vie{p}{}{},\Vie{p}{0}{}) \tau_{\varepsilon,\mu \nu}^{(r) *} (\Vie{p}{}{},\Vie{p}{0}{}) \Bigg) \nonumber \\
+ & \sigma_\varepsilon^2 \hat{W}_{\varepsilon \parallel} (\Vie{p}{}{}-\Vie{p}{0}{}) \, I\big(\ell_{\varepsilon \perp}, L, \alpha^-(\Vie{p}{}{},\Vie{p}{0}{}), \alpha^-(\Vie{p}{}{},\Vie{p}{0}{}) \big) \, |\tau_{\varepsilon,\mu \nu}^{(d)} (\Vie{p}{}{},\Vie{p}{0}{})|^2  \nonumber\\
+ & \sigma_\varepsilon^2 \hat{W}_{\varepsilon \parallel} (\Vie{p}{}{}-\Vie{p}{0}{}) \, I \big(\ell_{\varepsilon \perp}, L, \alpha^+(\Vie{p}{}{},\Vie{p}{0}{}), \alpha^+(\Vie{p}{}{},\Vie{p}{0}{}) \big) \, |\tau_{\varepsilon,\mu \nu}^{(r)} (\Vie{p}{}{},\Vie{p}{0}{})|^2 \Bigg] \: .
\end{align}
Here we have introduced
\begin{subequations}
\begin{align}
C^{(r)} (\Vie{p}{}{},\Vie{p}{0}{}) = \frac{\varepsilon_1^{1/2} k_0 \, \Re \big( \alpha_1(\Vie{p}{}{}) \big)^2}{4 (2 \pi)^2 \, |\alpha_2(\Vie{p}{}{})|^2 \, \alpha_1(\Vie{p}{0}{})}  \: ,\\
C^{(t)} (\Vie{p}{}{},\Vie{p}{0}{}) = \frac{\varepsilon_2^{1/2} k_0 \, \Re \big( \alpha_2(\Vie{p}{}{}) \big)^2}{4 (2 \pi)^2 \, |\alpha_2(\Vie{p}{}{})|^2 \, \alpha_1(\Vie{p}{0}{})} \: .
\end{align}
\end{subequations}

\section{Derivation of the asymptotics of $I$ and $J$ (Table~\ref{tab1})} \label{App:asymptotics}

Let us start with the asymptotic expression of $J$ when $\ell_{\varepsilon \perp} \ll L$. We will use the short hand notation $\alpha = \pm \alpha_{2}(\Vie{p}{}{}) + \alpha_2(\Vie{p}{0}{})$ which we assume to be real. We have

\begin{align}
J &= \int_{-L}^0 \exp \left[ - \frac{(x_3+d)^2}{\ell_{\varepsilon \perp}^2}\right] \, \exp \left( i \alpha \, x_3 \right) \, \mathrm{d}x_3 \nonumber \\
&= \ell_{\varepsilon \perp} \: \int_{(-L+d) / \ell_{\varepsilon \perp}}^{d/\ell_{\varepsilon \perp}} \exp \left( - u^2 \right) \, \cos \left( \alpha \ell_{\varepsilon \perp} \, u - \alpha d \right) \, \mathrm{d}u \nonumber \\
&= \ell_{\varepsilon \perp} \: \cos (\alpha d)  \: \int_{(-L+d) / \ell_{\varepsilon \perp}}^{d/\ell_{\varepsilon \perp}} \exp \left( - u^2 \right) \, \cos \left( \alpha \ell_{\varepsilon \perp} \, u \right) \, \mathrm{d}u + \ell_{\varepsilon \perp} \: \sin (\alpha d)  \: \int_{(-L+d) / \ell_{\varepsilon \perp}}^{d/\ell_{\varepsilon \perp}} \exp \left( - u^2 \right) \, \sin \left( \alpha \ell_{\varepsilon \perp} \, u \right) \, \mathrm{d}u \: , \label{eq:Jcossin}
\end{align}
where we have made the change of variable $u = \frac{x_3+d}{ \ell_{\varepsilon \perp}}$. For $d = 0$, we obtain
\begin{align}
J &= \ell_{\varepsilon \perp}  \: \int_{-L / \ell_{\varepsilon \perp}}^{0} \exp \left( - u^2 \right) \, \cos \left( \alpha \ell_{\varepsilon \perp} \, u \right) \, \mathrm{d}u \nonumber\\
&= \frac{\ell_{\varepsilon \perp}}{2} \: \int_{-L / \ell_{\varepsilon \perp}}^{L/\ell_{\varepsilon \perp}} \exp \left( - u^2 \right) \, \cos \left( \alpha \ell_{\varepsilon \perp} \, u \right) \, \mathrm{d}u \: ,
\end{align}
where we have used the fact that the integrand is an even function. Since $\ell_{\varepsilon \perp} \ll L$ we can approximate the integral over $[-L /  \ell_{\varepsilon \perp},L /  \ell_{\varepsilon \perp}]$ by an integral over $\mathbb{R}$ and we obtain
\begin{equation}
J \sim \frac{\ell_{\varepsilon \perp}}{2}  \: \int_{\mathbb{R}} \exp \left( - u^2 \right) \, \cos \left( \alpha \ell_{\varepsilon \perp} \, u \right) \, \mathrm{d}u  = \frac{\sqrt{\pi}}{2} \ell_{\varepsilon \perp} \: \exp \left(-\frac{\alpha^2 \, \ell_{\varepsilon \perp}^2 }{4} \right) \: .
\end{equation}
For $\ell_{\varepsilon \perp} \ll d$ and $\ell_{\varepsilon \perp} \ll L - d$, we can replace the integration over $[(-L+d)/\ell_{\varepsilon \perp}, d/\ell_{\varepsilon \perp}]$ by an integration over $\mathbb{R}$ in Eq.~(\ref{eq:Jcossin}) and we get
\begin{equation}
J \sim \sqrt{\pi} \ell_{\varepsilon \perp} \: \exp \left(-\frac{\alpha^2 \, \ell_{\varepsilon \perp}^2 }{4} \right) \: \cos(\alpha d) \: ,
\end{equation}
where the second integral vanishes since the integrand is an odd function.

In the same asymptotic regime, the $I$ integral yields for $\alpha, \beta \in \mathbb{R}$
\begin{align}
I(\ell_{\varepsilon \perp},L,\alpha,\beta)  &= \int_{-L}^0 \int_{-L}^0 \exp \left[ - \frac{(x_3-x_3^\prime)^2}{\ell_{\varepsilon \perp}^2} \right] \: \exp \Big[ - i \alpha \, x_3 + i \beta x_3^{\prime} \Big] \, \mathrm{d}x_3 \, \mathrm{d}x_3^\prime \nonumber \\
&\approx \frac{1}{2} \, \ell_{\varepsilon \perp}^2 \:  \int_{-L/\ell_{\varepsilon \perp}}^{L /\ell_{\varepsilon \perp} } \int_{-L/\ell_{\varepsilon \perp}}^{L / \ell_{\varepsilon \perp}} \exp \left[ - (u-v)^2 \right] \: \exp \Big[ - i \alpha \, \ell_{\varepsilon \perp} \, u  + i \beta \, \ell_{\varepsilon \perp} \, v  \Big] \, \mathrm{d}u \, \mathrm{d}v \nonumber\\
&\sim \frac{1}{2} \, \ell_{\varepsilon \perp}^2 \:  \int_{-L/\ell_{\varepsilon \perp}}^{L/\ell_{\varepsilon \perp}} \int_{-\infty}^\infty \exp \left[ - w^2 \right] \: \exp \Big[ - i \alpha \, \ell_{\varepsilon \perp} \, w \Big] \, \mathrm{d}w \, \exp \Big[ i (\beta-\alpha) \ell_{\varepsilon \perp} v \Big] \,  \mathrm{d}v \nonumber\\
&= \sqrt{\pi} \, \ell_{\varepsilon \perp}^2 \: \exp \left(-\frac{\alpha^2 \ell_{\varepsilon \perp}^2}{4} \right) \, \frac{\sin \big( (\beta-\alpha) L \big)}{(\beta-\alpha) \ell_{\varepsilon \perp}} \: .
\end{align}
Taking the limit $\beta \to \alpha$, we also have
\begin{equation}
I(\ell_{\varepsilon \perp},L,\alpha,\alpha) \sim \sqrt{\pi} \, \ell_{\varepsilon \perp} \, L \: \exp \left(-\frac{\alpha^2 \ell_{\varepsilon \perp}^2}{4} \right) \: .
\end{equation}
If in addition $\ell_{\varepsilon \perp} \ll \lambda$, we get
\begin{equation}
I(\ell_{\varepsilon \perp},L,\alpha,\alpha) \sim \sqrt{\pi} \, \ell_{\varepsilon \perp} \, L \: .
\end{equation}
Now for $L \ll \ell_{\varepsilon \perp}$ we have

\begin{equation}
J = \int_{-L}^0 \exp \left[ - \frac{(x_3+d)^2}{\ell_{\varepsilon \perp}^2}\right] \, \cos \left( \alpha \, x_3 \right) \, \mathrm{d}x_3  \sim \int_{-L}^0 1 \, \cos \left(\alpha \, x_3 \right) \, \mathrm{d}x_3 = \frac{\sin \left( \alpha L \right)}{\alpha} \: .
\end{equation}
Here we have approximated the exponential $ \exp \left( - \frac{(x_3+d)^2}{\ell_{\varepsilon \perp}^2}\right) \approx 1$ for small arguments $\frac{(x_3+d)^2}{\ell_{\varepsilon \perp}^2} \leq \frac{L^2}{\ell_{\varepsilon \perp}^2} \ll 1$. If in addition $\lambda \gg L$, we have

\begin{equation}
J \sim L \: .
\end{equation}
In the same asymptotic regime, the $I$ integral reads
\begin{align}
I(\ell_{\varepsilon \perp},L,\alpha,\beta)  &= \int_{-L}^0 \int_{-L}^0 \exp \left[ - \frac{(x_3-x_3^\prime)^2}{\ell_{\varepsilon \perp}^2} \right] \: \exp \Big[ - i \alpha \, x_3 + i \beta x_3^{\prime} \Big] \, \mathrm{d}x_3 \, \mathrm{d}x_3^\prime \nonumber \\
&\sim \int_{-L}^0 \int_{-L}^0 1 \: \exp \Big[ - i \alpha \, x_3 + i \beta x_3^{\prime} \Big] \, \mathrm{d}x_3 \, \mathrm{d}x_3^\prime \nonumber \\
&= \int_{-L}^0 \exp \Big( - i \alpha \, x_3 \Big) \, \mathrm{d}x_3 \, \int_{-L}^0 \exp \Big( i \beta \, x_3^\prime \Big) \, \mathrm{d}x_3^\prime  \nonumber \\
&= \frac{4}{\alpha \beta} \: \sin \big( \alpha L / 2\big) \, \sin \big( \beta L / 2\big) \, \exp \Big( i(\beta -\alpha) L / 2 \Big) \: .
\end{align}
For $\beta = \alpha$, we have
\begin{equation}
I(\ell_{\varepsilon \perp},L,\alpha,\alpha) = \frac{4}{\alpha^2} \: \sin^2 \big( \alpha L / 2\big) \: .
\end{equation}
If in addition we assume $L \ll \lambda$ then
\begin{equation}
I(\ell_{\varepsilon \perp},L,\alpha,\alpha) \sim L^2 \: .
\end{equation}


\section{Scattering mean free path}\label{App:mean_free_path}

We recall here the derivation of the scattering mean free path $\ell_s$ for an infinite medium with a fluctuating dielectric function $\varepsilon (\Vie{x}{}{}) = \varepsilon_0 + \Delta \varepsilon(\Vie{x}{}{})$ characterized by
\begin{subequations}
\begin{align}
\left\langle \Delta \varepsilon (\Vie{x}{}{}) \right\rangle &= 0 \\
\left\langle \Delta \varepsilon (\Vie{x}{}{}) \Delta \varepsilon (\Vie{x}{}{\prime}) \right\rangle &= \sigma_\varepsilon^2 \: W_\varepsilon (\Vie{x}{}{}-\Vie{x}{}{\prime}) = \sigma_\varepsilon^2 \: \exp \Bigg[- \sum_{j=1}^3 \frac{(x_j-x_j^\prime)^2}{\ell_{\varepsilon j}^2} \Bigg] \: .
\end{align}
\end{subequations}
Here $\varepsilon_0$ and $k_0$ denote respectively the average dielectric constant (homogeneous background) and the corresponding wave number. For a continuously fluctuating dielectric function, the scattering mean free path is linked to the three-dimensional Fourier transform of the autocorrelation function of $\Delta \varepsilon$ as follows \cite{Akkermans:Montambaux}
\begin{equation}
\ell_s^{-1} (\mathbf{u}) = \frac{k_0^4}{16 \pi^2} \, \int_{4 \pi} \sigma_\varepsilon^2 \, \hat{W}_\varepsilon \left[ k_r \big( \Vie{u}{}{\prime} - \Vie{u}{}{}) \right] \: \mathrm{d}\Omega^\prime \: ,
\label{eq:ls}
\end{equation}
where $k_r$ is the wave number in the effective medium, $\Vie{u}{}{}$ and $\Vie{u}{}{\prime}$ are vectors on the unit sphere and the integration is over $\Vie{u}{}{\prime}$. We have used Eq.~(\ref{eq:ls}) to evaluate numerically the scattering mean free path for $\Vie{u}{}{} = \Vie{\hat{e}}{3}{}$ and deduce the optical thickness for anisotropic dielectric fluctuations in the examples shown in the present paper. In the case of isotropic correlation, i.e., $\ell_1 = \ell_2 = \ell_3$ and $\hat{W}(\Vie{k}{}{}) = \hat{W}(|\Vie{k}{}{}|)$, the scattering mean free path is independent of $\Vie{u}{}{}$, and by the use of the angle $\theta$ between  $\Vie{u}{}{}$ and $\Vie{u}{}{\prime}$ and of a change of variables $q = k_r | \Vie{u}{}{\prime} - \Vie{u}{}{} |$, the scattering mean free path reads
\begin{equation}
\ell_s^{-1} = \frac{\sigma_\varepsilon^2 k_0^4}{8 \pi k_r^2} \, \int_{0}^{2 k_r}  \, \hat{W}_\varepsilon (q) \: q \, \mathrm{d}q \: .
\end{equation}
For a Gaussian correlation function we thus have
\begin{equation}
\ell_s^{-1} = \frac{\sigma_\varepsilon^2 k_0^4}{8 \pi k_r^2} \, \int_{0}^{2 k_r}  \, \pi^{3/2} \, \ell_\varepsilon^3 \, \exp \left( - \frac{q^2 \ell_\varepsilon^2}{4} \right) \: q \, \mathrm{d}q = \frac{\pi^{1/2} \sigma_\varepsilon^2 k_0^4 \ell_\varepsilon}{4 k_r^2} \: \Big[ 1 - \exp \left(- k_r^2 \ell_\varepsilon^2 \right) \Big] \: ,
\end{equation}
which in the regime $k_r \ell_\varepsilon \ll 1$ leads to
\begin{equation}
\ell_s^{-1} = \frac{\pi^{1/2}}{4} \,  \sigma_\varepsilon^2 k_0^4 \ell_\varepsilon^3 \: .
\end{equation}

\end{widetext}
%
\bibliography{biblio}

\end{document}